\documentclass[aps,prb,showpacs,twocolumn,superscriptaddress]{revtex4-2}
\usepackage{bm,color,amsmath,amssymb,mathrsfs,latexsym,graphicx,psfrag,mathtools}
\usepackage{color}
\usepackage[dvipsnames]{xcolor}
\usepackage[hidelinks,colorlinks,linkcolor=blue,
citecolor=blue,urlcolor=blue]{hyperref}
\usepackage{dsfont}
\usepackage{float}
\usepackage{hyperref}
\usepackage{empheq}

\newcommand{\bsl}[1]{\boldsymbol{#1}}

\renewcommand{\mod}{\,\mathrm{mod}\,}

\newcommand{\bra}[1]{\langle #1|}
\newcommand{\ket}[1]{|#1 \rangle}
\newcommand{\braket}[2]{\left\langle #1 | #2  \right\rangle}

\DeclareRobustCommand{\Eq}[1]{Eq.~(\ref{#1})}
\newcommand{\ii}{\mathrm{i}}

\newcommand{\Tr}{\mathop{\mathrm{Tr}}}

\newcommand{\eqnref}[1]{Eq.\,\eqref{#1}}
\newcommand{\figref}[1]{Fig.\,\ref{#1}}
\newcommand{\figsref}[1]{Figs.\,\ref{#1}}
\newcommand{\tabref}[1]{Tab.\,\ref{#1}}
\newcommand{\secref}[1]{Sec.\,\ref{#1}}
\newcommand{\appref}[1]{App.\,\ref{#1}}
\newcommand{\refcite}[1]{Ref.\,\cite{#1}}

\newcommand{\eq}[1]{\begin{equation} #1 \end{equation}}

\newcommand{\eqa}[1]{\begin{align}\begin{split} #1 \end{split}\end{align}}
\DeclareRobustCommand{\Tab}[1]{Table~\ref{#1}}

\let\oldAA\AA
\renewcommand{\AA}{\text{\normalfont\oldAA}}

\newcommand{\ie}{{\emph{i.e.}}}
\newcommand{\eg}{{\emph{e.g.}}}

\DeclareRobustCommand{\App}[1]{App.~\ref{#1}}
\DeclareRobustCommand{\Sec}[1]{Sec.~\ref{#1}}

\newcommand{\K}{\text{K}}

\newcommand{\bea}{\begin{equation} \begin{aligned}}
\newcommand{\eea}{\end{aligned} \end{equation} }

\newcommand{\al}{\alpha}
\newcommand{\be}{\beta}
\newcommand{\bpm}{\begin{pmatrix}}
\newcommand{\epm}{\end{pmatrix}}
\newcommand{\eps}{\epsilon}

\renewcommand{\th}{\theta}

\newcommand{\lp}{\left(}
\newcommand{\rp}{\right)}
\newcommand{\del}{\partial}
\newcommand{\mbf}[1]{\mathbf{#1}}

\DeclareRobustCommand{\Fig}[1]{Fig.~\ref{#1}}
\usepackage{braket}
\usepackage{amsmath}
\usepackage{amssymb}
\usepackage{wasysym}
\usepackage{mathrsfs}

\newcommand{\meV}{\text{meV}}
\newcommand{\nm}{\text{nm}}

\usepackage{comment}

\usepackage{commath}    
\usepackage{amsmath}

\begin{document}

\title{Moir\'e Fractional Chern Insulators II: First-principles Calculations and Continuum Models of Rhombohedral  Graphene Superlattices}

\author{Jonah Herzog-Arbeitman}
\thanks{These authors contributed equally.}
\affiliation{Department of Physics, Princeton University, Princeton, New Jersey 08544, USA}

\author{Yuzhi Wang}
\thanks{These authors contributed equally.}
\affiliation{Beijing National Laboratory for Condensed Matter Physics and Institute of physics,
Chinese academy of sciences, Beijing 100190, China}
\affiliation{University of Chinese academy of sciences, Beijing 100049, China}

\author{Jiaxuan Liu}
\thanks{These authors contributed equally.}
\affiliation{Beijing National Laboratory for Condensed Matter Physics and Institute of physics,
Chinese academy of sciences, Beijing 100190, China}
\affiliation{University of Chinese academy of sciences, Beijing 100049, China}

\author{Pok Man Tam}
\thanks{These authors contributed equally.}
\affiliation{Princeton Center for Theoretical Science, Princeton University, Princeton, NJ 08544}

\author{Ziyue Qi}
\thanks{These authors contributed equally.}
\affiliation{Beijing National Laboratory for Condensed Matter Physics and Institute of physics,
Chinese academy of sciences, Beijing 100190, China}
\affiliation{University of Chinese academy of sciences, Beijing 100049, China}

\author{Yujin Jia}
\affiliation{Beijing National Laboratory for Condensed Matter Physics and Institute of physics,
Chinese academy of sciences, Beijing 100190, China}
\affiliation{University of Chinese academy of sciences, Beijing 100049, China}

\author{Dmitri K. Efetov}
\affiliation{Faculty of Physics, Ludwig-Maximilians-University Munich, Munich 80799, Germany}
\affiliation{Munich Center for Quantum Science and Technology (MCQST), Ludwig-Maximilians-University Munich, Munich 80799, Germany}

\author{Oskar Vafek}
\affiliation{National High Magnetic Field Laboratory, Tallahassee, Florida, 32310, USA}
\affiliation{Department of Physics, Florida State University, Tallahassee, Florida 32306, USA}

\author{Nicolas Regnault}
\affiliation{Department of Physics, Princeton University, Princeton, New Jersey 08544, USA}
\affiliation{Laboratoire de Physique de l’Ecole normale sup\'erieure,
ENS, Universit\'e PSL, CNRS, Sorbonne Universit\'e,
Universit\'e Paris-Diderot, Sorbonne Paris Cit\'e, 75005 Paris, France}

\author{Hongming Weng}
\affiliation{Beijing National Laboratory for Condensed Matter Physics and Institute of physics,
Chinese academy of sciences, Beijing 100190, China}
\affiliation{University of Chinese academy of sciences, Beijing 100049, China}
\affiliation{Songshan Lake Materials Laboratory, Dongguan, Guangdong 523808, China}

\author{Quansheng Wu}
\email{quansheng.wu@iphy.ac.cn}
\affiliation{Beijing National Laboratory for Condensed Matter Physics and Institute of physics,
Chinese academy of sciences, Beijing 100190, China}
\affiliation{University of Chinese academy of sciences, Beijing 100049, China}

\author{B. Andrei Bernevig}
\email{bernevig@princeton.edu}
\affiliation{Department of Physics, Princeton University, Princeton, New Jersey 08544, USA}
\affiliation{Donostia International Physics Center, P. Manuel de Lardizabal 4, 20018 Donostia-San Sebastian, Spain}
\affiliation{IKERBASQUE, Basque Foundation for Science, Bilbao, Spain}

\author{Jiabin Yu}
\email{jiabinyu@princeton.edu}
\affiliation{Department of Physics, Princeton University, Princeton, New Jersey 08544, USA}

\begin{abstract}
The experimental discovery of fractional Chern insulators (FCIs) in rhombohedral pentalayer graphene twisted on hexagonal boron nitride (hBN) has preceded theoretical prediction. Supported by large-scale first principles relaxation calculations at the experimental twist angle of $0.77^\circ$, we obtain an accurate continuum model of $n=3,4,5,6,7$ layer rhombohedral graphene-hBN moir\'e systems. Focusing on the pentalayer case, we analytically explain the robust $|C|=0,5$ Chern numbers seen in the low-energy single-particle bands and their flattening with displacement field, making use of a minimal two-flavor continuum Hamiltonian derived from the full model. We then predict nonzero valley Chern numbers at the $\nu = -4,0$ insulators observed in experiment. Our analysis makes clear the importance of displacement field and the moir\'e potential in producing localized "heavy fermion" charge density in the top valence band, in addition to the nearly free conduction band. Lastly, we study doubly aligned devices as additional platforms for moir\'e FCIs with higher Chern number bands.
\end{abstract}
\maketitle

\section{Introduction}

Fractional Chern insulators (FCIs) have long been defined as topologically ordered states arising from partial filling of a Chern insulator \cite{neupert,sheng,regnault}. While it was once thought that such physics could not be observed without nonzero magnetic field to stabilize the state \cite{2018Sci...360...62S,2021Natur.600..439X}, now two separate moir\'e platforms, first in MoTe$_2$~\cite{park2023observation,zeng2023integer,Xu2023FCItMoTe2,cai2023signatures} and second in pentalayer graphene~\cite{Ju2023PentalayerGraphenehBN}, have shown that innate band topology combined with spontaneous spin/valley polarization can reproduce some -- but not yet all --  of the FCI phase diagram in its original setting at strictly zero magnetic field. These FCI states have been shown to exhibit a fractional quantum anomalous Hall effect in transport experiments. Importantly, experiments show deviations from the typical FCI phase diagram caused by the moir\'e potential, which favors competing states at different fillings~\cite{Yu2023FCI}, and allows many different phases and phase transitions to be observed in the same device tuned by filling and displacement field. This offers an unprecedented opportunity to make and check theoretical predictions~\cite{reddy2023fractional,wang2023fractional,Dong2023CFLtMoTe2,Goldman2023CFLtMoTe2,Reddy2023GlobalPDFCI,Xu2023MLWOFCItTMD,Zaletel2023tMoTe2FCI,Yu2023FCI,Fu2023BandMixingFCItMoTe2,Fengcheng2023tMoTe2HFnum1,Zhang2023MoTe2}. In a prior publication \cite{Yu2023FCI} regarding FCIs in twisted bilayer MoTe$_2$, we have shown that band mixing effects are important to resolve the competition (including FCI, spin polarization, etc) in the phase diagram, and accurate models derived from ab-initio calculations are required. Many sets of single-particle parameters~\cite{Wu2019TIintTMD,reddy2023fractional,wang2023fractional,Xu2023MLWOFCItTMD,Zhang2023MoTe2} exist, and hence establishing good single-particle models is the initial step of an interacting analysis. While our interacting calculations \cite{Yu2023FCI} can capture several (though not all) key features of the FCIs and spin polarizations present in the experiments~\cite{park2023observation,zeng2023integer,Xu2023FCItMoTe2,cai2023signatures} for one existing set of parameters \cite{wang2023fractional} of the known model~\cite{Wu2019TIintTMD}, we showed in the first paper of the current series \cite{firstMFCI} that there could be changes to the single-particle parameters and model, which may be crucial for a full understanding of the experimental phase diagram of twisted bilayer MoTe$_2$.

This paper focuses on another moir\'e system, rhombohedral pentalayer graphene twisted on hexagonal boron nitride (hBN)\cite{Ju2023PentalayerGraphenehBN}.
FCIs were recently observed in this system after their initial observation in MoTe$_2$, as evidenced by the fractional quantum anomalous Hall effect in transport measurements and fractional slopes down to zero magnetic field in the Wannier diagram. 
It is the purpose of this paper to develop a reliable model of the single-particle band structure in preparation for many-body calculations. Hence, we start from large-scale first principles calculations of rhombohedral graphene at the commensurate angle $\th = 0.76715^\circ$ consistent with the experimental value~\cite{Ju2023PentalayerGraphenehBN}. This step is crucial~\cite{firstMFCI} to obtain correctly relaxed structures, a feature so far not included in existing calculations~\cite{PhysRevB.89.205414,PhysRevB.90.155406,PhysRevB.99.075127,Park2023RMGhBNChernFlatBands,dong2023theory,zhou2023fractional,dong2023anomalous}.
Based on the relaxed structure, we then use the Slater-Koster (SK) tight-binding model~\cite{Slater1954LCAO} to calculate the band structure of the rhombohedral $n$-layer graphene twisted on hBN (R$n$G/hBN), where the SK parameters are fit to density functional theory (DFT) bands of pristine rhombohedral $n$-layer graphene. We perform these DFT+SK calculations for $n=3,4,..,7$ layers, in two distinct stacking configurations of the hBN, and in various displacement fields. Lastly, we undertake similar DFT+SK calculations on rhombohedral $n$-layer graphene encapsulated by two nearly-aligned (twist angle $0.76715^\circ$) hBN (hBN/R$n$G/hBN) with parallel moir\'e patterns.
We find that the the low-energy moir\'e bands of the relaxed structure have as much as $\sim$10meV differences from those of the rigid structures, which means that relaxation is not negligible.

We further revisit the continuum model of R$n$G/hBN and hBN/R$n$G/hBN proposed in Refs.~\cite{PhysRevB.89.205414,Park2023RMGhBNChernFlatBands}, which is a continuum model acts on the $2n$ R$n$G orbitals (after integrating out the hBN) for $n$ layers.
We find that the non-uniform potential of the moir\'e pattern can be reduced to have only one single complex parameter under the first-harmonic approximation, owing to the layer and sublattice polarization of the low-energy states.
Our finding justifies the simple form of the moir\'e potential.
We determine the values of the parameters in the $2n\times 2n$ continuum model through the Fourier transformation of the SK hoppings and fitting to the DFT+SK bands. With these  parameter values, the dispersion of the $2n\times 2n$ continuum model matches the DFT+SK bands remarkably well.
We further plot the single-particle phase diagram (as function of the twist angle and displacement field) of the $2n\times 2n$ continuum model in one valley (and one spin), and find that $\pm 1$ Chern numbers of the lowest conduction or the highest valence bands can only appear in the hBN/R$n$G/hBN structures.
The R$n$G/hBN structures only have $0$ or $n$ Chern numbers in their the lowest conduction or the highest valence bands in a considerable range of angles and displacement fields.
In particular, in the large displacement fields regime that is relevant to Chern insulators and FCIs observed in R5G/hBN~\cite{Ju2023PentalayerGraphenehBN}, our model suggests that the low-energy conduction bands feel little effect from the moir\'e potential, which is consistent with \refcite{zhou2023fractional,dong2023anomalous}. However, the moir\'e potential has strong effect on the highest valence band, making it trivial atomic with a localized charge distribution reminiscent of a heavy fermion \cite{2022PhRvL.129d7601S}.
Finally, we build a $2\times 2$ effective continuum model by applying perturbation theory to the two lowest energy states of R$n$G $2n\times 2n$ model, which exhibit perfect sublattice polarization, exponentially good layer polarization, and a holomorphic/anti-holomorphic structure due to chiral symmetry. This basis provides a direct understanding of our numerical results, and allows us to obtain an analytic understanding of the topology of the low-energy bands.

In the remainder of this paper, we first present our DFT+SK calculations on R$n$G/hBN structures in \Sec{sec:DFT} before discussing the $2n\times 2n$ continuum model (with the hBN integrated out) in \Sec{fullmodels}. From this model, we derive an $2\times 2$ effective two-flavor model built on the low energy chiral R$n$G states, which we use to explain the single particle phase diagram analytically in \Sec{PDsec}. We further discuss the DFT+SK calculations and the continuum models for hBN/R$n$G/hBN structures, which can isolate $\pm 1$ Chern bands at the single particle level, in \Sec{doublyaligned}.
We conclude our paper in \secref{sec:conclusion}, and provide details of the paper in a series of appendices.
Throughout the work, we will neglect spin unless specified otherwise.

\section{First Principles Results for moir\'e Structures}
\label{sec:DFT}

We first discuss the experimental setup of the R5G/hBN device based on which the first principle calculations are performed. The full data set available in Ref.\,\cite{Ju2023PentalayerGraphenehBN} is consistent with only a single moir\'e pattern coupling to the graphene: all gapped states at positive and negative displacement field appear at commensurate fillings set by the moir\'e unit cell size, where negative and positive field point away from and toward the nearly-aligned hBN, respectively. As argued in Ref.\,\cite{Ju2023PentalayerGraphenehBN}, the moir\'e unit cell and twist angle $\th = 0.77^\circ$ can be determined by assuming that the strongest gap observed at filling $\nu = +4$ for negative displacement field (which points toward hBN) is given by the single-particle gap (possibly enhanced by interaction while having no valley/spin polarization). 

\begin{figure*}
    \centering
\includegraphics[width=1.0\linewidth]{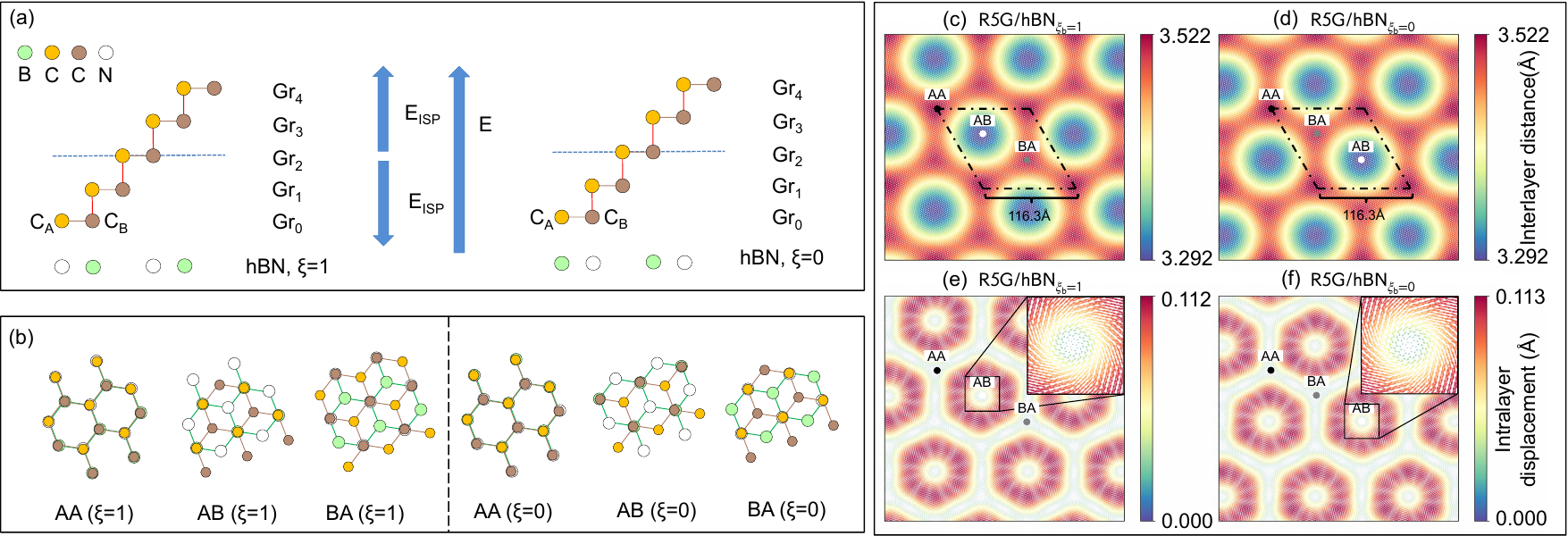}

    \caption{(a) Two different stacking configurations for R5G/hBN. In the $\xi=1$ setup, nitrogen (N) atoms align with carbon C$_A$ of the closest graphene layer (Gr$_0$) in the AA region; in the $\xi=0$ setup, boron (B) atoms align with C$_A$ of Gr$_0$ in the AA region. $E$ indicates the applied electrical ﬁeld (with arrow labeling the positive direction) and ISP indicates the direction internal symmetrical polarization due to the different chemical environment of outer and inner atoms in the graphene.  (b) shows different regions in the moir\'e pattern for each stacking configuration. In the AA region, both boron and nitrogen are aligned with carbon atoms in lowest layer (denoted Gr$_0$). In AB regions, only boron atoms are aligned with carbon atoms in Gr$_0$. In BA region, only nitrogen atoms are aligned with the carbon atoms in Gr$_0$. (c) and (d) show the inter-layer distance between hBN and Gr$_0$ for $\xi =1$ and $\xi = 0$ respectively. (e) and (f) show the intra-layer displacement of Gr$_0$ for $\xi =1$ and $\xi = 0$ respectively
    }
    \label{relaxation}
\end{figure*}

The presence of one relevant moir\'e pattern within the doubly encapsulated device is consistent with a single nearly-aligned hBN layer at $\th \approx 0.77^\circ$ with the opposing layer unaligned and electronically decoupled. Even with this assumption, there are two microscopically distinct structures dependent on the stacking configurations of graphene on hBN due to the broken $C_{2}$ symmetry (with axis perpendicular to the sample) of hBN as well as rhombohedral graphene. These configurations are labeled by $\xi = 0,1$, where $\xi = 0$ (resp. $\xi = 1$) mean that the carbon A/B sublattice of the lowest graphene layer is on top of boron/nitrogen (resp. nitrogen/boron) in the AA region of the moir\'e pattern  as shown in \figref{relaxation}(a). 
We will first focus on these two R$n$G/hBN configurations to study the effects of relaxation. Our work builds on that of Refs.\,\cite{PhysRevB.89.205414} and \cite{Park2023RMGhBNChernFlatBands} where relaxation in the full moir\'e unit cell has not been considered.

\begin{figure}
    \centering
\includegraphics[width=1.0\linewidth]{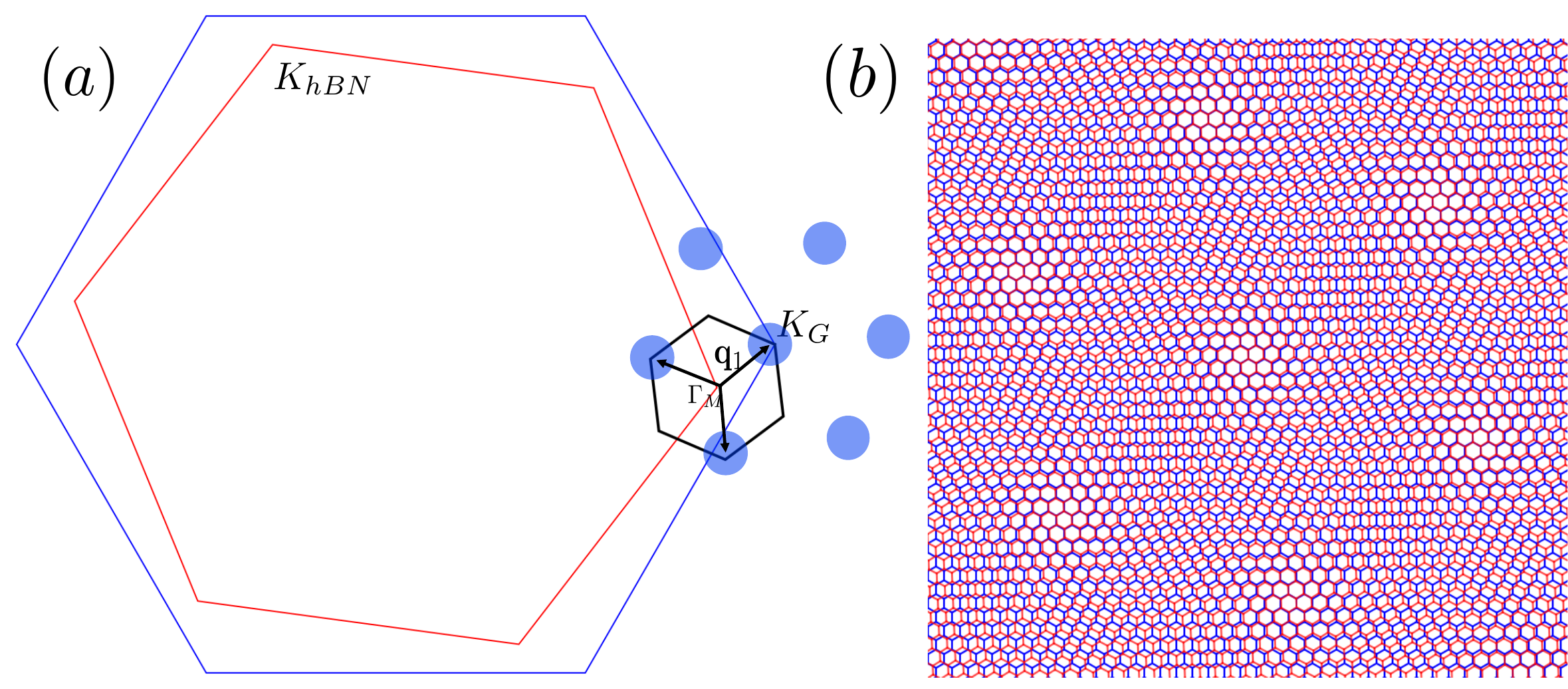}
    \caption{$(a)$ Crystalline and moir\'e BZs for rhombohedral graphene (blue) and hBN (red). The $K$ points of both materials are labeled, and the blue disks highlight the first shell of moir\'e reciprocal lattices around $K_G$. The $\mbf{q}_1$ vector and its $C_3$ partners are shown in orange about the moir\'e $\Gamma_M$ point.
    $(b)$ Real space lattice for 7 moir\'e unit cells.}
\label{moireconventions}
\end{figure}

In our setup, we take the rhombohedral graphene with lattice constant $a_G = 2.46\AA$ to be situated on top of the hBN with lattice constant to be $a_{hBN} = 2.50\AA = (1+\eps_{rigid})a_G$ with $\eps_{rigid} = 0.0163$. The difference between the graphene $\K$ point and hBN $\K$ point rotated clockwise by $\th$ is
\bea
\label{eq:qvecmain}
\mbf{q}_1 = \mbf{K}_G - \mbf{K}_{hBN} = \frac{4\pi}{3 a_G} (1 - \frac{R(-\th)}{1+\eps})\hat{x}
\eea
with its $C_3$ partners $\mbf{q}_{j+1} = R(\frac{2\pi}{3}) \mbf{q}_j$, where $\th$ is the twist angle and $R(\th)$ is a rotation matrix. The moir\'e reciprocal lattice vectors are $\mbf{b}^M_j = \mbf{q}_3 - \mbf{q}_j$ for $j =1,2$. The commensuration condition $m \mbf{b}^M_1 + n \mbf{b}^M_2 = \mbf{b}^G_1$ for $\mbf{b}^G_1 = \frac{4\pi}{\sqrt{3} a_G} \hat{y}$ the graphene reciprocal lattice vector results in configurations labeled by $m,n \in \mathds{Z}$  (for $\eps >0$):
\bea
1+ \eps &= \Big(1 + \frac{2n+m+1}{m^2+m n +n^2} \Big)^{-1/2}  \\
\tan (-\th) &= \frac{\sqrt{3} m}{2(m^2+m n +n^2) +2n + m} \\
\eea
and we pick $(m,n) = (34,-54)$ so that $\eps = 0.01673$ (close to $\eps_{rigid}$) for $\th = 0.76715^\circ$, the experimental angle. 
Using the formulae, we find that the graphene $\K$ point is folded to
\bea
\label{eq:K_G_folding}
\mbf{K}_G = \left[ \frac{m-n}{3} \mod\ 1 \right](\mbf{b}^M_1 +\mbf{b}^M_2)
\eea
within the moir\'e BZ. For the commensurate configuration $(m,n) = (34,-54)$ used here, $\mbf{K}_G$ folds onto the moir\'e $\K_M$ point (see \Fig{moireconventions}).

\begin{figure}
    \centering    \includegraphics[width=1.0\linewidth]{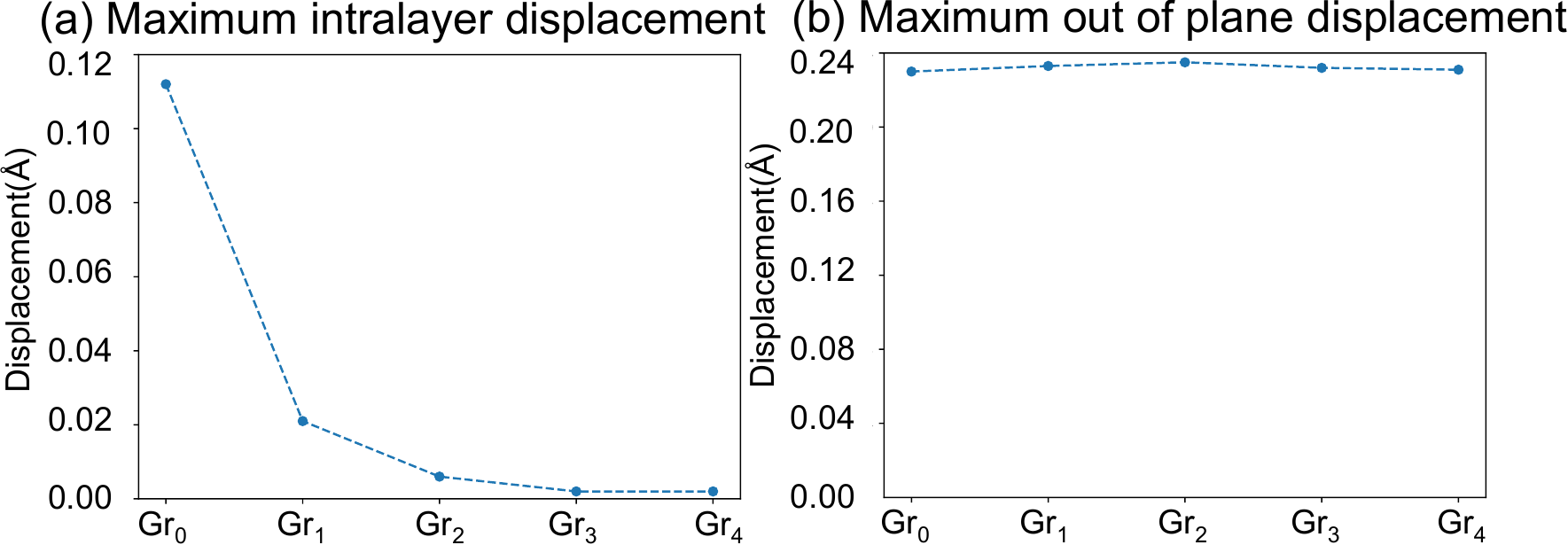}
    \includegraphics[width=1.0\linewidth]{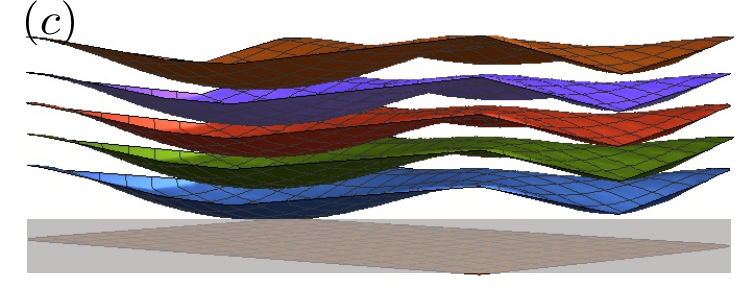}
     \caption{Relaxation of R5G/hBN for $\xi = 1$. $(a)$ The maximum in-plane displacement vector magnitude of each layer, which decays rapidly each layer away from the hBN. $(b)$ Unlike the intra-layer relaxation, the out-of-plane displacement, which measures the displacement along the $z$ direction, remains constant. $(c)$ Depiction of the relaxed pentalayer graphene on hBN (gray) with the out-of-plane relaxation \emph{increased} by a factor of 20 for visibility. We see that the profile of the out-of-plane relaxation remains throughout the device so that the inter-layer distances \emph{between} neighboring graphene layers stay roughly constant.}
    \label{displacement}
    \label{outofplane}
\end{figure}

We start by performing first-principle structural relaxation of the R$n$G/hBN commensurate superlattice with a classical force field as implemented in LAMMPS~\cite{thompson_lammps_2022}. During the relaxation, we held the hBN layer fixed to simulate a thick substrate, and the moir\'e unit cell was preserved. Two empirical interatomic potentials are used to perform the relaxation. For intra-layer interaction within graphene layers, we used the reactive empirical bond order potential~\cite{brenner_second-generation_2002}. For inter-layer interaction, we used an inter-layer potential developed for graphene and hBN systems\cite{ouyang_nanoserpents_2018}. 

The relaxation results of the the lowest graphene layer which has direct contact with the hBN is shown in \Fig{relaxation}(b-f) for configurations $\xi = 0,1$. The out-of-plane (inter-layer) relaxation is shown in \Fig{relaxation}(c,d) and the in-plane (intra-layer) relaxation is shown in \Fig{relaxation}(e,f). 
The inter-layer distance between the hBN and lowest graphene layer is in the range of 3.29\AA-3.52\AA. The distance is minimal at the AB point and reaches a maximum at AA region (with a similar value at the BA region). From \Fig{relaxation}(e,f), we see the intra-layer displacement shows that the Carbon atoms near AB region tend to rotate in the clockwise direction with respect to hBN, against the global twist (counterclockwise). This enlarges the AB region. Both trends indicate that the local stacking of $C_A$, the A-sublattice carbon, on top of the boron atom is energetically favorable, and the bilayer relaxes to maximize this stacking at the expense of the AA and BA regions. This is expected because of the polarity of the boron-nitrogen bond, where the more electro-negative nitrogen accepts boron's valence electrons to fill its $p$-orbital shell. Thus the lowest unoccupied molecular orbital is localized to the boron. Furthermore, as we will show in \Sec{fullmodels}, the low-energy wavefunctions of the pentalayer graphene have exponentially larger weight on the $C_A$ than $C_B$ of the lowest graphene layer. Thus the AA and BA regions, where $C_A$ is aligned to nitrogen, is disfavored, and the AB region, where $C_A$ is on top of boron, is favored. 

\begin{figure}[!htpb]
    \centering
\includegraphics[width=1.0\linewidth]{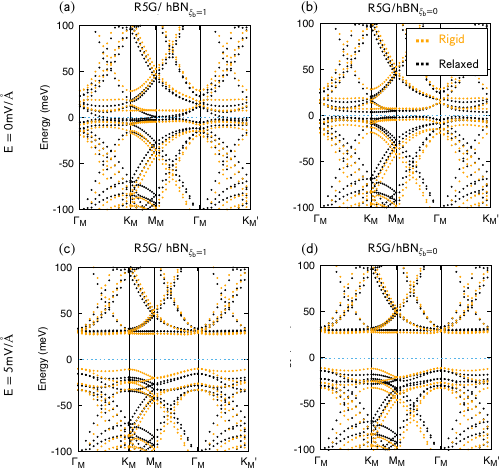}
    \caption{Comparison between rigid SK and relaxed DFT+SK band structures for 0.76715$^\circ$ twist R5G/hBN in the $\xi=0$ and $\xi=1$ configurations under different external electric field strengths, depicted with orange dotted lines for rigid structures and black dotted lines for relaxed structures. (a) $E = 0$ and R5G/hBN$_{\xi_b=1}$, (b) $E = 0$ and R5G/hBN$_{\xi_b=0}$, (c) $E = 5$ mV/\AA\ and R5G/hBN$_{\xi_b=1}$, (d) $E = 5$ mV/\AA\ and R5G/hBN$_{\xi_b=0}$, respectively. Here, both K and K' valley bands are included. 
    }
    \label{fig:rigid-relaxed-5Gr-main}
\end{figure}
We find that the graphene intra-layer relaxation decays quickly away from the hBN substrate (see \App{app:relaxation}). However, the out-of-plane relaxation pattern of the bottom-most layer persists throughout the device (see \Fig{outofplane}) so that the inter-layer graphene distances remain constant for each $\mbf{r}$. 
This is consistent with recent measurements in the graphene-graphite moir\'e system~\cite{2023Natur.620..750W}, where it was found that the moir\'e potential affects all layers of the graphitic thin film.
Note that out-of-plane modulation $\mbf{r} \to \mbf{r} + h(\mbf{r})\hat{z}$ of the graphene sheet does not couple at first order to the Dirac cones due to the effective mirror symmetry of a single graphene sheet \cite{2010PhR...496..109V}, and therefore the effect of spatially varying $h(\mbf{r})$ can be neglected. We have verified that this pattern extends out to $n=7$ layers in the relaxed structures. 

A comparison of the relaxed and rigid band structures can be found in \figref{fig:rigid-relaxed-5Gr-main} for 5 layers. (See the comparison for other layers in Figs.\,\ref{compare-rigid-relaxed-3Gr-with-t2.pdf}-\ref{compare-rigid-relaxed-7Gr-with-t2.pdf} in App. \ref{app:band_structure}.) While the qualitative features of the bands remain similar between the relaxed and rigid structures, there are non-negligible quantitative changes of the band structure, \eg, $\xi = 1$ case shows a significantly reduced gap at charge neutrality (the change is around $\sim $10meV). Finally, we have also computed the valley-resolved band structures by making use of the emergent valley symmetry at low energy \cite{PhysRevResearch.2.033357}, with the results and details of construction summarized in \App{app:valley_resolved}. 
To better understand these band structures and elucidate the physics behind them, we now turn the analysis of a continuum model.

\section{Continuum Model}
\label{fullmodels}

We now discuss the moir\'e continuum Hamiltonian for R$n$G/hBN.
The continuum model of the moir\'e system takes of form of a Bistritzer-MacDonald (BM) Hamiltonian \cite{2011PNAS..10812233B}
From \Eq{eq:qvecmain}, the moir\'e scattering momentum $ |\mbf{q}_1| \approx \frac{4\pi}{3 a_G} \sqrt{\eps^2 + \th^2}$ becomes small in the $\eps,\th \to 0$ limit, and the two low-energy valleys $\K,\K' = \pm \mbf{K}_G$ of the rhombohedral graphene become decoupled. In this limit, valley becomes a good quantum number and we can build seperate models for the $\K$ and $\K'$ valley bands. We only discuss $H_{\K, \xi}(\mbf{r})$, the Hamiltonian for the $\K$ valley states (given a stacking configuration $\xi = 0,1$) since $H_{\K', \xi}(\mbf{r}) = H^*_{\K, \xi}(\mbf{r})$ is obtained from the spin-less time-reversal symmetry of graphene. For reference, $\eps/ \sqrt{\eps^2 + \th^2} = 0.8$ at the experimental $\th = 0.77^\circ$, showing that the effect of the twist angle is to enlarge the moir\'e lattice constant by about $20\%$. 

The continuum model of the system takes the following Bistritzer-MacDonald form \cite{2011PNAS..10812233B}:
\bea
H_{\K,\xi} &= \bpm
H_{\K}(-i \pmb{\nabla}) & \tilde{T}^\dag(\mbf{r}) \\
 \tilde{T}(\mbf{r}) & \sigma_1^\xi H_{BN} \sigma_1^\xi \\
\epm, \qquad \xi = 0,1
\eea
where $H_{BN} = \text{diag}(V_B, V_N)$ is the Hamiltonian of the nearly aligned hBN with all $\mbf{k}$-dependence neglected in comparison to the large potentials \bea
V_B = 3352\text{ meV}, \quad V_N = -1388\text{meV}, 
\eea
$H_{\K}$ is the $\mbf{k} \cdot \mbf{p}$ Hamiltonian of pentalayer graphene to be discussed momentarily, and $[\tilde{T}(\mbf{r})]_{l} = \delta_{l,0}T(\mbf{r})$ is the moir\'e coupling between bottom graphene layer and hBN\cite{PhysRevB.89.205414,Park2023RMGhBNChernFlatBands}. Here $l = 0,\dots, n-1$ indexes the layers, and $l=0$ is the lowest layer in contact with the hBN.

\begin{figure*}
    \centering
\includegraphics[width=1.0\linewidth]{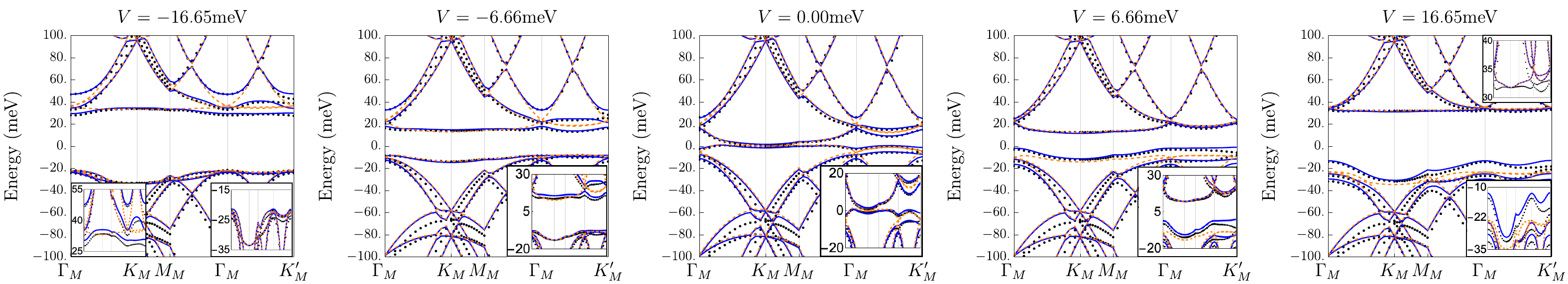}
    \caption{
    Band structures of R5G/hBN in the $\xi=1$ configuration at twist angle $\theta=0.76715^\circ$ for inter-layer potential energy differences $V=-16.65,-6.66,0,6.66,16.65\;\text{meV}$ in $\K$ valley. The black dots are from the first-principles calculation, and the lines are from the $10\times 10$ continuum model with (blue solid) and without (orange dashed) moir\'e potential. For $V=16.65$\text{meV}, it is clear that the conduction bands are almost 
    free from the moir\'e. In fact, the average overlap of the projectors onto the lowest conduction band with and without moir\'e is $\langle \Tr[P_c(\mbf{k}) P_c^{(0)}(\mbf{k})] \rangle = 0.999$, with the average computed over region of the mBZ where the lowest conduction band has a least 0.5 meV direct gap to nearby bands such that the projectors are well-defined. The highest valence band, on the other hand, responds strongly to the moir\'e potential.
    }
    \label{fig:main_2n}
\end{figure*}

Due to the large chemical potentials of the $N$ and $B$ orbitals, we use second order perturbation theory to integrate them out~\cite{PhysRevB.89.205414,PhysRevB.90.155406} and obtain the potential term
\bea
\tilde{V}_\xi(\mbf{r}) &= - \tilde{T}^\dag(\mbf{r}) \frac{1}{\sigma_1^\xi H_{BN} \sigma_1^\xi} \tilde{T}(\mbf{r})
\eea
acting only on the bottom graphene layer. We obtain a graphene-only model in the form
\bea \label{eq:fullmodel}
H_{\K,\xi} &= H_{\K}(-i \pmb{\nabla}) + \tilde{V}_\xi(\mbf{r}) \ .
\eea 
We will now derive the form of $H_{\K}$ and $\tilde{V}_\xi(\mbf{r})$. We first discuss $H_{\K}$ in \Sec{sec:pentapure} to derive its low energy structure, which organizes our understanding of the system. In \Sec{sec:moirecouplingmin}, we present the form of $T(\mbf{r})$ and show that only the A-sublattice of graphene coupling significantly influences the band structures, due to the sublattice and layer polarization of the low energy bands. This will justify the simple form of our model, despite the low symmetry of the system and the large number of fitting parameters used in other works \cite{PhysRevB.89.205414,Park2023RMGhBNChernFlatBands}. As a summary of our results in this section, we compute the continuum model band structures of R5G/hBN with parameter values determined by the SK hopping function and fitting to the DFT+SK band structure; the resultant band structures are shown for $\xi=1$ and the inter-layer potential energy differences $V = -16.7,-6.66,0,6.66,16.65$meV in Fig. \ref{fig:main_2n}. The continuum model band structures for $n=3,4,6,7$ layers are also shown in App. \ref{app:fitting}.

\subsection{Rhombohedral $n$-layer Graphene}
\label{sec:pentapure}

\begin{figure}
    \centering
    \includegraphics[width=.85\linewidth]{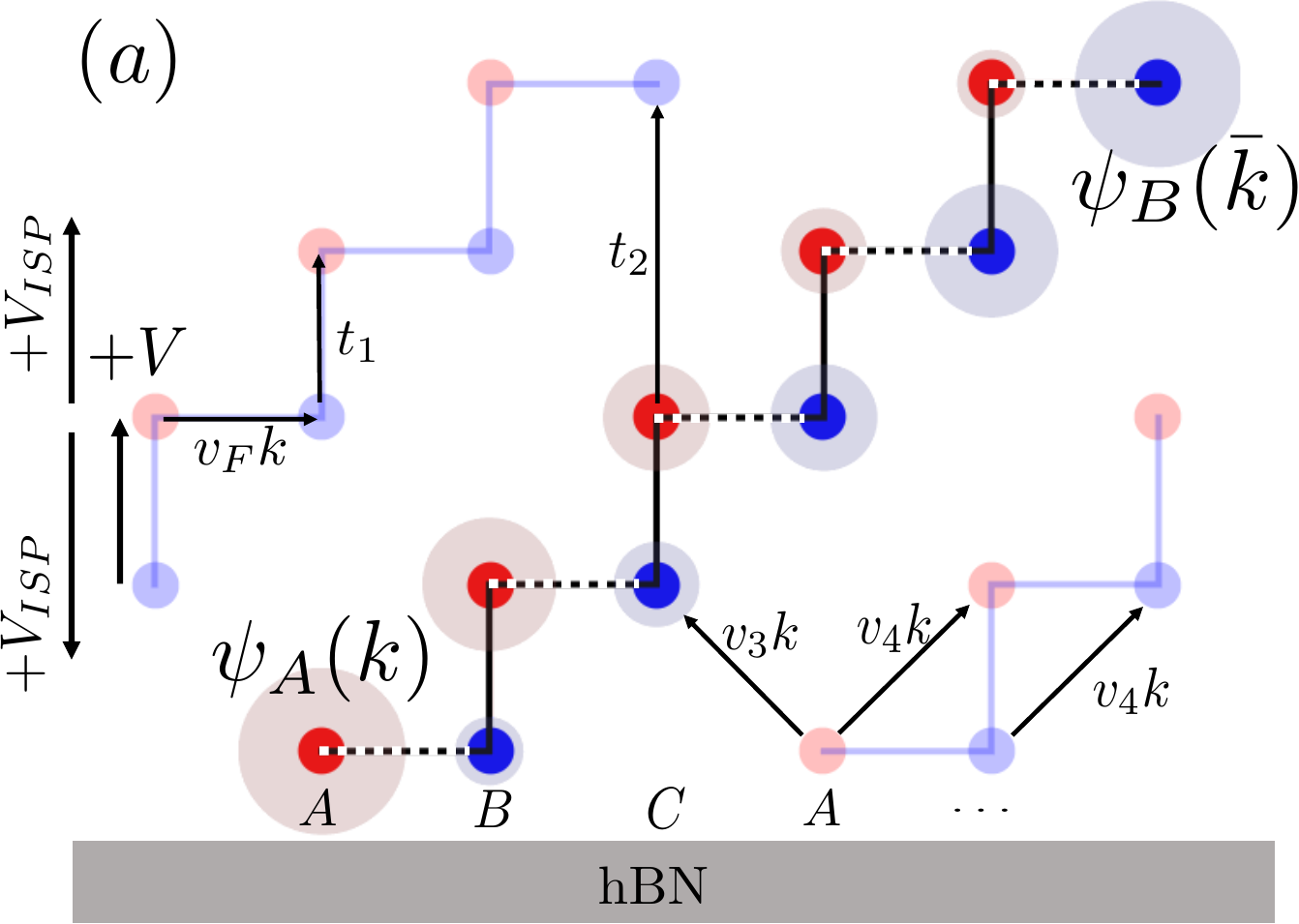}
\includegraphics[width=1.0\linewidth]{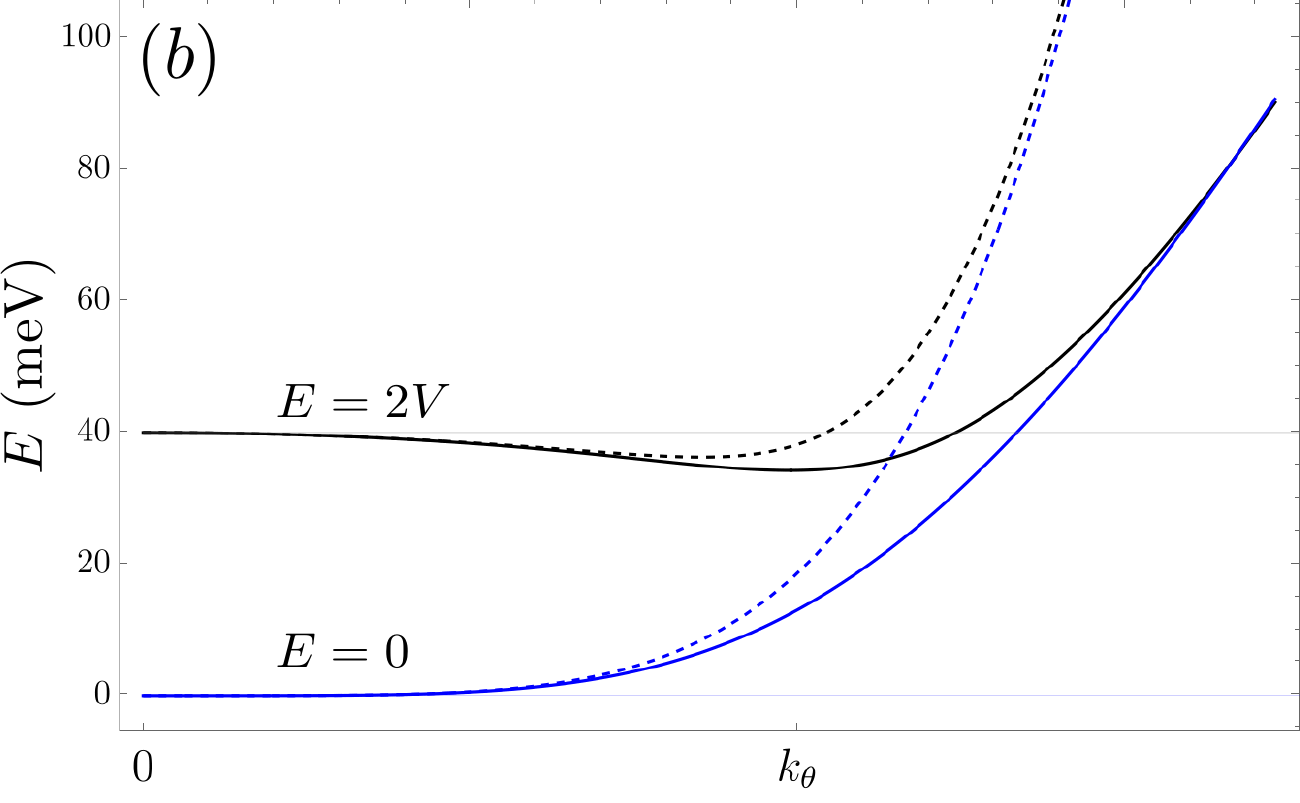}

    \caption{Pristine R5G. $(a)$ The unit cell of rhombohedrally stacked pentalayer graphene is shown with a depiction of the sublattice-polarized chiral states $\psi_A(k)$ and $\psi_B(\bar{k})$ that make up the low energy Hilbert space (other unit cells shown faded).  They are bound to the outermost orbitals of the unit cell and decay exponentially within the crystal. The in-plane coupling is $v_F k$ shown with a dotted line, and $t_1, v_3,v_4$ are the out-of-plane couplings. Note that $V$ is the inter-layer potential energy difference of electrons due to the applied field. The arrows show the positive directions of $V$ and $V_{ISP}$. $(b)$ The spectrum (solid lines) of the low-energy states $\pm E(|k|)$ is shown in the limit $v_3=v_4 = 0$ at two values of $V$. At $V=0$ (blue), the bands are monotonically increasing, but $V=20$ meV (black), they are non-monotonic. Dashed lines show analytical approximations (see \appref{app:symmSO2}) capturing these features.}
    \label{holomorphic}
\end{figure}

A schematic of the lattice structure of rhombohedral $n$-layer graphene is shown in \Fig{holomorphic}(a). Around the graphene $\mbf{K}$ point, the Hamiltonian can be expanded to take the form
\bea \label{eq: H_K}
H_{\K} &= \bpm
v_F\mbf{k} \cdot \pmb{\sigma} & t^\dag(\mbf{k}) & t'^\dagger &   &\\
t(\mbf{k}) & \ddots & \ddots & t'^\dagger \\
t' & \ddots & v_F\mbf{k} \cdot \pmb{\sigma} & t^\dagger(\mbf{k})\\
& t' & t(\mbf{k})  & v_F\mbf{k} \cdot \pmb{\sigma}
\epm + H_{ISP}, \\ 
t(\mbf{k}) &= -\bpm v_4 k & -t_1 \\ v_3\bar{k} &  v_4 k \epm, \qquad  \qquad t' = \bpm 0 & 0 \\ t_2 & 0 \epm
\eea
where $k = k_x + i k_y$ is the holomorphic momentum ($\bar{k}$ is its complex conjugate), $v_F$ is the Fermi velocity, and $t_1,v_3,v_4$ are inter-layer hopping parameters (see \Fig{holomorphic}). The local chemical potential of each layer is set by the inversion symmetric potential $[H_{ISP}]_{l l'} = V_{ISP} \delta_{ll'} | l - \frac{n-1}{2}|$ where $V_{ISP} \sim 16.65$meV and $l = 0,\dots, n-1$. $H_{ISP}$ acts as a confining potential within the graphene. 

To expose some simple features of the spectrum, we set $v_3=v_4 =t_2 = V_{ISP} = 0$ where the model has full $SO(2)$ rotation symmetry and the anti-commuting sublattice/chiral symmetry (see \App{app:symmSO2}). Even in this limit, it cannot be solved exactly for all $n$, in particular for $n = 5$ where its characteristic polynomial is tenth order. Nevertheless, we can analyze the model for a general number of layers $n$ in the low energy limit. The model is trivially solvable at $\mbf{k}=0$, showing two $E=0$ modes with eigenvectors $(1,\dots,0)$ and $(0,\dots, 1)$ and the other $n-2$ energies at $\pm t_1$ (this is true even if $v_3,v_4 \neq 0$). Doing degenerate perturbation theory, we see these null modes are connected at order $t_1^n$, so the low energy bands can be seen (\App{app:symmSO2}) to form a high-degree node  $E_\pm(\mbf{k}) = \pm (v_F|k|)^n / t_1^{n-1} + \dots $. Their eigenstates are also solvable to order $O(|k|^{n+1})$. It is convenient to write them in the chiral basis\cite{PhysRevB.81.125304} with the holomorphic/anti-holomorphic form (up to normalization) 
\bea
\label{eq:eigholo}
\null [\psi_A]_{l \al}(k) &= \Big( \frac{-v_F k}{t_1}\Big)^l \delta_{\al,A}, \quad [\psi_B]_{l \al}(\bar{k}) = \Big( \frac{-v_F \bar{k}}{t_1} \Big)^{\bar{l}} \delta_{\al,B}
\eea
with $l = 0,\dots, n-1$, $\bar{l} = n-1-l$, and $\al = A,B$ to index the sublattice. These states are sublattice polarized, exchanged under the combined space-time inversion symmetry $D[\mathcal{I}\mathcal{T}] = \delta_{l+l',n-1} [\sigma_1]_{\al \be}K$ (whose center is the middle of the third layer, i.e., $l = 2$ for $n=5$), and decay exponentially from one side to the other with a momentum-dependent length scale set by $v_F |k| / t_1$. This perturbation theory is valid for $v_F |k| / t_1 \ll 1$, and we can gauge its validity by picking $|k| = |\mbf{q}_1|$ which is the largest momentum in the first moir\'e BZ. We find that $v_F |\mbf{q}_1| / t_1 \leq 0.65$ for $\th \leq 1^\circ$, demonstrating that the states in \Eq{eq:eigholo} are valid. This basis provides a powerful tool for exposing the low-energy features of the model when the moir\'e potential and displacement field are added. 

In particular, the observation of FCIs and correlated Chern insulators occurs at large displacement field $D$ \cite{Ju2023PentalayerGraphenehBN} described by the Hamiltonian
\bea
\label{eq:H_D}
\null [H_D]_{l \al,l' \be} = V (l - \frac{n-1}{2})\delta_{ll'} \delta_{\al \be}
\eea
where $V \propto -|e| d_0 D$ is the inter-layer potential difference which is proportional to the displacement field, electric charge $-|e|$, and the inter-layer distance $d_0 \sim 3.33\AA$. The proportionality constant depends on the effective screening which is not directly computed in this work, although attempts have been made to estimate it from experiment \cite{2022Sci...375.1295Y}, resulting in an effective screening constant $\eps_r \sim 5.5$ in trilayer devices. For this reason, we use $V$ in this work. 

\begin{figure}
    \centering
    \includegraphics[width=1.\linewidth]{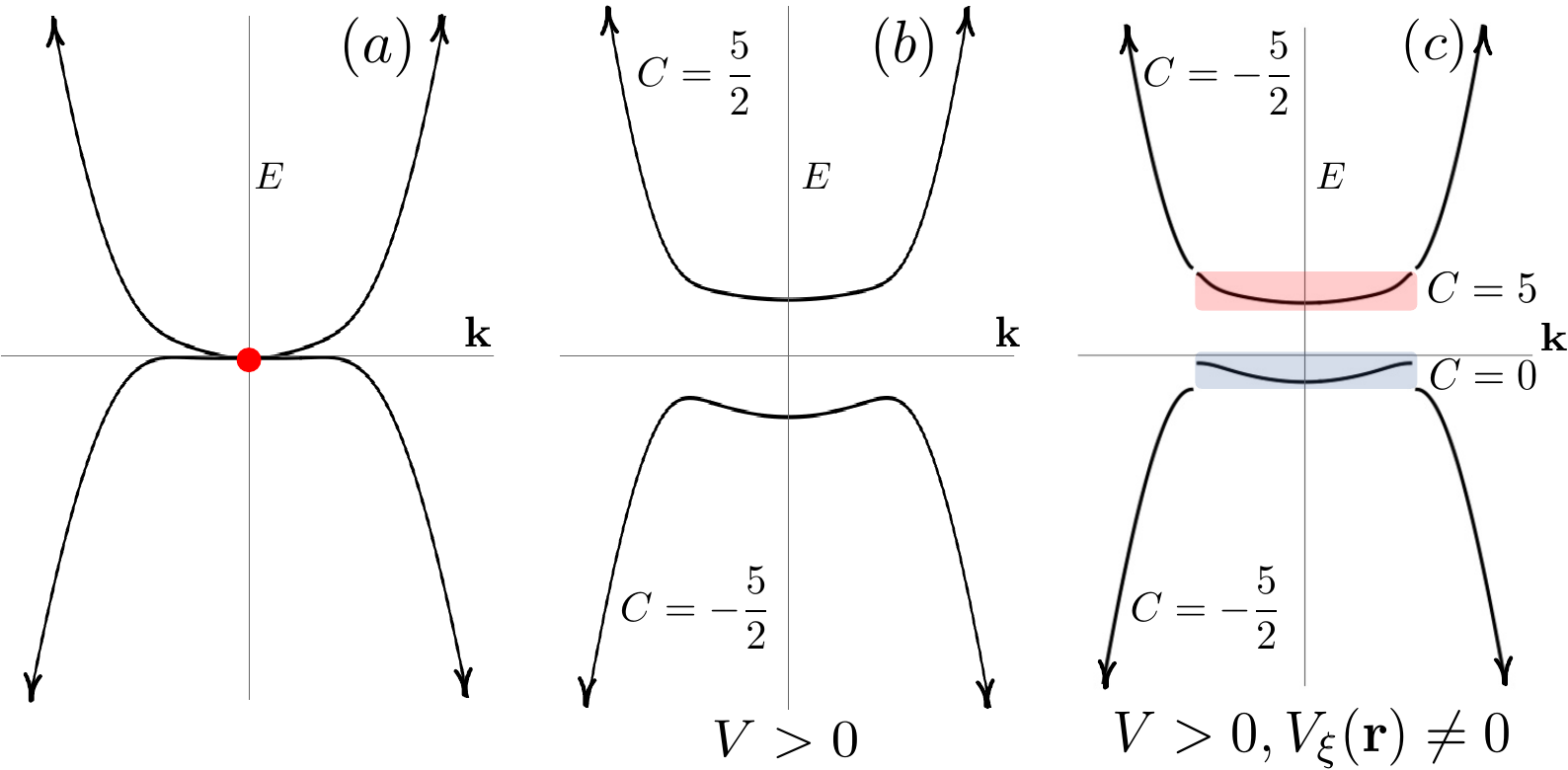}
    \caption{Evolution of topology in pentalayer on hBN in the graphene $K$ valley (per spin). $(a)$ Pristine R5G has a Berry curvature monopole corresponding to $C = \pm \frac{5}{2}$ due to the five Dirac cones. $(b)$ Adding displacement field only will split the bands at charge neutrality, and the Chern number will change by $5$ going through the gap. The valence bands carry $C=-\frac{5}{2}$ and the conduction bands carry $C=\frac{5}{2}$. Note that including both valleys causes the total Chern number at charge neutrality to vanish. $(c)$ Adding the moir\'e potential opens small gaps and isolates the lowest conduction band with $C=5$ and highest valence band with $C=0$. The gap at charge neutrality is not closed, and still carries a nontrivial valley Chern number. }
    \label{valleyChern}
\end{figure}

Collecting the chiral states into the column matrix $\Psi(\mbf{k}) = [\psi_A(k),\psi_B(\bar{k})]/\sqrt{\mathcal{N}(\mbf{k})}$ where $\mathcal{N}(\mbf{k}) = |\psi_A(k)|^2 = |\psi_B(\bar{k})|^2$ is the normalization, we find (for now setting $v_3=v_4 =t_2 = V_{ISP} = 0$) 
\bea
\label{eq:HeffV}
\Psi^\dag(\mbf{k})(H_{\K}+H_D) \Psi(\mbf{k}) &=  \frac{v_F^n}{t_1^{n-1}} \bpm 0 & \bar{k}^n \\ k^n & 0 \epm \\
& \!\!\!\!\!\!\!\! \!\!\!\!\!\!\!\!\!\!\!\!\!\!\!\! \!\!\!\!\!\!\!\! +V \Big( \frac{1-n}{2} +\sum_{m=1}^{2m\leq n}\big(v_F|k|/t_1\big)^{2m} \Big) \sigma_3
\eea 
which is a $2\times 2$ matrix in Pauli form (see \App{app:effmodelderivation} for a derivation.) Two essential features are revealed. Firstly, at $V=0$, the Hamiltonian describes the well-known high degree node with a $n\pi$ Berry curvature monopole. Turning on $V$ splits the node, breaking the $n \pi$ monopole and distributing the Berry curvature $F(\mbf{k})$ to the valence bands (opposite to the conduction bands) according to 
\bea
F(|k|) = - \frac{n^2(v_F^n/t_1^{n-1})^2}{2} \frac{|k|^{2(n-1)} \Delta}{\big((v_F^n/t_1^{n-1})^2|k|^{2n}+\Delta^2\big)^{3/2}}
\eea 
where $\Delta = V \frac{n-1}{2}$ is the gap at $k=0$, the graphene $\K$ point. Integrating $F(|k|)$ over all $\mbf{k}$ gives $ - n \pi \text{ sign}(V)$, corresponding to a half-quantized Chern number $C = - \text{ sign}(V) n/2$ for the valence bands. Introducing a moir\'e potential will open gaps at the boundary of the first moir\'e BZ, isolating the highest valence band and the lowest conduction band (see \Fig{valleyChern} for a schematic of gap openings driven by displacement field and the moir\'e potential). 

For $n=5$ and large $V>0$, we will show by symmetry and confirm numerically in \Sec{eq:chern} that the Chern number of the conduction band is $C=5$, while the valence band is trivial with $C=0$, and has the symmetry representation of an atomic limit. This means the remaining conduction and valence bands each carry a half-quantized $C = -5/2$, which we check numerically. The $K'$ valley is related by time-reversal has opposite Berry curvature, so the total Chern number at charge neutrality is zero. However, we predict a nonzero valley Chern number of 
\bea
C_V = \sum_{s = \uparrow,\downarrow} \frac{C_{\K,s}-C_{\K',s}}{2} = -5,
\eea
including spin, at the charge neutrality point for large $V>0$. The valley Chern number may be measured in multi-terminal transport experiments \cite{2022NatPh..18...48W}, or throughout gap closings in the Hofstadter spectrum  \cite{2020arXiv200614000B,2020PhRvL.125w6804H} which is accessible up to one flux due to the large moir\'e unit cell \cite{PhysRevLett.128.217701,2022PhRvL.129g6401H}. Since the highest valence band band below charge neutrality also has $C=0$, we predict a valley Hall effect with $C_V = 5$ at $\nu = -4$ (\ie, the Fermi energy in between the top and second top valence band in each valley/spin) as well in the non-interacting limit. 

The second important effect of $V$ is to flatten the bands. The full $\mbf{k}$-dependence in $\sigma_3$ term in \Eq{eq:HeffV} is crucial for explaining this effect since the constant term $\frac{n-1}{2}$ and $\mbf{k}$-dependent terms have opposite sign.  This results in a non-monotonic dispersion that can lead to a flattened band. An example is shown in \Fig{holomorphic}(b) comparing monotonically increasing $|k|^5$ dispersion at $V=0$ to the non-monotonic dispersion at $V = 20$meV. We can estimate the optimal $V_c$ through the band flatness criterion $E(\mbf{0}) = E(\mbf{q}_1)$. In \App{app:symmSO2}, we estimate that this criterion is satisfied for $n=5$ when
\bea
v_F^2 |\mbf{q}_1(\th)|^2 \sim t_1 \sqrt{2 V_c t_1} \ .
\eea
Using the experimental twist angle, we estimate that $V_c \sim 17$meV results in the optimally flattened band, corresponding to a top-to-bottom potential difference of $\sim 70$meV. 

This completes our discussion of R$n$G. We now consider the addition of a moir\'e potential. 

\subsection{Moir\'e Coupling}
\label{sec:moirecouplingmin}

 The form of the moir\'e coupling can be derived from the Bistrizter-MacDonald two-center approximation~\cite{PhysRevB.90.155406} by keeping only the lowest harmonic terms.
 We note that the lowest-harmonic approximation for moir\'e coupling between R$n$G and hBN is not necessarily quantitatively accurate here due to the large gap of hBN, which makes its low-energy dispersion flattish.
Nevertheless, we can still use the lowest-harmonic moir\'e coupling between R$n$G and hBN to derive the form of the effective moir\'e potential after integrating out hBN, since it is reasonable to keep only the lowest-harmonic terms in the effective moir\'e potential since it only couples graphene degrees of freedom. (See \appref{app:microscopic_model} for details.)
Therefore, in this part, we will still use the lowest harmonic approximation

Under the lowest-harmonic approximation, the moir\'e coupling between the lowest graphene and hBN reads
\bea
\label{eq:equalamp}
T(\mbf{r}) &=  \sum_{j=1}^3 e^{i \mbf{q}_{j} \cdot \mbf{r}} T_j, \quad T_{j+1} = w \bpm 1 & e^{i \frac{2\pi j}{3}} \\ e^{-i \frac{2\pi j}{3}} & 1\epm
\eea
where the single moir\'e coupling parameter $w$ results from the assumption of equal hoppings between the carbon and hBN orbitals at all positions of the moir\'e lattice. 
The assumption of equal hoppings is reasonable due to the exponentially accurate layer and sublattice polarization of the low energy states in \Eq{eq:eigholo}, as we now show. 

\begin{figure}
    \centering
    \includegraphics[width=1.\linewidth]{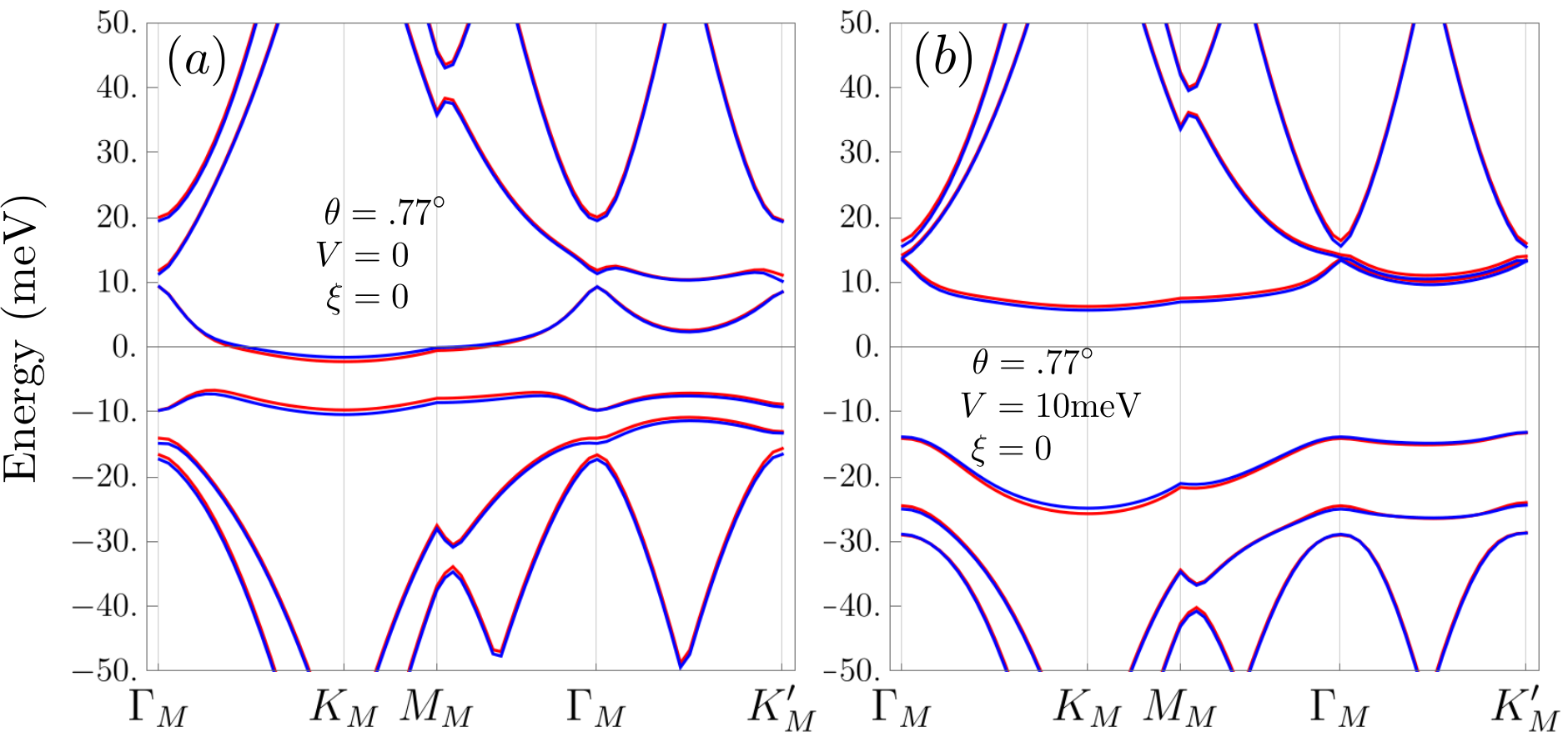}
    \caption{Comparison of equal-amplitude $T(\mbf{r})$ (blue) and one-orbital-only coupling in \Eq{eq:onlyA} (red) for $V=0$ in $(a)$ and $V=10$meV in $(b)$ for $\xi = 1$ in the $K$ valley. We conclude that only the moir\'e coupling to the outer orbital, which is sublattice polarized due to the low-energy holomorphic basis, plays a role in the band structure. In both plots we have shifted the chemical potential to the minimum of the conduction band and boosted both valleys to the moir\'e $\Gamma_M $ point.}
    \label{AAonly}
\end{figure}

The most general $C_3$-allowed moir\'e coupling matrix takes the form
\bea
\label{eq:T_j_before_perturbation}
T_j &= \bpm t_{B,C_A} & e^{-i \frac{2\pi}{3} j}  t_{B,C_B}  \\
 e^{i \frac{2\pi}{3} j} t_{N,C_A} & t_{N,C_B} \epm
\eea
where $C_\al$ are the carbon orbitals on the bottom layer and $B,N$ are the hBN orbitals beneath. However, we check that --- because the low-energy states have exponentially small weight on the $C_B$ atoms on the bottom layer --- one can set $t_{B,C_B} = t_{N,C_B} = 0$ with negligible change to the band structure, as shown in \figref{AAonly}. This is because the second column of $T_j$ has very small overlap on the low-energy bands, so it can be set to zero without incurring error.

With $t_{B,C_B} = t_{N,C_B} = 0$, we can now integrate out the hBN and get the effective coupling via the second order perturbation theory.
As a result, we get 
\bea
\label{eq:onlyA}
V_{eff, \xi}(\mbf{r})&= - T^\dag(\mbf{r}) \frac{1}{\sigma_1^\xi H_{BN} \sigma_1^\xi} T(\mbf{r}) \\
&= \big(V_0 + 2V_1 \sum_{j=1}^3 \cos(\mbf{g}_j \cdot \mbf{r} + \psi_\xi) \big) \bpm 1 & 0 \\ 0 & 0 \epm\\
\eea
which couples only to the active $A$ sublattice as only $A$ sublattice is active in the bottom layer, and the parameters are given by (with $\omega=e^{2\pi i/3}$, see \App{app:moirecoupling})
\bea
\label{eq:psi_determination}
V_0 &= -3 \bpm |t_{B,C_A}|^2 & |t_{N,C_A}|^2 \epm \sigma_1^\xi \bpm V_B^{-1} \\ V_N^{-1} \epm \\
V_1 e^{i \psi} &= - \bpm |t_{B,C_A}|^2 & \omega^* |t_{N,C_A}|^2 \epm \sigma_1^\xi \bpm V_B^{-1} \\  V_N^{-1} \epm \ .
\eea
Therefore, although there are two independent parameters $t_{B,C_A}$ and $t_{N,C_A}$ in \eqnref{eq:T_j_before_perturbation}, they only contribute to one complex parameter in the non-uniform part of the moir\'e potential after integrating out hBN. 
The form of the potential is similar to the one proposed in Ref. \cite{PhysRevLett.122.086402} for twisted transitional metal dichalcogenides. We have found it convenient to define $\mbf{g}_j = R(\frac{2\pi}{3}(j-1)) (\mbf{q}_2-\mbf{q}_3)$ for $j = 1,2,3$ related by $C_3$ symmetry. This simplified model agrees with \Eq{eq:equalamp} when the latter is also restricted to the active $A$ sublattice only, as we have shown numerically holds to good approximation in \Fig{AAonly}. Thus, \Eq{eq:onlyA} shows that although the most general form of the potential contains many more parameters than the equal amplitude case of \Eq{eq:equalamp}, \Eq{eq:equalamp} is sufficient to fit the data. Thus, we arrive at the effective hBN potential
\bea
\label{eq:Vxifinal}
V_\xi(\mbf{r})&= - T^\dag(\mbf{r}) \sigma_1^\xi H_{BN}^{-1} \sigma_1^\xi T(\mbf{r}) \\
&= V_0 + \left[V_1 e^{i\psi_\xi}\sum_{j=1}^3 e^{i \mbf{g}_j\cdot\mbf{r}}\bpm 1& \omega^{-j} \\ \omega^{j+1} &\omega \epm + h.c.\right]
\eea
which couples only to the bottom graphene layer. 

\subsection{Fitting Results}

As discussed at the beginning of \eqnref{sec:moirecouplingmin}, the lowest-harmonic form of the effective moir\'e potential in \eqnref{eq:Vxifinal} is reasonable since it only couples to the graphene degrees of freedom, even if we include higher-harmonics terms in the $T(\bsl{r})$ in \eqnref{eq:equalamp}. 
What can be quantitatively inaccurate is the expression of $V_0$ and $V_1 e^{\ii \psi_\xi}$ in \eqnref{eq:psi_determination}, since they rely on the lowest-harmonic approximation of $T(\bsl{r})$ in \eqnref{eq:equalamp}.
Therefore, in practice, we should treat $V_0$, $V_1$ and $\psi_{\xi}$ as tuning parameters to fit the DFT+SK band structure.
The resultant parameter values are listed in \Tab{tab:parameters_full}. (See more details of the fitting in \appref{app:fitting}.) 
The $2n\times 2n$ continuum model in \eqnref{eq:fullmodel} with the potential form in \eqnref{eq:Vxifinal} and the parameter values in \tabref{tab:parameters_full} can reproduce the DFT+SK band structure remarkably well, as exemplified in \figref{fig:main_2n} for $n=5$ and $\xi=1$. (The fitting for all the configurations can be found in Figs. \ref{fig:fitting_3}-\ref{fig:fitting_7} in \appref{app:fitting}.)
Interestingly, we find that fixing $\psi_\xi$ by its value in the lowest harmonic $t_{N,C_A} = t_{B,C_A}$ case can actually provide very good fitting.

\begin{table}[t]
\centering
\begin{tabular}{ c|c c c c| c c c } 
 & $v_F$& $v_3$  & $t_1$  & $t_2$  & $V_0$ & $V_1$ & $\psi_{\xi}$ \\ 
 \hline
 \text{$n=3$, $\xi $=1} & 542.1 & 34. & 355.16 & -7 & 0 & 5.54 & \text{16.55${}^{\circ}$} \\
 \text{$n=4$, $\xi $=1} & 542.1 & 34. & 355.16 & -7 & 1.44 & 6.91 & \text{16.55${}^{\circ}$} \\
 \text{$n=5$, $\xi $=1} & 542.1 & 34. & 355.16 & -7 & 1.50 & 7.37 & \text{16.55${}^{\circ}$} \\
 \text{$n=6$, $\xi $=1} & 542.1 & 34. & 355.16 & -7 & 1.56 & 7.80 & \text{16.55${}^{\circ}$} \\
 \text{$n=7$, $\xi $=1} & 542.1 & 34. & 355.16 & -7 & 1.47 & 7.93 & \text{16.55${}^{\circ}$} \\
 \hline
 \text{$n=3$, $\xi $=0} & 542.1 & 34. & 355.16 & -7 & 6.13 & 5.95 & \text{-136.55${}^{\circ}$} \\
 \text{$n=4$, $\xi $=0} & 542.1 & 34. & 355.16 & -7 & 7.16 & 6.65 & \text{-136.55${}^{\circ}$} \\
 \text{$n=5$, $\xi $=0} & 542.1 & 34. & 355.16 & -7 & 7.19 & 7.49 & \text{-136.55${}^{\circ}$} \\
 \text{$n=6$, $\xi $=0} & 542.1 & 34. & 355.16 & -7 & 7.12 & 7.16 & \text{-136.55${}^{\circ}$} \\
 \text{$n=7$, $\xi $=0} & 542.1 & 34. & 355.16 & -7 & 7.00 & 7.37 & \text{-136.55${}^{\circ}$} \\
 \hline
\end{tabular}
 \caption{Parameter values of the full model for $n=3,4,5,6,7$ layers. Here $v_F, v_3=v_4$ are reported in meV$\cdot$nm, while $t_1, t_2, V_0$ and $V_1$ are in meV.  
 }
    \label{tab:parameters_full}
 \end{table}

\section{Phase diagram and Topology}
\label{PDsec}

\begin{figure}
    \centering
    \includegraphics[width=1.\linewidth]{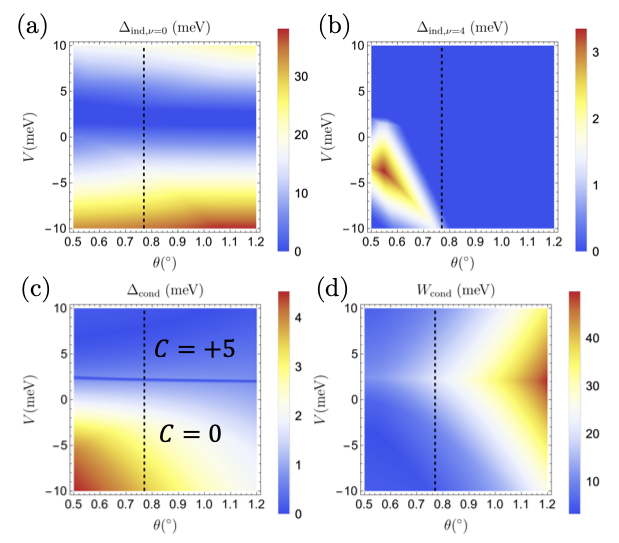}
    \caption{Phase diagrams for the lowest conduction band in pentalayer graphene on a single hBN substrate in the $\xi=0$ stacking configuration. (a) $\Delta_{\text{ind},\nu=0}$ denotes the indirect gap below the conduction band at filling $\nu = 0$ and (b) $\Delta_{\text{ind},\nu=4}$ denotes the indirect gap above the conduction band at filling $\nu=4$ (including spin/valley degeneracy). (c) $\Delta_{\text{cond}}$ is the direct gap around the conduction band, from which the topological phase transition is visible. Chern number of each phase is indicated. (d) $W_{\text{cond}}$ denotes the bandwidth. The dashed line refers to the experimental twist angle at $\theta=0.77^\circ$.
    }
    \label{hBN180_5G_PD}
\end{figure}

With the $2n \times 2n$ single-particle R$n$G/hBN moir\'e model constructed in the previous section (see Eqs.\,\eqref{eq:fullmodel} \eqref{eq: H_K} and \eqref{eq:Vxifinal}) we now move on to understand its spectrum and topology, first for general $n\leq 3$ then focusing on $n=5$. First we numerically compute phase diagrams for various inter-layer potential energy difference $V$ and twist angle $\th$. The results are shown in \Fig{hBN180_5G_PD} for $\xi = 0$ and $n=5$. The phase diagram for $\xi = 1$ is similar and can be found in \App{app:phase_diagrams}. The notable features are the following. Firstly, the Chern numbers accessible in $V$ are $C=0$ and $C=5$, as can be expected from the fifth degree node, and the direct gap between the lowest conduction band and the upper bands becomes extremely small for large $V$. Clearly, interactions dominate the phase diagram at large $V$\cite{dong2023anomalous,zhou2023fractional,dong2023theory} to produce correlated gaps, as is known to be the case in rhombohedral trilayer graphene \cite{chen2020tunable}. The interacting phase diagram is not the focus on this paper. Instead, we will now go on to  derive these key features analytically and cement an understanding of the phase diagram. In fact, we will show the topology and symmetry features are insensitive to parameter tuning around the fitted values.

\begin{figure*}
    \centering
    \includegraphics[width=1.\linewidth]{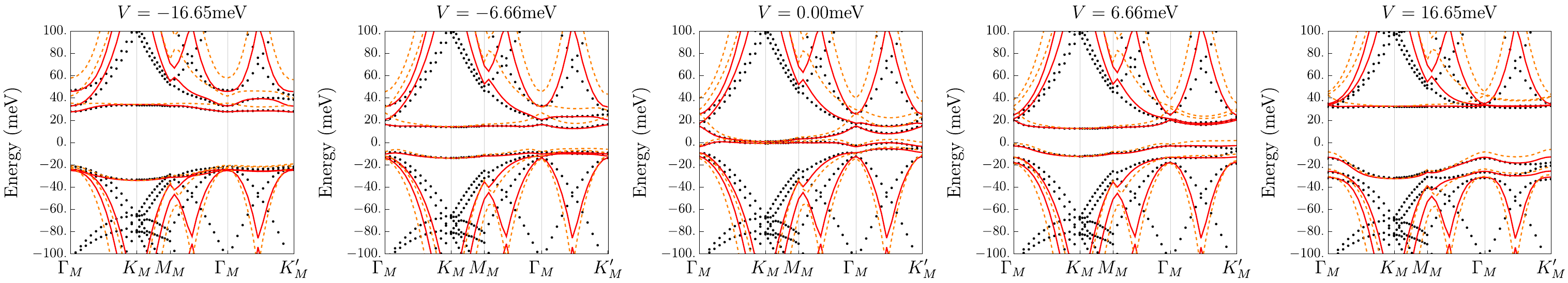}
    \caption{Band structure of the effective model for pentalayer graphene on a single hBN in the $\xi=1$ stacking. The red solid (orange dashed) lines show the band structure of the effective $2\times 2$ model in the graphene $\K$ valley with fitted parameters (with parameters perturbatively derived from the $2n\times 2n$ continuum model). 
     Black dots are the first principles calculations.
    The match to the low energy bands is very good. }
    \label{fig:main_eff}
\end{figure*}

\subsection{Effective Model}
\label{app:effmodel}

We now reduce the full $2n\times 2n$ moir\'e model of $n$-layer rhombohedral graphene to a $2\times 2$ model built on the chiral basis. Our procedure is to project the full Hamiltonian with trigonal warping terms $v_3, v_4$ onto the chiral basis in \Eq{eq:eigholo}, from which we obtain

\begin{equation}\label{eq:Heff}
    H_{eff} = h_{eff}(-i\nabla) + V_{eff,\xi}(\mbf{r})
\end{equation}
with $V_{eff,\xi}(\mbf{r})$ given in \Eq{eq:onlyA} and
\eqa{
& h_{eff}(\mbf{k}) \\
&\ = \bpm H_0(|k|) + V_3(|k|) & \bar{k}^{n-3}(\beta \bar{k}^3+\gamma |k|^2 +\delta) \\ k^{n-3}(\beta k^3+\gamma |k|^2 +\delta) & H_0(|k|) -V_3(|k|)\epm.
}
where
\begin{subequations}\label{eq:V3_def}
\begin{align}
   H_0(\mbf{k}) &= \alpha |k|^2- V_{ISP} \Big[\frac{1-n}{2} + \sum_{m=1}^{2m < n}\big(\frac{v_F|k|}{t_1}\big)^{2m}\Big],\\
   V_3(\mbf{k}) &= V \Big[\frac{1-n}{2} + \sum_{m=1}^{2m\leq n}\big(\frac{v_F|k|}{t_1}\big)^{2m}\Big],
\end{align}
\end{subequations}
Here $\al,\be,\gamma, \delta$ as well as $V_0$, $V_1$ and $\psi_{\xi}$ in $V_{eff,\xi}(\mbf{r})$ can be computed in terms of the bare graphene parameters via perturbation theory, as listed in \tabref{tab:parameters_eff}. (See details in App.\ref{app:effmodelderivation}.) 
Focusing on the $\K$ valley, directly using the values derived from the perturbation theory can well match the low-energy features around $\K_M$ point, while we find worse match around other high-symmetry points such as $\Gamma_M$, as shown in \figref{fig:main_eff}.
To achieve a better match to the band structure, we treat $\{\alpha,\beta,\gamma, V_1\}$ as tuning parameters and optimize them around their values derived from the perturbation theory, while preserving the form of the model and keeping the values of other parameters fixed.
With the fitting parameter values listed in \tabref{tab:parameters_full}, we can improve the matching away from $\K_M$ point, especially at $\Gamma_M$ as examplified in \figref{fig:main_eff} for $n=5$ and $\xi=1$.
A more detailed comparison between the effective model and the DFT+SK calculation is shown in \appref{app:fitting}.

\begin{table*}[t]
\centering
\begin{tabular}{ c|c c c c | c c   c } 
 & $\alpha$ & $\beta$ & $\gamma$ & $\delta$  & $V_0$ & $V_1$ & $\psi_{\xi}$ \\ 
  \hline
 \text{$n=3$, $\xi $=1} & \text{66.52(103.79)} & \text{963.17(1263.00)} & \text{77.60(120.10)} & -7.00 & 0 & \text{5.83(5.54)} & \text{16.55${}^{\circ}$} \\
 \text{$n=4$, $\xi $=1} & \text{60.77(103.79)} & \text{-1390.30(-1927.70)} & \text{-210.73(-287.42)} & 21.37 & 1.44 & \text{5.38(6.91)} & \text{16.55${}^{\circ}$} \\
 \text{$n=5$, $\xi $=1} & \text{56.18(103.79)} & \text{2158.00(2942.40)} & \text{401.13(597.60)} & -48.92 & 1.50 & \text{5.71(7.37)} & \text{16.55${}^{\circ}$} \\
 \text{$n=6$, $\xi $=1} & \text{52.67(103.79)} & \text{-3291.20(-4491.10)} & \text{-859.67(-1154.70)} & 99.57 & 1.56 & \text{5.47(7.80)} & \text{16.55${}^{\circ}$} \\
 \text{$n=7$, $\xi $=1} & \text{49.72(103.79)} & \text{4631.40(6855.10)} & \text{1507.50(2132.70)} & -189.97 & 1.47 & \text{5.58(7.93)} & \text{16.55${}^{\circ}$} \\ \hline
 \text{$n=3$, $\xi $=0} & \text{57.57(103.79)} & \text{963.82(1263.00)} & \text{79.45(120.10)} & -7.00 & 6.13 & \text{4.40(5.95)} & \text{16.55${}^{\circ}$} \\
 \text{$n=4$, $\xi $=0} & \text{53.34(103.79)} & \text{-1443.10(-1927.70)} & \text{-185.88(-287.42)} & 21.37 & 7.16 & \text{4.06(6.65)} & \text{16.55${}^{\circ}$} \\
 \text{$n=5$, $\xi $=0} & \text{49.80(103.79)} & \text{2230.90(2942.40)} & \text{458.05(597.60)} & -48.92 & 7.19 & \text{4.16(7.49)} & \text{16.55${}^{\circ}$} \\
 \text{$n=6$, $\xi $=0} & \text{44.72(103.79)} & \text{-3592.70(-4491.10)} & \text{-910.79(-1154.70)} & 99.57 & 7.12 & \text{4.60(7.16)} & \text{16.55${}^{\circ}$} \\
 \text{$n=7$, $\xi $=0} & \text{37.61(103.79)} & \text{4113.00(6855.10)} & \text{1706.00(2132.70)} & -189.97 & 7.00 & \text{4.65(7.37)} & \text{16.55${}^{\circ}$} \\
 \hline
\end{tabular}
 \caption{Parameter values of the $2\times 2$ effective model for the R$n$G/hBN structure of $n=3,4,5,6,7$ layers. Here $\alpha,\beta,\gamma,\delta$ are reported in meV$\cdot$nm$^2$,meV$\cdot$nm$^n$, meV$\cdot$nm$^{n-1}$ and meV, respectively, while $V_0, V_1$ are in meV. If the parameter values are difference from those directly derived from the perturbation theory of $2n$-band model, the latter will be provided in the parentheses.}
    \label{tab:parameters_eff}
 \end{table*}

\subsection{Chern number of the Conduction Band}
\label{eq:chern}

\begin{figure}
    \centering
    \includegraphics[width=1.\linewidth]{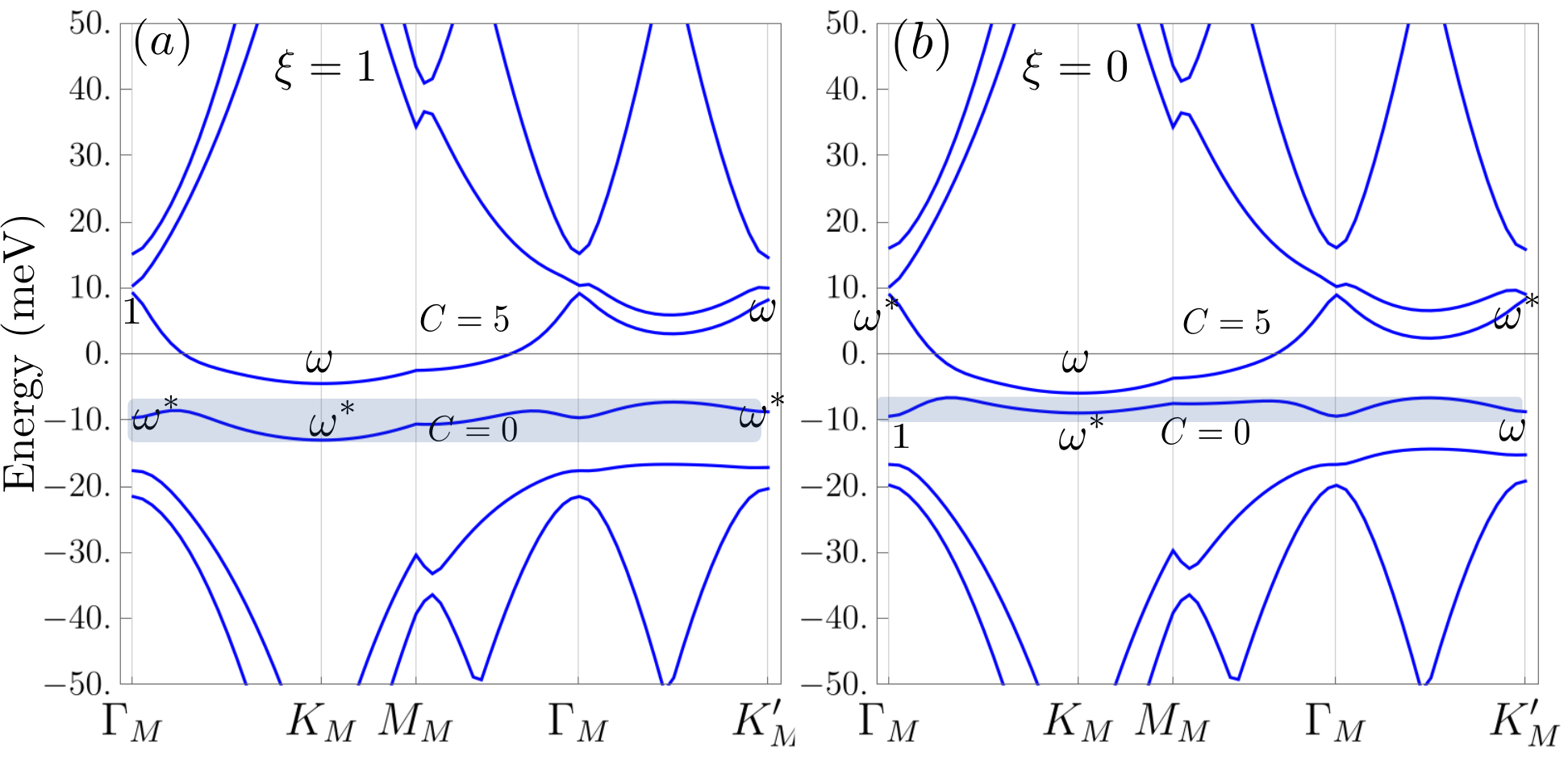}
    \includegraphics[width=1.\linewidth]{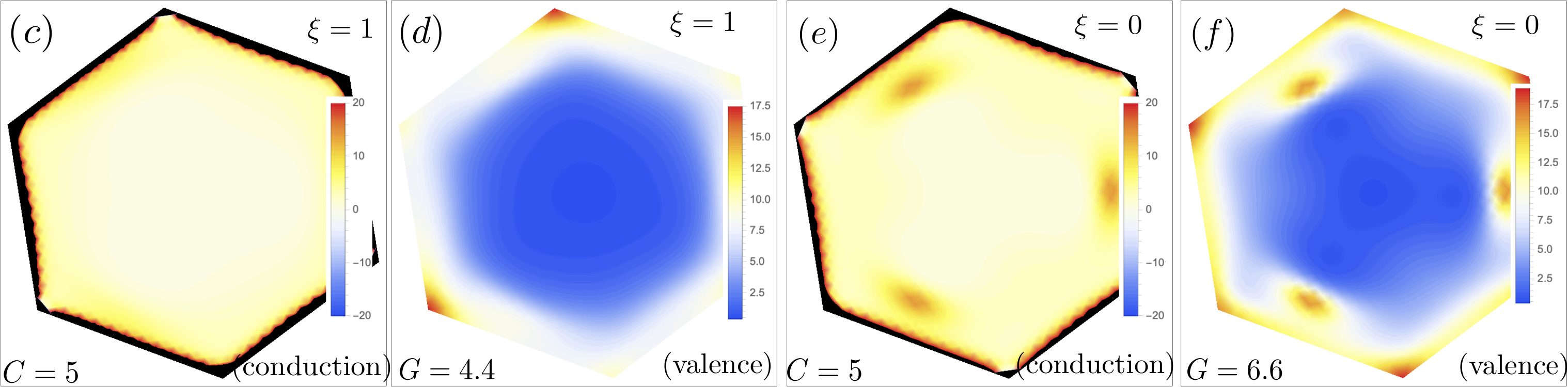}
    
    \caption{Topology, $C_3$ irreps, and quantum geometry at $V=3$meV for R5G/hBN at $\theta=0.76715^\circ$. The Chern number of the lowest conduction band is $C = 5$ consistent with the $C_3$ irreps for stacking $\xi =1$ in $(a)$ and $\xi = 0$ in $(b)$. The lowest bands show the irreps of a trivial atomic limit $p_x - ip_y$ orbital at the origin in $(a)$ and $p_x - ip_y$ at the moir\'e AB site in $(b)$. Here $\omega = e^{2\pi i/3}$ denotes the $C_3$ eigenvalues at the high-symmetry points. $(c)$ and $(e)$ show the Berry curvature distribution for the lowest conduction band and for $\xi =1$ and $\xi =0$ respectively, which is extremely peaked due to the small gaps (clipped regions are shown in black).  $(d)$ and $(f)$ show the Fubini-Study metric of the highest valence band for $\xi =1$ and $\xi =0$ respectively, which is large despite the triviality of the bands.  The integrated Fubini-Study metric $G$ \cite{2022PhRvL.128h7002H,Parameswaran13,RevModPhys.84.1419} lower bounds the Wannier spread. In $(c)$ through $(f)$, the center of BZ is the $\mbf{K}_G$ point. }
    \label{single_irreps}
\end{figure}

Equipped with the effective model in Eq.\,\eqref{eq:Heff}, we now analyze the $C_3$ symmetry eigenvalues at $C_3$ symmetric points in the continuum model moir\'e Brillouin zone for the graphene $\K$-valley, from which we deduce the Chern number (mod 3) in the lowest conduction band \cite{PhysRevB.86.115112}. Since for different commensurate twist angles the graphene $K_G$-point can be folded to either $\Gamma_M$, $K_M$ or $K'_M$ of the commensurate moir\'e BZ, as explained in Eq. \eqref{eq:K_G_folding}, here for simplicity of discussion we legislate that $K_G$ is folded onto the center of a shifted BZ defined for each graphene valley in the continuum model, denoted as $\tilde{\Gamma}_M$, and the $\tilde{K}_M$ ($\tilde{K}'_M$) point is situated at $\mbf{q}_1$ ($-\mbf{q}_1$) (see Fig. \ref{fig:severalMBZ} for a comparison of different BZs).

We will focus on the $n=5$ case in this section, and denote the corresponding $C_3$ eigenvalue at a $C_3$ symmetric $\mbf{k}$ as $\eta_\mbf{k}$. To proceed analytically, we consider a reduced Hamiltonian (akin to the tripod model or hexagon model \cite{2021PhRvB.103t5411B}) that involves only the reciprocal lattice points closest to the high symmetry point of interest. At $\tilde{\Gamma}_M$, we have
\begin{equation}
    H_{eff}(\mbf{\tilde{\Gamma}_M}) = (\frac{V_0}{2}+2V_{ISP}) \mathds{1} + (\frac{V_0}{2}-2V)\sigma_3. 
\end{equation}
The projected $C_3$ operator takes the form $D[C_3]=\text{diag}(\omega^*,\omega)$ for $n=5$ (see App. \ref{app:microscopic_model}), recalling $\omega\equiv e^{2i\pi/3}$. Thus,
\begin{equation}
\label{eq:etagamma}
    \eta_{\tilde{\Gamma}_M} = \begin{cases} \omega^*, \quad \text{for} \quad \frac{V_0}{2}-2V>0, \\ \omega, \quad \text{for}\quad \frac{V_0}{2}-2V<0 \end{cases} \ .
\end{equation}

At $\tilde{\mbf{K}}_M=\mbf{q}_1$, we need to consider the three closest reciprocal lattice vectors, which gives  

\begin{equation}
\begin{split}
    H_{eff}(\tilde{\mbf{K}}_M) &= V_0 \mathds{1}_{3\times 3}\otimes \sigma_{11} +\\
    &\bpm h_{eff}(\tilde{\mbf{K}}_M)  & V_1 e^{i\psi_\xi} \sigma_{11} & V_1 e^{-i\psi_\xi} \sigma_{11} \\ V_1 e^{-i\psi_\xi} \sigma_{11} & h_{eff}(C_3 \tilde{\mbf{K}}_M) & V_1 e^{i\psi_\xi} \sigma_{11} \\ V_1 e^{i\psi_\xi} \sigma_{11} & V_1 e^{-i\psi_\xi} \sigma_{11} &h_{eff}(C^2_3 \tilde{\mbf{K}}_M) \epm 
\end{split}
\end{equation}
where $\sigma_{11} \equiv (\sigma_0 +\sigma_3)/2$. The $C_3$ operator that commutes with $H_{eff}(\mbf{\tilde{K}_M})$ takes the form 
\begin{equation}
    \bpm 0 & 0 & 1\\ 1 & 0 & 0 \\ 0 & 1 & 0 \epm \otimes D[C_3], 
\end{equation}
which can be used to diagonalize  $H_{eff}$ into three 2 $\times$ 2 blocks that correspond to symmetry eigenvalues $1, \omega$ and $\omega^*$, respectively. The highest energy $\eps^{(c)}_{j}(\pm \mbf{\tilde{K}_M})$ in the symmetry sector with eigenvalue $\omega^j$ is found to be
\bea\label{eq:tripod_E}
\eps^{(c)}_{j-1}(\pm \mbf{\tilde{K}_M}) &= f_0 +  V_1 \cos\psi_{\xi, \mp j}\\
+&\sqrt{f^2_{\pm}+\big(\frac{V_0}{2}+V_3(|\mbf{\tilde{K}_M}|) +V_1 \cos \psi_{\xi,\mp j} \big)^2}
\eea
where $V_3$ is defined in Eq. \eqref{eq:V3_def},  and $\psi_{\xi,j} = \psi_\xi + \frac{2\pi}{3}(j-1)$. Here $f_0$ is a constant unimportant for determining the symmetry eigenvalues, $f_\pm \sim 20$ meV (for twist angle $\theta=0.77^\circ$), and all of their analytical forms can be found in App.\ref{app:tripod}. In the $v_3 = v_4 = t_2 = 0$ limit, the expression is simply $f_\pm = \be |\tilde{\mbf{K}}_M|^5$. For a wide range of parameters, the state of the lowest conduction band  at $\pm \mbf{\tilde{K}_M}$ can be identified by finding the minimum of $\cos \psi_{\xi, \mp j}$; the $C_3$ eigenvalue of the resultant state is consistent with a direct computation of symmetry eigenvalues using the $10 \times 10$ model as labeled in Fig. \ref{single_irreps}. Furthermore, independent of the value of $\psi_\xi$, the product of the $C_3$ eigenvalues at the $\mbf{\tilde{K}_M}$ and $\mbf{\tilde{K}_M'}$ points obeys $\eta_{\mbf{\tilde{K}_M}} \eta_{\mbf{\tilde{K}_M'}} = \omega$. Altogether, we find that the Chern number $C$ of the lowest conduction band obeys \cite{PhysRevB.86.115112}
\begin{equation}
\label{eq:cherncond3}
    \exp(\frac{2\pi i }{3} C) = \begin{cases} 1, \quad \text{for} \quad \frac{V_0}{2}-2V>0, \\ \omega^*, \quad \text{for}\quad \frac{V_0}{2}-2V<0\end{cases}
\end{equation}
and is insensitive to values of the graphene parameters, indicating the robustness of the Chern number to single-particle effects. This is confirmed by  the large $C=5$ and $C=0$ phases identified in Fig. \ref{hBN180_5G_PD}(c). 
It is clear from Eq. (\ref{eq:tripod_E}) that for very large \textit{positive} displacement field where $-(\frac{V_0}{2}+V_3) \gg f_{\pm}$ and $V_1$, the lowest three energies in the conduction band $(\epsilon^{(c)}_{-1}, \epsilon^{(c)}_{0}, \epsilon^{(c)}_{1})$ at $\mbf{\tilde{K}}_M$/$\mbf{\tilde{K}'}_M$ would stick together, which is confirmed in our band structure calculations using the 10 $\times$ 10 continuum model. 
It is consistent with fact that the conduction bands are nearly free (\ie, nearly having continuous translation symmetries) at large $V>0$ as shown in \figref{fig:main_2n} and the near degeneracy comes from the band folding.
We note that the nearly free conduction electrons at large $V>0$ were described in \refcite{zhou2023fractional,dong2023anomalous}.

\begin{figure}
    \centering
    \includegraphics[width=1.\linewidth]{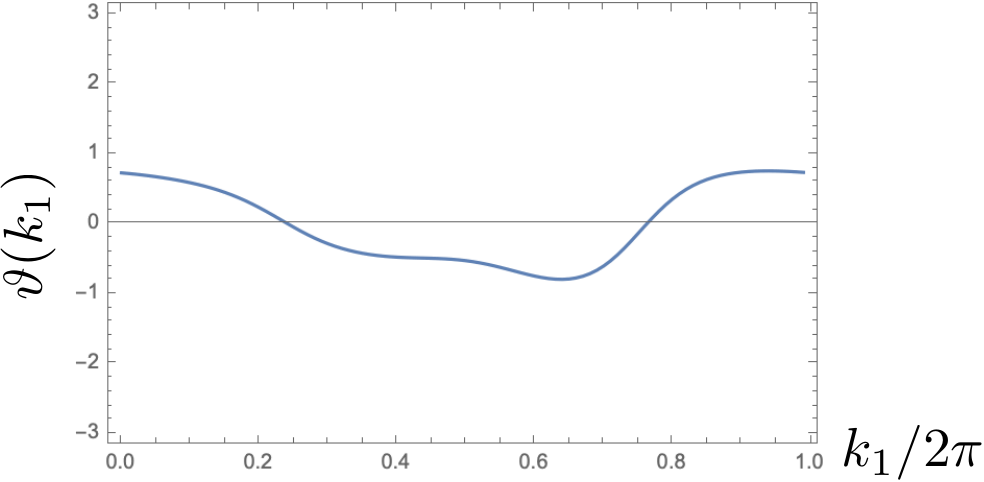}
    \caption{Wilson loop $W(k_1) = e^{i \vartheta(k_1)} = \exp i \int dk_2 A_2(\mbf{k}) $ at $V = -20$meV for $\xi = 1$, confirming the triviality of the band.}
    \label{wilson}
\end{figure}

\begin{figure}
    \centering
    
\includegraphics[width=1.\linewidth]{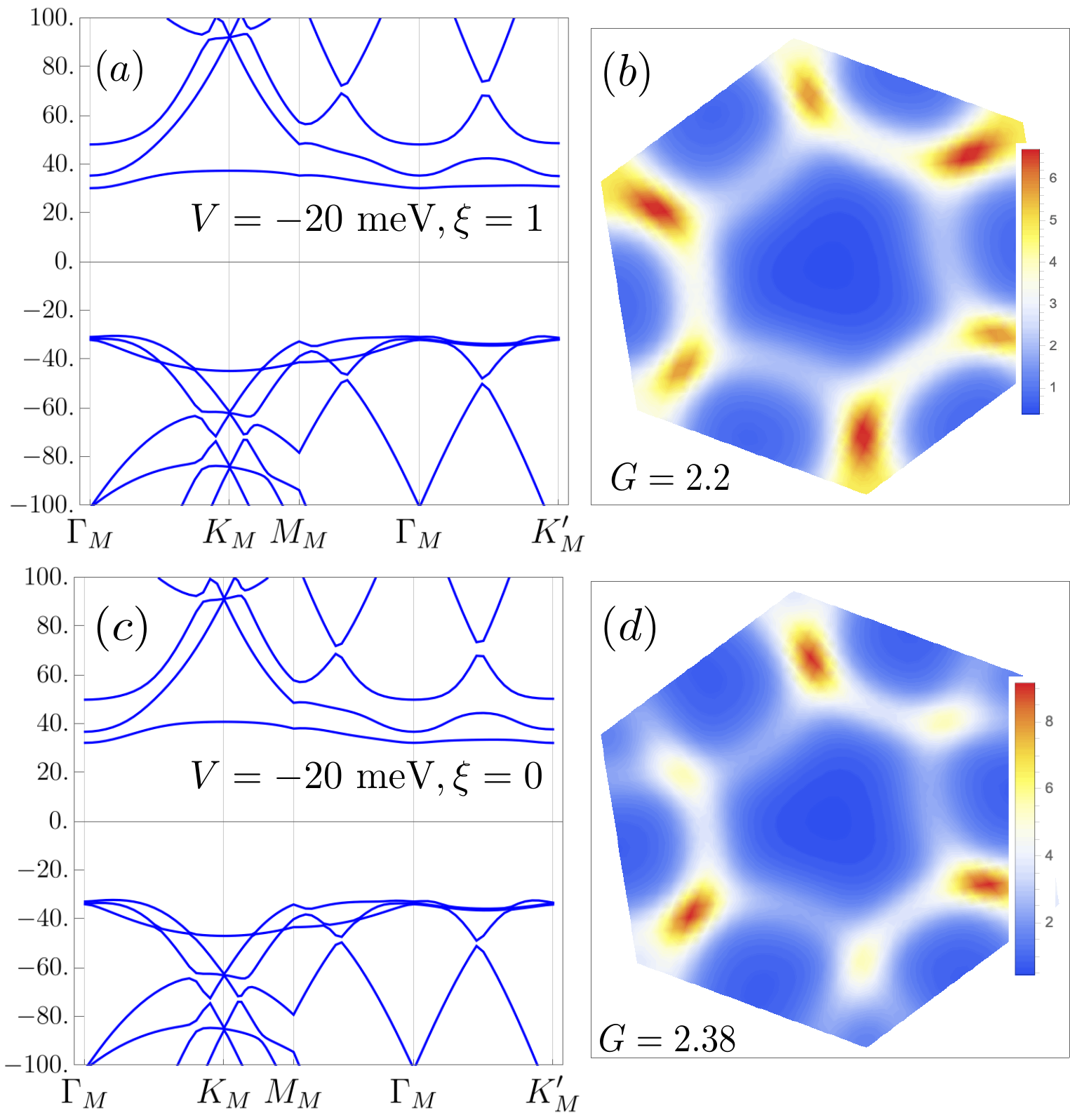}

    \caption{Quantum Geometry of the lowest conduction band at $V = -20$meV for $\xi = 1$ in $(a,b)$ and $\xi = 0$ in $(c,d)$. We find that the Chern number of the bands vanishes, but the quantum geometry of the bands remains nontrivial as measured by the integrated Fubini-Study metric $G$. \cite{Parameswaran13}.}
    \label{minusDcond}
\end{figure}

Lastly, we comment on the conduction bands for $V<0$. In this case, the potential pushes the electronic excitations onto the moir\'e potential, opening up gaps and isolating the conduction band. \Eq{eq:cherncond3} predicts the Chern number of the bands to be zero, which we confirm numerically from the Wilson loop\cite{2012arXiv1208.4234A} in \Fig{wilson}.
However, we show in \Fig{minusDcond} that the quantum geometry of the bands is nontrivial, as measured by the integrated Fubini-Study metric
\bea
\frac{G}{2\pi} = \int \frac{d^2k}{(2\pi)^2} \frac{1}{2} \Tr \del_i P \del_i P
\eea
where $P$ is the projector onto the conduction band. Note that even bands with trivial irreps may have nontrivial quantum geometry and even nontrivial lower bounds in lattice models \cite{2022PhRvL.128h7002H}.

\subsection{Atomic Nature of the Valence Band}

\begin{figure}
    \centering
    
    \includegraphics[width=1.\linewidth]{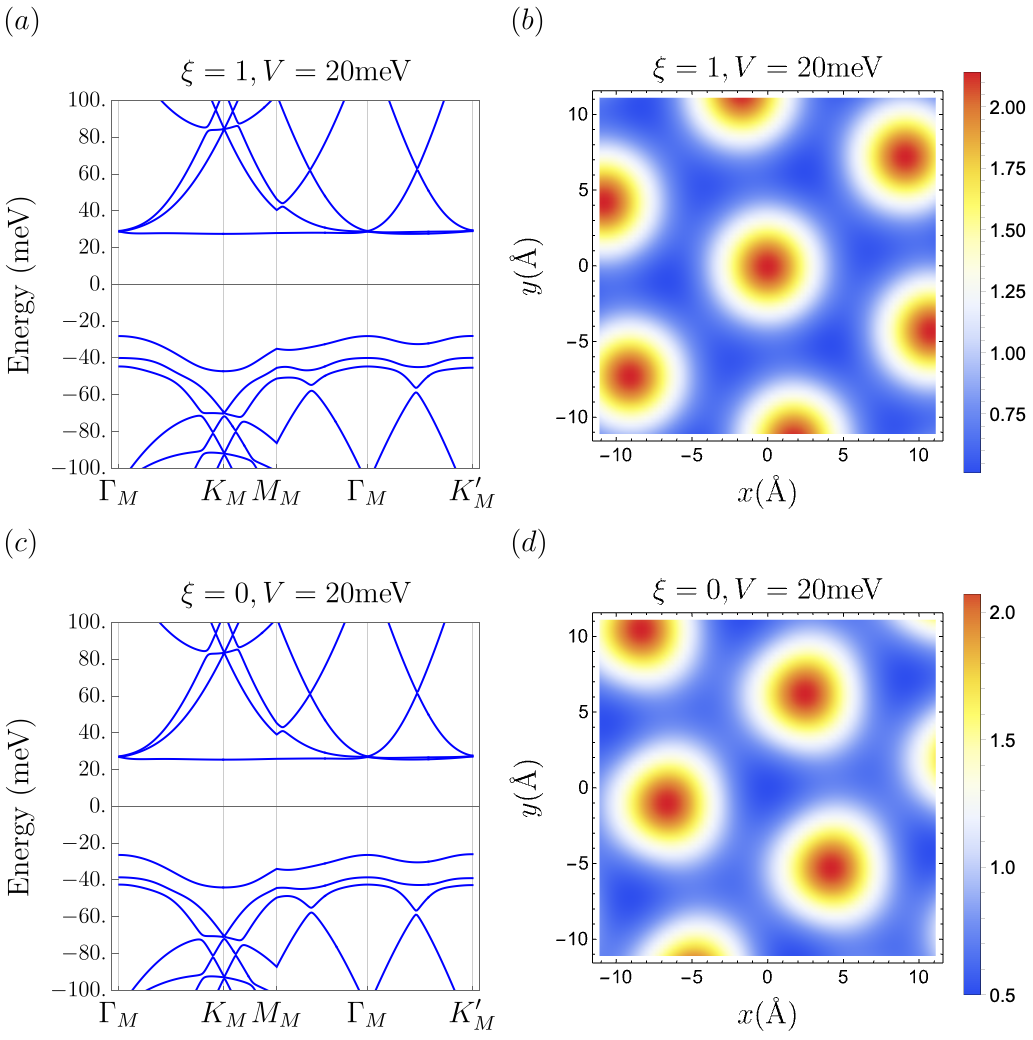}

    \caption{We plot the band structure at $V = 20$meV for $\xi= 1$ (a) and $\xi=0$ (c). 
    In $(b)$ and $(d)$ we compute the real-space density profile of the highest valence band for $\xi=1$ and $\xi=0$, respectively, which matches the Wannier center derived from the irreps. }
    \label{rhoforvalence}
\end{figure}

We can also apply the tripod model to the \emph{valence} bands, whose energies are
\bea
\eps^{(v)}_{j-1}(\pm \tilde{\mbf{K}}_M) &= f_0 + V_1 \cos\psi_{\xi,\mp j} \\
&\!\!\! - \sqrt{f_{\pm}+\big(\frac{V_0}{2}+V_3(|\tilde{\mbf{K}}_M|) +V_1 \cos \psi_{\xi, \mp j} \big)^2}\\
\eea
for the $\omega^j$ irrep, again showing that $\eta_{\tilde{\mbf{K}}_M} \eta_{\tilde{\mbf{K}}_M'} = \omega$ independent of the model parameters. At the $\tilde{\mbf{\Gamma}}_M$ point, \Eq{eq:etagamma} shows that the $\omega^*$ irrep is always occupied for $2V \geq V_0/2$, and thus that the flat bands have $C=0 \mod 3$ when $2V \geq V_0/2$. In fact, we see from \Fig{single_irreps} that their irreps correspond to trivial atomic limits formed of $p_x - i p_y$ orbitals transforming in the $\omega^*$ irrep using topological quantum chemistry \cite{2017Natur.547..298B}. For $\xi = 0$, this atomic orbital is located at the moir\'e unit cell corner (the AB site), and for $\xi = 1$ at the origin (the moir\'e AA site). Using \emph{Wannier90}~\cite{wannier90}, we find that for $V=20$meV, the localized Wannier function of the top valence band for $\xi = 0$  ($\xi = 1$) has square root of Wannier spread $0.63 a_M$ ($0.66 a_M$) where $a_M$ is the moir\'e lattice constant, and the hopping between the nearest-neighbor Wannier functions has amplitude $1.97$meV ($2.01$meV).
Therefore, they are localized Wannier functions with weak hopping among them, and the localization properties of those modes can also be seen in the charge density plots in \Fig{rhoforvalence}.
The location of the atomic orbitals can be understood from the effective model in \Eq{eq:onlyA}, whose minimum is at the origin for $|\psi_\xi| \leq \pi/3$ and at the moir\'e unit cell corners otherwise.

The Wannier spreads of the localized Wannier orbitals can be decreased if we mix the states of top valence band by those of second and third top valence bands \emph{away} from the high-symmetry momenta---such that we can keep the symmetry reps unchanged and thus can keep the Wannier center fixed. 
As a result, for $V=20$meV, we find a localized Wannier function for $\xi = 0$ ($\xi=1$) with the square root of Wannier spread being $0.45 a_M$ ($0.45a_M$) and  the nearest-neighbor hopping amplitude being $2.02$meV (2.14meV), and the Wannier function has $88\%$ ($85\%$) average probability overlap with the top valence band.
Our analysis suggests a potential heavy-fermion framework for the model presented here \cite{2022PhRvL.129d7601S,2023LTP....49..640C,PhysRevB.108.035129,2022PhRvL.129d7601S,PhysRevB.106.245129,2023arXiv230414064H,2023arXiv231002218S,2023arXiv230302670L,2023PhRvL.131b6502H,2023arXiv230908529R,zhang2023edge}.

\section{Doubly Aligned Rhombohedral Graphene on Hexagonal Boron Nitride}
\label{doublyaligned}  

\begin{table}[t]
    \centering
\begin{tabular}{c|cccccc}
 & $V_{b0}$ & $V_{b1}$ & $\psi_{b\xi}$ & $V_{t0}$ & $V_{t1}$ & $\psi_{t\xi}$ \\
\hline
 \text{$n=3$, $(\xi_b, \xi_t)$=(1,1)} & 0 & 3.20 & \text{16.55${}^{\circ}$} & 0 & 11.09 & \text{16.55${}^{\circ}$} \\
 \text{$n=4$, $(\xi_b, \xi_t)$=(1,1)} & 1.44 & 5.76 & \text{16.55${}^{\circ}$} & 5.40 & 7.08 & \text{16.55${}^{\circ}$} \\
 \text{$n=5$, $(\xi_b, \xi_t)$=(1,1)} & 1.50 & 7.29 & \text{16.55${}^{\circ}$} & 6.48 & 7.91 & \text{16.55${}^{\circ}$} \\
 \text{$n=6$, $(\xi_b, \xi_t)$=(1,1)} & 1.56 & 6.02 & \text{16.55${}^{\circ}$} & 7.52 & 7.78 & \text{16.55${}^{\circ}$} \\
 \text{$n=7$, $(\xi_b, \xi_t)$=(1,1)} & 1.47 & 5.45 & \text{16.55${}^{\circ}$} & 5.79 & 7.79 & \text{16.55${}^{\circ}$} \\ 
 \hline
 \text{$n=3$, $(\xi_b, \xi_t)$=(1,0)} & 0 & 6.72 & \text{16.55${}^{\circ}$} & 0 & 6.72 & \text{-136.55${}^{\circ}$} \\
 \text{$n=4$, $(\xi_b, \xi_t)$=(1,0)} & 1.44 & 7.65 & \text{16.55${}^{\circ}$} & 1.44 & 7.65 & \text{-136.55${}^{\circ}$} \\
 \text{$n=5$, $(\xi_b, \xi_t)$=(1,0)} & 1.50 & 5.43 & \text{16.55${}^{\circ}$} & 1.50 & 5.43 & \text{-136.55${}^{\circ}$} \\
 \text{$n=6$, $(\xi_b, \xi_t)$=(1,0)} & 1.56 & 7.80 & \text{16.55${}^{\circ}$} & 1.56 & 7.80 & \text{-136.55${}^{\circ}$} \\
 \text{$n=7$, $(\xi_b, \xi_t)$=(1,0)} & 1.47 & 7.22 & \text{16.55${}^{\circ}$} & 1.47 & 7.22 & \text{-136.55${}^{\circ}$} \\
\end{tabular}
    \caption{
    Parameter values of the $2n\times 2n$ model for $n=3,4,5,6,7$ layers for the hBN/R$n$G/hBN structures. The R$n$G parameters $v_F,v_3,t_1,t_2$ are the same as those in \tabref{tab:parameters_full}.
    $V_{b0}$, $V_{b1}$, $V_{t0}$ and $V_{t1}$ are in meV. 
    }
    \label{table:double_parameters}
\end{table}

\begin{figure*}[t!]
    \centering
    \includegraphics[width=1.\linewidth]{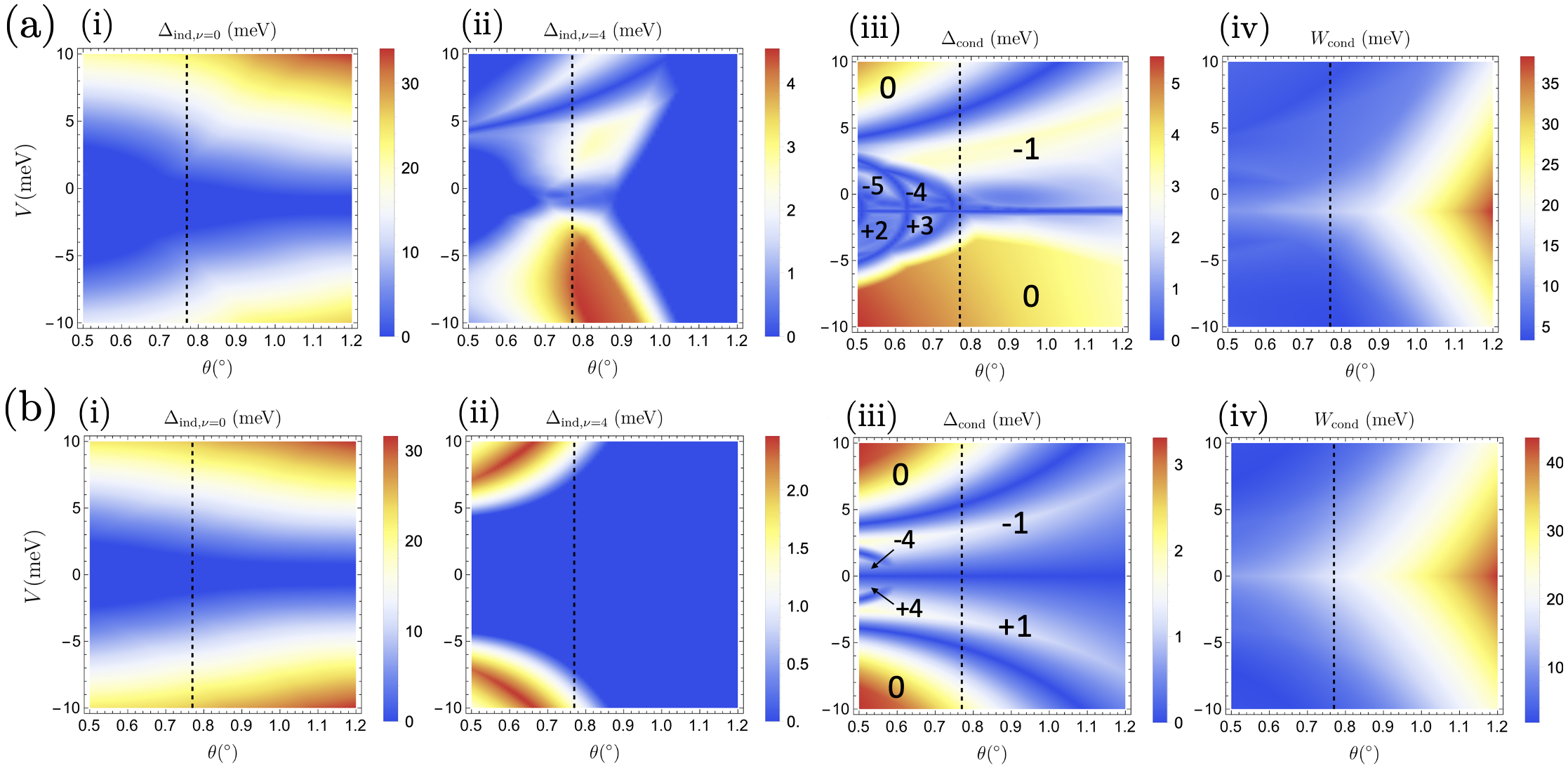}
    \caption{Phase diagrams of hBN/R5G/hBN with stacking orientation $(\xi_b,\xi_t)=(1,1)$ in (a) and $(\xi_b,\xi_t)=(1,0)$ in (b). Panel (i) shows the indirect gap at filling $\nu=0$ (at charge neutrality), (ii) shows the indirect gap at $\nu=4$, (iii) shows the minimal direct gap around the lowest conduction band, and (iv) shows the bandwidth of the lowest conduction band. Chern number of the lowest conduction band is indicated in panel (iii) where boundaries of topologically distinct phases can be seen as the direct gap closes. The dashed line is a reference to the twist angle at $\theta=0.77^\circ$.}
    \label{double_encap_PD}
\end{figure*}

\begin{figure}
    \centering
    \includegraphics[width=1.\linewidth]{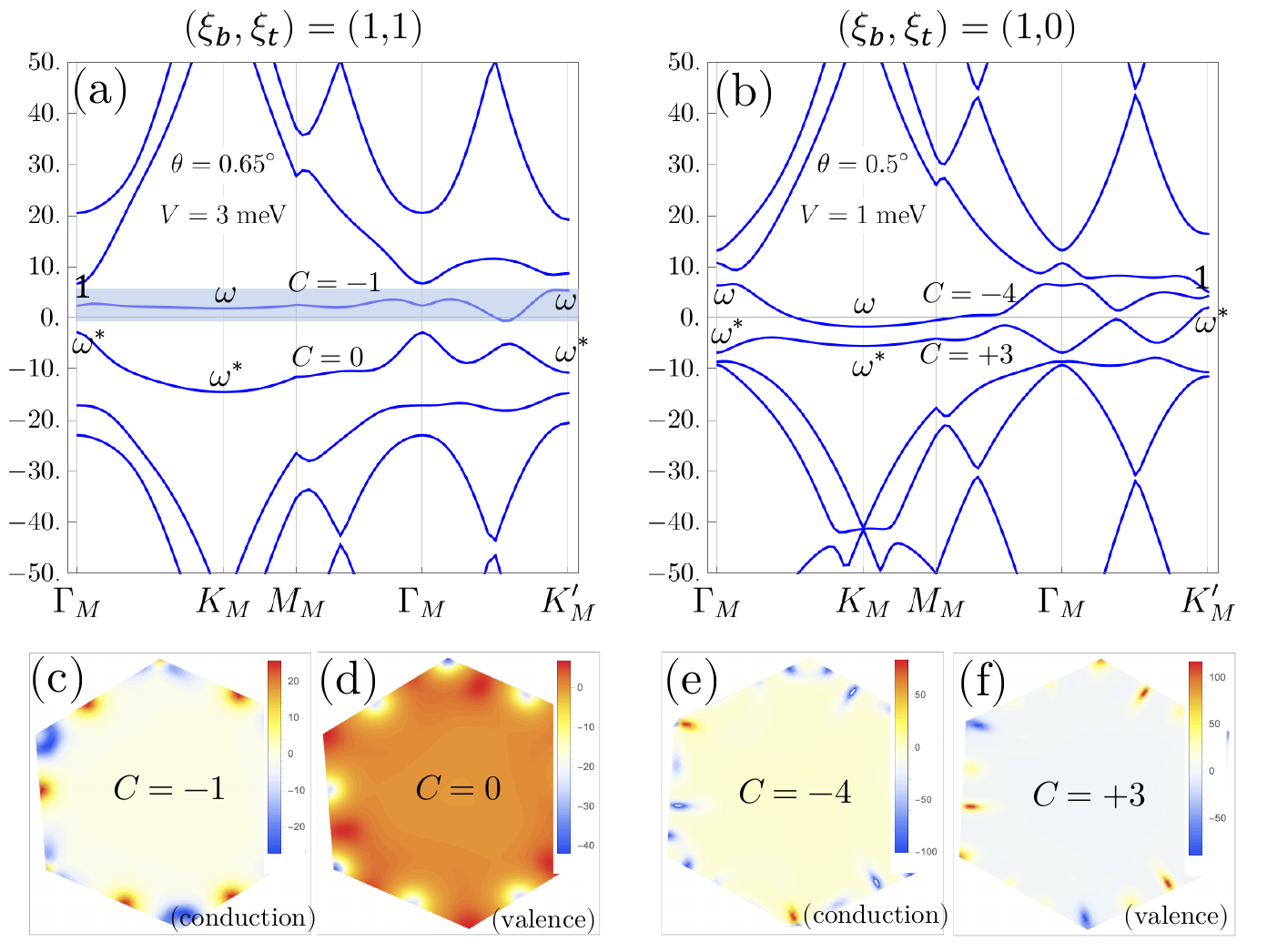}
    \caption{Topology and $C_3$ irreps for hBN/R5G/hBN. (a) shows the band structure for stacking orientation $(\xi_b,\xi_t)=(1,1)$ where there is a $\sim 1$ meV indirect gap at filling $\nu=4$ for the lowest conduction band with Chern number $C=-1$. (b) shows the band structure for stacking orientation $(\xi_b,\xi_t)=(1,0)$ with a $C=-4$ lowest conduction band. (c-f) show the Berry curvature distribution for the bands labeled in (a) and (b).}
    \label{double_irreps}
\end{figure}

We have shown that the form of the continuum Hamiltonian fit to the first principles band structure cannot reproduce the Chern number $|C|=1$ found in experiment, and thus relies on interaction effects. We now propose hdoubly aligned hBN-encapsulated devices (hBN/R$n$G/hBN) as an alternative platform for moir\'e FCIs where various Chern numbers (in the $K$ valley) can be obtained even at the single-particle level. 

We perform first principles calculations to get the relaxed structures for hBN/R$n$G/hBN, where the hBN layers are parallel on top and bottom. With the relaxed structure, its band structure is then computed in the SK method. To fit the DFT+SK band structure with a moir\'e model, we add a second moir\'e potential $V_{\xi,t}(\mbf{r})$ acting only on the top layer. The full Hamiltonian can be written
\bea
H_{\K,\xi_t,\xi_b} &= \bpm
\sigma_1^{\xi_t} H_{BN} \sigma_1^{\xi_t} & \tilde{T}_n(\mbf{r}) & \\
\tilde{T}_n^\dag(\mbf{r}) & H_{\K}(-i \pmb{\nabla}) & \tilde{T}^\dag(\mbf{r}) \\
 & \tilde{T}(\mbf{r}) & \sigma_1^{\xi_b} H_{BN} \sigma_1^{\xi_b} \\
\epm \\
\eea
where $\xi_t, \xi_b \in \{0,1\}$ are the stacking orders of the top and bottom hBN, and $[\tilde{T}_n(\mbf{r})]_{l} = \delta_{l,n} T_n(\mbf{r})$ with the shifted hopping matrix
\bea
T_n(\mbf{r}) &=  \sum_{j=1}^3 e^{i \mbf{q}_{j} \cdot \mbf{r}} e^{- \frac{2\pi i}{3}(j-1)(n-1)} T_j
\eea
as shown in \App{app:moirecoupling}. 
The parameter values are listed in \tabref{table:double_parameters}, and the match between the model and the DFT+SK calculation is shown in Figs.\ref{fig:fitting_3}-\ref{fig:fitting_7} of \appref{app:fitting}, where we also provide the details on the fitting procedure.
The nature of hBN/R$n$G/hBN is rather different since both chiral modes feel a moir\'e potential as well as hybridizing strongly for small $V$. The phase diagram of the system for $n=5$ is shown in Fig.\,\ref{double_encap_PD}, and representative band structures are shown in Fig.\,\ref{double_irreps}. (A complete set of phase diagrams for hBN/R$n$G/hBN  with $n=3,4,...,7$ are shown in Figs.\ref{3LG_phasediag}-\ref{7LG_phasediag}.) We find that hBN/R$n$G/hBN  structures showcase a much richer landscape of Chern numbers than the R$n$G/hBN case.

The $(\xi_b, \xi_t) = (1,1)$ case shows \textit{isolated} $C=-1$ conduction bands accessible at small positive $V$ which provides an alternative parent state for moir\'e FCIs. By adjusting both $V$ and $\theta$, one can now access $C$ for all integer values between 0 and 5, owing to the multitude of gap closing transitions in this system: starting from $V \sim $ 1 meV and $\theta \gtrsim 0.8^\circ$ (where $C = -1$) and decreasing the twist angle first leads to three $C_3$-related gap closings between the conduction and valence bands near the middle of the $\Gamma_M$-$K'_M$ line, which change the Chern number by 3 (to $C=-4$). Further decreasing the twist angle would lead to another gap closing between the conduction and valence bands at $\Gamma_M$, which changes the Chern number by 1 (to $C=-5$). A similar sequence of gap closing transitions happen for small negative $V$, which change $C=0$ to $C=3$, and then to $C=2$. 

The $(\xi_b,\xi_t)=(1,0)$ case also hosts $|C|=1,4$ conduction bands, while $|C|=3$ can be found in the valence bands. There are again three $C_3$-related gap closings that happen near the middle of the $\Gamma_M$-$K'_M$ line, which connect the $C=\pm 4$ phase to the $C=\pm 1$ phase. The gap closing transitions connecting $C=4$ to $C=-4$ (as well as those connecting $C=2$ to $C=-5$, or $C=3$ to $C=-4$, in the $(\xi_b,\xi_t)=(1,1)$ configuration) are more complicated, and will be saved for future works.

The $(\xi_b,\xi_t)=(1,0)$ case has a notable feature that for a fixed twist angle $\theta$, the Chern number flips sign under the reversal of $V$. This can be simply understood from the fact that the $(\xi_b,\xi_t)=(1,0)$ configuration has inversion symmetry at $V=0$, while $V$ and $-V$ are related by inversion. Combined with time-reversal symmetry such that we remain in the original valley sector, we see that the sign-reversal of $V$ is accompanied the sign-reversal of $C$. This explains the symmetry of the phase diagram in Fig. \ref{double_encap_PD} (b)(iii).

These rich double-aligned structures may serve as a more versatile platform for realizing FCIs from higher Chern bands, which are not adiabatically connected to Fractional Quantum Hall states. Our results show that stacking, alignment, and twist angle are all useful tuning knobs for the R$n$G-hBN family, most of which remains to be investigated. Finally, we have also derived the $2\times 2$ moir\'e effective model for the doubly-aligned structure, and determined the model parameters by fitting the DFT+SK bands, which are discussed in \appref{app:fitting}.
We find that the $2\times 2$ effective moir\'e model can capture most of the low-energy features of the DFT+SK bands.

\section{Conclusion}
\label{sec:conclusion}

In this work, we studied the single-particle band structure and topology of R$n$G/hBN and hBN/R$n$G/hBN for $n=3,\dots, 7$ layers. Starting with large-scale DFT+SK calculations of the relaxed moir\'e band structure, we find that the relaxation can cause non-negligible quantitative changes of the bands structure (changes as large as $\sim 10$meV).
We then adapted the continuum model proposed in Ref.\,\cite{PhysRevB.89.205414}, finding changes in the parameters due to relaxation effects within the graphene lattice induced by the hBN.
This model has size of $2n \times 2n$ at each position, but we showed that a minimal $2\times 2$ continuum model built on the chiral low-energy modes could faithfully reproduce the low energy bands. 
Using this model to study the experimentally relevant R5G/hBN, we find the Chern numbers of the lowest conduction and highest valence bands are 0 or 5, and we analytically explain them and their robustness against parameter tuning. We found that for R5G/hBN, a large displacement field pointing away from hBN leads to nearly free conduction electrons, while the lowest valence band forms an isolated Wannierizable band.
For a large displacement field pointing toward hBN, the conduction bands of R5G/hBN are topologically trivial, but nevertheless show nontrivial quantum geometry. 

Reference \cite{Ju2023PentalayerGraphenehBN} has already demonstrated the presence of interaction-induced phase beyond the single-particle phase diagram studied here, including correlated trivial and Chern insulators and the discovery of moir\'e FCIs. 
Interestingly, for a large displacement field pointing toward hBN, the cascade of insulators were observed in R5G/hBN in Ref.\cite{Ju2023PentalayerGraphenehBN} in the conduction bands at $\nu = 2,3,4$ (with faint signatures at $\nu =1$ for $V<0$, which is similar to the phenomenology of twisted bilayer graphene, where correlated insulators appear at integer filling~\cite{Cao2018TBGMott,Cao2018TBGSC,2020PhRvX..10c1034B,PhysRevLett.122.246401,2022PhRvL.129d7601S, PhysRevX.11.041063,PhysRevLett.128.156401,2023Natur.620..525N} of the flat band manifold.
Besides the CI and FCI phases in R$5$G/hBN, various spontaneous symmetry breaking phases, as well as superconductivity, have also been found in the R$n$G family with and without moir\'e coupling \cite{ref29, ref43, ref44, ref45, ref46, ref47,ref48,ref49,ref50,ref51,ref52,ref53,ref54,ref55,ref56}.
Taken together, this family of systems provides an unprecedented opportunity to study the interplay among quantum geometry, topology, and strong electronic correlation~\cite{PhysRevLett.124.167002,2021Natur.600..641L,yankowitz2022moire,2022arXiv220702312Z,mak2022semiconductor,mai2023interaction,2023arXiv230914340S,PhysRevB.103.205415,2022PhRvL.129d7601S,Bultinck2019GroundSA,2023PhRvB.107x5145S,2022arXiv221200030H,Phillips2023DQMCHaldaneModel,2023arXiv230105588W,Phillips2023DQMC, ding2023particle,mai2023interaction,fu2021flat,crepel2023topological,PhysRevLett.123.237002,PhysRevLett.124.167002,PhysRevB.101.060505,rossi2021quantum}.

\section{Acknowledgments}
The authors thank W.Q. Miao for helpful discussions.
J.Y., J. H.-A., P.M.T. are grateful for conversations with Yves Kwan and Chao-Xing Liu. 
This work was supported by the Ministry of Science and Technology of China (Grant No. 2022YFA1403800), the Science Center of the National Natural Science Foundation of China (Grant No. 12188101) and the National Natural Science Foundation of China (Grant No.12274436). H.W. acknowledge support from the Informatization Plan of the Chinese Academy of Sciences (CASWX2021SF-0102). 
B. A. B.’s work was primarily supported by the DOE Grant No. DE-SC0016239 and the Simons Investigator Grant No. 404513. N.R. also acknowledges support from the QuantERA II Programme that has received funding from the European Union’s Horizon 2020 research and innovation programme under Grant Agreement No 101017733 and from the European Research Council (ERC) under the European Union’s Horizon 2020 Research and Innovation Programme (Grant Agreement No. 101020833). 
J. Y. is supported by the Gordon and Betty Moore Foundation through Grant No. GBMF8685 towards the Princeton theory program and through the Gordon and Betty Moore Foundation’s EPiQS Initiative (Grant No. GBMF11070) and NSF-MERSEC DMR-2011750.
P.M.T. is supported by a postdoctoral research fellowship at
the Princeton Center for Theoretical Science and a Croucher Fellowship.
J. H.-A. is supported by a Hertz Fellowship, with additional support from DOE Grant No. DE-SC0016239 by the Gordon and Betty Moore Foundation through Grant No. GBMF8685 towards the Princeton theory program, Office of Naval Research (ONR Grant No. N00014-20-1-2303), BSF Israel US foundation No. 2018226 and NSF-MERSEC DMR-2011750, Princeton Global Scholar and the European Union’s Horizon 2020 research and innovation programme under Grant Agreement No 101017733 and from the European Research Council (ERC). 

\bibliography{fcitmds}
\bibliographystyle{apsrev4-1}

\cleardoublepage

\tableofcontents

\appendix
\onecolumngrid

\section{Model Hamiltonians for Rhombohedral Graphene twisted on hBN}\label{app:microscopic_model}

In this Appendix, we present a bottom-up derivation of the continuum Hamiltonian for the active bands of $n$-layer rhomohedral graphene twisted on top of or encapsulated by hexagonal boron nitride (hBN). We provide three models: (1) a fully microscopic model with hoppings onto the hBN layer(s), (2) a carbon-only model with the hBN layer(s) integrated out to the leading order, and (3) an effective model obtained by projection onto two gapless states of isolated rhombohedral graphene. The parameters of this minimal model are then optimized numerically to match the \emph{ab-initio} band, as elaborated in \appref{app:fitting}. 

\subsection{Rhombohedral Graphene}

First we discuss the Hamiltonian of rhombohedral $n$-layer graphene (R$n$G). To set our conventions, we begin with the minimal two-orbital tight-binding model of graphene with nearest-neighbor hopping $t_0$ between the $p_z$ carbon orbitals on the two sublattice sites:
\bea
\label{eq:h_1}
h_1(\mbf{k}) &= - t_0 \bpm 0& \sum_{j} e^{i \mbf{k} \cdot \pmb{\delta}_j} \\
 \sum_{j} e^{-i \mbf{k} \cdot \pmb{\delta}_j} & 0\epm
\eea
where $\pmb{\delta}_1 = (0,a_G/\sqrt{3}), \pmb{\delta}_2 = C_3 \pmb{\delta}_1, \pmb{\delta}_3 = C_3 \pmb{\delta}_2$ and $a_G = 0.246$nm is the graphene lattice constant. The Dirac points are located at 
\bea
\mbf{K} = 2\pi \lp \frac{2}{3 a_G}, 0\rp, \quad \mbf{K}' = - \mbf{K} \ ,\\
\eea
which are related by time-reversal. We will focus on the low-energy states near the $\mbf{K}$ point, which we refer to as the $\K$ valley. Expanding the Hamiltonian and defining $v_F =  \frac{3}{2} t_0 a_G $, we find
\bea
h_1(\mbf{K} + \mbf{k}) &= v_F \mbf{k} \cdot \pmb{\sigma} + \dots \ . \\
\eea

We now consider R$n$G where each layer is shifted by $\pmb{\delta}_1$ relative to the one below, resulting in a $2n\times 2n$ Hamiltonian whose basis is layer $\otimes$ 
sublattice. Expanding in the $\K$-valley, this model takes the form
\bea
\label{eq:Hrhomb}
H_{n}(\mbf{K}+\mbf{k}) &= \bpm
v_F\mbf{k} \cdot \pmb{\sigma} & t^\dag(\mbf{k}) & \\
t(\mbf{k}) & \ddots & t^\dag(\mbf{k}) \\
 & t(\mbf{k})  & v_F\mbf{k} \cdot \pmb{\sigma}\\
\epm + H_{ISP} + H_{t_2}, \qquad t(\mbf{k})= \bpm - v_4 k & t_1 \\ -v_3 \bar{k} & - v_4 k \epm, \quad k , \bar{k} = k_x \pm i k_y \\
\eea
where $t(\mbf{k})$ is the inter-layer coupling matrix. The inter-layer AB hopping is $t_1$, and the next-nearest inter-layer hoppings yield the effective velocity terms $v_3$ (inter-layer AB coupling) and $v_4$ (inter-layer AA/BB coupling), see Fig. \ref{holomorphic} in the main text. The term $H_{ISP}$, which reads
\eq{
[H_{ISP}]_{l l'} = V_{ISP} \delta_{ll'} | l - \frac{n+1}{2}|\ ,
}
with $l = 1,\dots, n$, describes the differences in the local chemical environment on the internal versus external graphene layers due to an effective inversion-symmetric potential. The value $V_{ISP}/c = 5\meV/\AA$ is fit from the DFT on pristine pentalayer graphene and reflects the higher chemical potential of the outer layers, where $c$ is the interlayer distance between graphene layers. 
Finally, 
\eq{
[H_{t_2}]_{ll'} = \delta_{l,2+l'} t_2 \frac{\sigma_1 - i \sigma_2}{2}
}
is a coupling between $A$ and $B$ carbon orbitals 2 layers apart. Although this coupling is small ($t_2 = -7$meV) it is important to include it in the $n=3$ case, where $t_2$ directly couples the zero energy states at $\mbf{k}=0$ (see \Fig{layer-graphene-vasp}a), and opens up a gap there. For consistency, we include it for all number of layers. 
We now derive the values of these couplings.

To do so, we employ a generalized Slater-Koster (SK) approach which parameterizes the hoppings between any two orbitals by their SO(3) character (spherical harmonic) and the distance between them $\mbf{r}$ (the so-called two-center approximation). We use the following form of the SK hopping between $p_z$ orbitals with parameters fit to match ab-initio:
\bea
\label{eq:SK}
t_{SK}(\mbf{r}) &= V_{pp\pi} (1 - \frac{z^2}{r^2}) e^{q_\pi (1 - r/a_\pi)}/(1+ e^{(r-r_c)/l_c}) + V_{pp\sigma} \frac{z^2}{r^2} e^{q_\sigma (1 - r/a_\sigma)}/(1+ e^{(r-r_c)/l_c}), \quad r = |\mbf{r}|, \ z = \hat{z} \cdot \mbf{r}
\eea
and the parameters are
\bea
V_{pp\pi} &= -2810 \text{meV}, \ V_{pp\sigma} = 480 \text{meV}, \ l_0 = 3.364 \AA, \ q_\pi = 3.1451, \ a_\pi = 1.418 \AA, \\ 
q_\sigma &= 7.428, \ a_\sigma = 3.349\AA, \ r_c = 6.14 \AA, l_c = 0.265\AA \ .
\eea
This parameterization is fit to the DFT calculation of the pristine R$n$G for $n=3,4,5,6,7$. The resulting graphene tight-binding model parameters can be computed by performing the SK sums
\bea
h_{l \al, l' \be}(\mbf{p}) = \sum_{\mbf{R}} t(\mbf{r}_{l,\al} - \mbf{r}_{l',\be} -\mbf{R}) e^{-i (\mbf{r}_{l,\al} - \mbf{r}_{l',\be} -\mbf{R}) \cdot \mbf{p}}
\eea
which converges exponentially in $\mbf{R}$. 
We note that $v_3 = v_4$ in the SK two-center approximation since the nearest-neighbor distance between the AB and AA/BB inter-layer orbitals is the same. Throughout this work, we keep $v_3 = v_4$ since we find that allowing them to differ does not noticeably improve the fits. 

This completes our derivation of the microscopic rhombohedral graphene Hamiltonian. In the next section, we study its symmetries and low-energy spectrum.

\subsection{Band Flattening with Displacement Field in Rhombohedral Graphene}
\label{app:symmSO2}

The rhombohedral band structure is tunable in experiment by displacement field, creating an inter-layer potential $V$. In this subsection, we study the behavior of the R$n$G bands in $V$ to understand the flattening of the low energy spectrum. 

In order to first understand the essential physics, we set $v_3 = v_4 = 0$ and $t_2=0$ (and we will restore them for a full analysis in the main text, as well as in App. \ref{app:effmodelderivation}), in which case the Hamiltonian is fully isotropic and the spectrum is a function of $|\mbf{k}|$ only. Explicitly, the model in this limit, which is called $h_n(\mbf{k})$, reads
\bea
\label{eq:hchiralcomp}
\null [h_n(\mbf{k})]_{ij} = v_F \delta_{ij} \mbf{k} \cdot \pmb{\sigma} + t_1 \delta_{i,j+1} \sigma^+ + t_1 \delta_{i,j-1} \sigma^-
\eea
which has the chiral symmetry $\Sigma h_n(\mbf{k})\Sigma^\dag = - h_n(\mbf{k})$ with $\Sigma = \mathds{1} \otimes \sigma_3$ where $\mathds{1}$ is the identity on the $n$ layers. The other important symmetry obtained by this model is $SO(2)$ rotation, which takes the form
\bea\label{eq: SO2rotation}
h_n(R_\th \mbf{k}) = D_\th h_n(\mbf{k}) D^\dag_\th, \qquad [D_\th]_{ll'} = \delta_{ll'} e^{i \th (l-1 - \lfloor \frac{n}{2} \rfloor)} e^{-i \th \sigma_3/2}
\eea
corresponding to the angular momenta $-n/2, -n/2+1,-n/2+1,-n/2+2,\dots, n/2$. Of course, the realistic model with $v_3 \neq 0$ only possesses $C_3$ symmetry, which we obtain from $D[C_3] = -D_\frac{2\pi}{3}$ (with the $-1$ phase determined by requiring $D[C_3]^3 = \mathds{1}$). There is also spacetime-inversion symmetry $D_{ll'}[\mathcal{I} \mathcal{T}] = \delta_{l,-l'} \sigma_1 \mathcal{K}$ which is intra-valley because inversion and time-reversal both flip the valley. This symmetry is broken by the displacement field. While $C_{2z}\mathcal{T}$ is not broken by the displacement field, it is broken by stacking structure of the R$n$G for $n>1$.

\begin{figure}
    \centering
    \includegraphics[width=.5\linewidth]{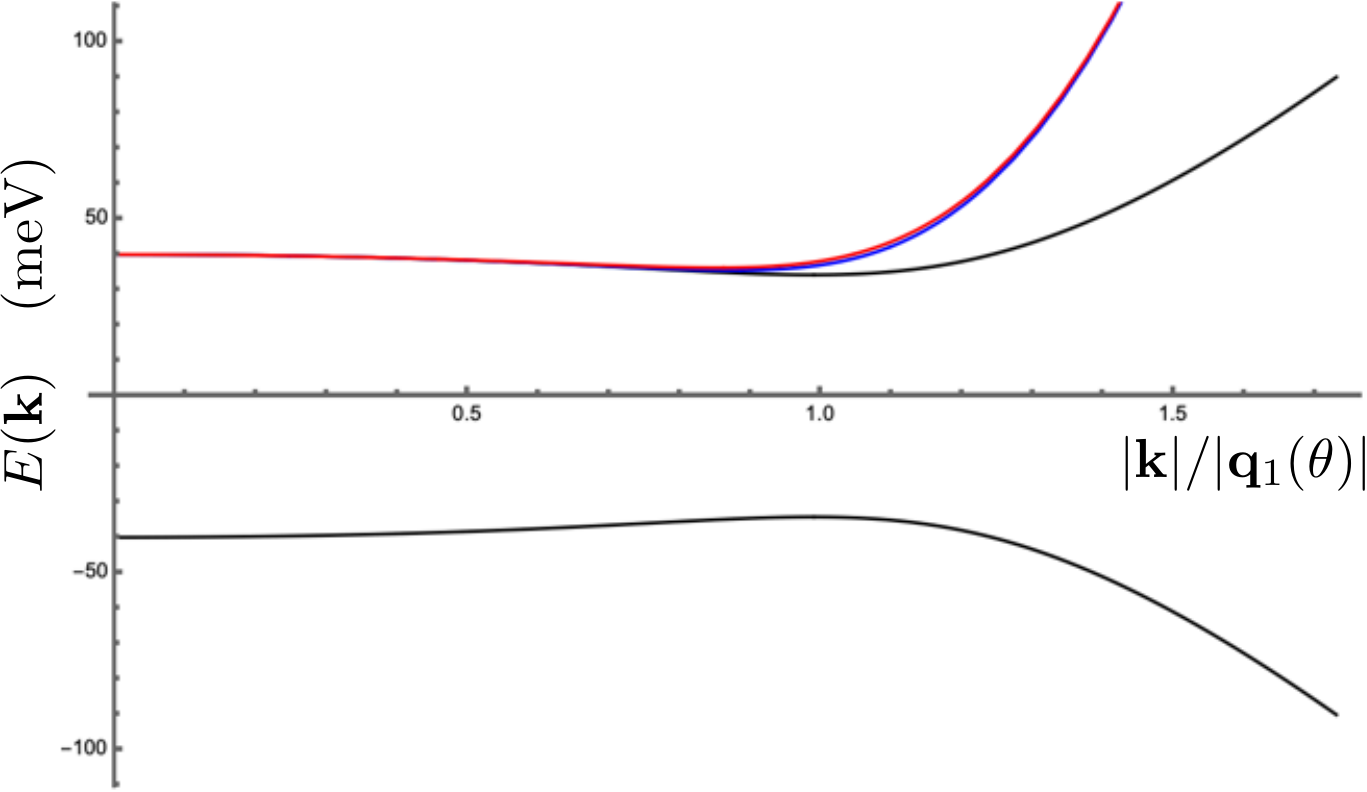}
    \caption{Dispersion $\pm E(\mbf{k})$ of the $10\times 10$ pentalayer Hamiltonian is shown in black, in comparison with the analytical energies of the projected model of \Eq{eq:projmodel} (blue) at $V = 10$meV. The red curve is the asymptotic approximation $\sqrt{(2V)^2 + |v_F k|^{10}/t_1^8 - 4 V^2 |v_F k|^2 /t_1^2}$ for $\th = 0.767^\circ$. }
    \label{fig:EkvFcomparison}
\end{figure}

We start in the $v_3 = v_4 = 0$ and $t_2=0$ limit where we can expose some simple analytical results.  The $SO(2)$ symmetry requires that the spectrum $E_n(\mbf{k})$ depends only on $|\mbf{k}|$. Then we can expand the characteristic polynomial of $h_n(\mbf{k})$ in $|\mbf{k}|$ to find that
\bea
E_n(\mbf{k}) = \pm \frac{(v_F|\mbf{k}|)^n}{t_1^{n-1}}, \quad \pm t_1 + \dots
\eea
where the $\pm t_1$ eigenvalues are $(n-1)$-fold degenerate and correspond to bonding/anti-bonding inter-layer dimers hybridized by $t_1$. By direct substitution, one can verify that the eigenspace of the $O(|k|^n)$ eigenvalues is spanned by the holomorphic states (up to normalization):
\bea\label{eq:lowE_basis}
\psi_{A}(k) &= \{1,0,(-v_F k/t_1),0,(-v_Fk/t_1)^2,0,\dots \} + O(k^{n}) \\
\psi_{B}(\bar{k}) &= \{0,(-v_F\bar{k}/t_1)^{n-1},0,(-v_F\bar{k}/t_1)^{n-2},\dots , 0, 1\} + O(\bar{k}^n) \\
\eea 
or in components $[\psi_{A}(k)]_{A_l} = (-v_F k/t_1)^{l-1}$ and $[\psi_{B}(\bar{k})]_{B_l} = (-v_F \bar{k}/t_1)^{n-l}$. Here $k = k_x + i k_y = |k| e^{i \phi}, \bar{k} = k_x - i k_y = |k| e^{-i \phi}$ are the usual holomorphic coordinates, and $\psi_{A,B}$ are sublattice polarized since they diagonalize the chiral operator $\Sigma$, which is sublattice diagonal. Given that $v_F|k|/t_1 < 1$, $\psi_A(k)$ has its maximum weight on the top layer and $\psi_{B}(\bar{k})$ on the bottom, and they exponentially decay away from the top and bottom layers respectively. Indeed, $v_F |\mbf{q}_1(\th)| /t_1 \sim 0.5$ at $\th = 0.767^\circ$ ($\mbf{q}_1(\th)$ is the relevant BZ scale at the experimental moir\'e angle) so that the perturbation theory will be qualitatively valid across the first moir\'e BZ.  We gather the chiral states (which we note are not \emph{eigenstates}) into the column vector $\Psi(\mbf{k}) = [\psi_{A}(k), \psi_{B}(\bar{k})] / |\psi_A(k)|$ (where $|\psi_A(k)| = |\psi_B(k)|$ is the normalization). From Eq.\,\ref{eq:hchiralcomp}, it can readily be verified that
\bea
\Psi^\dag(\mbf{k}) h_n(\mbf{k}) \Psi(\mbf{k}) &= \frac{v_F^n}{t_1^{n-1}} \bpm 0 & \bar{k}^n \\ k^n & 0 \epm + O(|k|^{n+1}) \\
\eea
which directly shows the $\pm |k|^n$ dispersion from coupling the chiral modes. Now we add a displacement field $[H_V]_{ll'} = (l -(n+1)/2) V \delta_{ll'}$ ($l=1,2,...,n$) with inter-layer potential difference $V$. We find
\bea
\label{eq:projmodel}
\mathcal{U}^\dag(\mbf{k}) \big( h_n(\mbf{k}) +H_V \big)\mathcal{U}(\mbf{k}) &= \frac{v_F^n}{t_1^{n-1}} \bpm 0 & \bar{k}^n \\ k^n & 0 \epm +V \Big[\frac{1-n}{2} + \sum_{m=1}^{2m\leq n}\big(\frac{v_F|k|}{t_1}\big)^{2m}\Big]\sigma_3
\eea
The closed form expression for the projection of the displacement field term can be found in Eq. \eqref{eq:closedform}. This Hamiltonian is in Pauli form and can be immediately diagonalized to yield (setting $n=5$)
\bea
\label{eq:Eplusminus}
E_\pm(k) = \pm \sqrt{ |v_F k|^{10}/t_1^8 + V^2 (2 - v_F^2 |k|^2/t_1^2 - v_F^4 |k|^4/t_1^4)^2}
\eea
which compares very well with the numerically diagonalized energies shown in \figref{fig:EkvFcomparison}. The most important feature of \Eq{eq:Eplusminus} is its non-monotonicity appearing from the $ - v_F^2 |k|^2/t_1^2 - v_F^4 |k|^4/t_1^4$ terms. Hence at small $\mbf{k}$, the energies will initially decrease, while at large $\mbf{k}$ they must approach infinity. Thus we can define a flatness condition
\bea
\label{eq:flatness_condition}
E_+(\mbf{k} =0) = 2V  = E_+(\mbf{k} =\mbf{q}_1(\th))
\eea
set by the scale of the moir\'e momentum. Here we used that at $\mbf{k}=0$, the full $2n \times 2n$ Hamiltonian can be diagonalized to yield $E(\mbf{0}) = 2V$ (assuming $t_2 = 0$). 

We can now obtain an estimate for the critical $V$ that satisfies the flatness condition in \eqnref{eq:flatness_condition}.
Using \Eq{eq:Eplusminus} to evaluate $E(\mbf{k} =\mbf{q}_1(\th))$ is straightforward, but it results in a high order polynomial equation to solve for the critical $V$. To get an analytical solution, we keep only the lowest $O(k^2)$ terms to capture the non-monotonicity and highest $O(k^{10})$ terms to capture the large $\mbf{k}$ behavior. This approximation is validated in \Fig{fig:EkvFcomparison}. Then we find
\bea
E_+(\mbf{k}) \sim  \sqrt{(2V)^2 + |v_F k|^{10}/t_1^8 - (2 V)^2 |v_F k|^2 /t_1^2}
\eea
leading to the flat-band condition being 
\bea
v_F^2 |\mbf{q}_1(\th)|^2 = t_1 \sqrt{2 V t_1} \ .
\eea
While $v_3, v_4, t_2, V_{ISP}$  will modify this result (see \App{app:effmodelderivation} for their effect), it serves to identify a maximally flattened region tuned by $D$ field, at least at the single-particle level.

\subsection{Moir\'e Hamiltonian of R$n$G/hBN and hBN/R$n$G/hBN}
\label{app:moirecoupling}

In this section, we derive the moir\'e Hamiltonian of the superlattices formed by R$n$G and hBN. This Hamiltonian has three parts: the intra-layer Hamiltonians for R$n$G and hBN, and the moir\'e coupling between the two as caused by their lattice mismatch and relative twist. We will derive the form of the moir\'e coupling from one graphene layer to the hBN, and then build the full Hamiltonian for the variety of possible configurations and encapsulations shown in \Fig{lattice_define}. 

In the experiment~\cite{Ju2023PentalayerGraphenehBN}, the R$n$G is encapsulated by two hBNs; however, only one of them is twisted at a small angle and thus nearly aligned, while the other does not contribute to the electronic structure of the system. To model this case, we consider the configuration where R$n$G is on top of one nearly-aligned hBN (R$n$Gg/hBN) without any hBN on the other side.
We will also consider the case where R$n$G is encapsulated by two nearly-aligned hBN that generate the same moir\'e pattern (hBN/R$n$G/hBN).

The $\mbf{k} \cdot \mbf{p}$ Hamiltonian of R$n$G is discussed in \eqnref{eq:Hrhomb}. We approximate the hBN Hamiltonian as
\bea
H_{BN, \xi} &= \sigma_1^\xi \bpm V_B  & \\ & V_N \epm \sigma_1^\xi, \qquad V_B = 3352 \text{meV}, \quad V_N =-1388 \text{meV}
\eea
for the stacking configuration $\xi = 0,1$ (see Main Text) where carbon-A,B is nearly vertically aligned with B,N or N,B respectively in the AA region. (See Fig.\,\ref{relaxation} in Main Text.) Here $V_B$ and $V_N$ are the chemical potentials for boron and nitrogen, respectively. Here we have neglected the $\mbf{k}$-dependence of the hBN altogether, which is acceptable because $V_B, V_N \sim 1000$ meV, and corrections from hBN dispersion will not affect the low energy graphene bands, which are near chemical potential $0$.

 The R$n$G/hBN devices have the following Hamiltonians in the $\K$ valley:
\bea
\label{eq:signelmodels}
H_{\K, \xi=0}(\mbf{r}) &= \bpm
H_{\K}(- i \pmb{\nabla})  & \tilde{T}_b^\dag(\mbf{r}) \\
 \tilde{T}_b(\mbf{r}) & H_{BN}\\
\epm, \qquad H_{\K, \xi=1}(\mbf{r}) = \bpm
H_{\K}(- i \pmb{\nabla}) & \tilde{T}_b^\dag(\mbf{r}) \\
 \tilde{T}_b(\mbf{r}) & \sigma_1 H_{BN}  \sigma_1 \\
\epm\\
\eea
describing the two possible stackings of the bottom hBN layer on the graphene. They are exchanged by a $C_2$ rotation of the hBN, but the models are not symmetry-related because R$n$G is not $C_2$-symmetry (it is inversion symmetric, which exchanges the top and bottom layers). The moir\'e coupling $\tilde{T}(\mbf{r})$ acts only on the bottom layer, meaning
\bea
\null [\tilde{T}_b(\mbf{r})]_{l} = \delta_{l,0} T_b(\mbf{r})
\eea
where $l = 0,\dots, n-1$ indexes the layers of R$n$G where $l=0$ is the layer that is closest to hBN. $T_b(\mbf{r})$ is a $2\times 2$ matrix as shown in \eqnref{eq:VhBNbottom}.

Secondly, we discuss the hBN/R$n$G/hBN models
\bea
H_{\K, \xi_b,\xi_t}(\mbf{r}) &= \bpm
\sigma_1^{\xi_t} H_{BN} \sigma_1^{\xi_t}& \tilde{T}_t(\mbf{r}) &  \\
\tilde{T}_t^\dag(\mbf{r}) & H_{\K}(- i \pmb{\nabla}) & \tilde{T}_b^\dag(\mbf{r}) \\
 & \tilde{T}_b(\mbf{r}) & \sigma_1^{\xi_b} H_{BN} \sigma_1^{\xi_b}\\
\epm, \qquad (\xi_t,\xi_b) = (1,1), (1,0)
\eea
which differ in the relative orientation of the N and B stackings on top and bottom. $H_{\K, \xi_b = 1, \xi_t=1}$ is strongly inversion asymmetric since the low-energy modes couple to different hBN orbitals on opposite sides, recalling that the bottom layer A-sublattice and top layer B-sublattice are the dominant orbitals in the chiral basis (see \Eq{eq:lowE_basis}).  In contrast,  $H_{\K, \xi_b = 1, \xi_t=0}$ is exactly inversion symmetric (nearly symmetric if the inversion center of two hBN deviates slightly from that of the R$n$G), since the bottom layer A-sublattice and top layer B-sublattice are aligned with the same atom. The moir\'e couplings $[\tilde{T}_t(\mbf{r})]_l = \delta_{l,n-1} T_t(\mbf{r}),\tilde{T}_b(\mbf{r}) = \delta_{l,0} T_b(\mbf{r})$ again connect hBN to the nearest graphene layer only. 

We will now derive the inter-layer moir\'e coupling $T_{t,b}(\mbf{r})$ using the Bistritzer-MacDonald (BM) two-center approximation \cite{2011PNAS..10812233B} following the appendices of \refcite{2018arXiv180710676S}. We consider the coupling between graphene layer $l$ with orbitals
\bea
\label{eq:ralell}
\mbf{R}+\mbf{r}_{\al,l}, \qquad \mbf{r}_{\al,l} = \al \pmb{\delta}_{1} + l (\pmb{\delta}_1 + d_0 \hat{z} )
\eea
where $\mbf{R}$ indexes the graphene unit cells, $\al = 0,1$ corresponds to the positions of the carbon $A,B$ sublattices (see \Fig{holomorphic}a of the Main Text), $\bsl{\delta}_1$ is defined under \eqnref{eq:h_1}, and $l = 0,\dots,n-1$ labels the rhomobohedrally stacked layers which are spaced $d_0 \sim 3.36\AA$ apart in the rigid structure. The hBN layers have orbitals at 
\bea
M(\mbf{R}+\mbf{r}_{\be,l'})
\eea
where $M = 1 + i \th \sigma_2 + \mathcal{E} + \dots$, $l'$ corresponds to the top and bottom encapsulated layers, and $\be$ here labels the B and N orbitals in the top/bottom layers. The twist angle $\theta$ can be tuned in device construction and, in this work, we take $\mathcal{E} = \eps \sigma_0$ ($\epsilon>0$) to describe the enlarged lattice constant of hBN $a_{hBN} = (1+\eps)a_G$. 

The inter-layer Hamiltonian hopping hBN onto graphene is given by
\bea
H^{inter}_{\al l, \be l'}(\mbf{p}, \mbf{p}') &= \frac{1}{\mathcal{N}} \sum_{\mbf{R},\mbf{R}'} e^{-i (\mbf{R} + \mbf{r}_{\al,l}) \cdot \mbf{p} + i (M (\mbf{R}' + \mbf{r}_{\be,l'})) \cdot \mbf{p}'} \braket{\mbf{R} + \mbf{r}_{\al,l} | H | M (\mbf{R}' + \mbf{r}_{\be,l'}) }  \\
\eea
where $H$ is the underlying microscopic Hamiltonian. We emphasize that $\mbf{R},\mbf{R}'$ are lattice vectors in the unrotated graphene layer. To proceed, we assume that the matrix element of the Hamiltonian is only dependent on the distance between orbitals (the ``two-center" approximation) leading to
\bea
\braket{\mbf{R} + \mbf{r}_{\al,l} | H | M (\mbf{R}' + \mbf{r}_{\be,l'})} &= \frac{1}{N\Omega} \sum_{\mbf{q} \in BZ} \sum_{\mbf{G}} t_{\mbf{q} + \mbf{G}} e^{i (\mbf{q} + \mbf{G}) \cdot \big( \mbf{R} + \mbf{r}_{\al,l}  -  (M (\mbf{R}' + \mbf{r}_{\be,l'}) ) \big)} 
\eea 
and $t_{\mbf{q}+\mbf{G}}$ is the momentum-space matrix element for the hopping between the orbitals labeled by $\al,\be$, $\mathcal{N}$ is the number of graphene unit cells with area $\Omega$. Here we have made the  SK approximation
\bea
t_{\mbf{q} + \mbf{G}} &= \int d^2r\, e^{- i (\mbf{q}+\mbf{G}) \cdot \mbf{r}} t_{SK}(\mbf{r}) 
\eea
which treats the boron-carbon and nitrogen-carbon hoppings identically, since both are $p_z$-$p_z$ orbital overlaps. Although this is an approximation, we argue in the Main Text that the resulting form of the Hamiltonian is general enough to fit the band structure obtained from the large-scale numerical calculations. (See \appref{app:fitting}.) Plugging this expression into the inter-layer Hamiltonian and using $M \mbf{r} \cdot \mbf{k} = (M\mbf{r})^T \mbf{k} = \mbf{r} \cdot M^T \mbf{k}$
gives
\bea
H^{inter}_{\al l,\be l'}(\mbf{p},\mbf{p}') &= \frac{1}{N^2\Omega} \sum_{\mbf{q} \in BZ} \sum_{\mbf{G}} \ t_{\mbf{q} + \mbf{G}} e^{i \mbf{r}_{\al,l} \cdot (\mbf{q} + \mbf{G} -\mbf{p}) - i \mbf{r}_{\be,l'} \cdot M^{T}(\mbf{q} + \mbf{G}-\mbf{p}')}  \sum_{\mbf{R},\mbf{R}'} e^{i \mbf{R} \cdot (\mbf{q} + \mbf{G} -\mbf{p}) - i \mbf{R}' \cdot M^{T} (\mbf{q} + \mbf{G} -\mbf{p}')}  \\
&= \frac{1}{\Omega} \sum_{\mbf{q} \in BZ} \sum_{\mbf{G}} \  t_{\mbf{q} + \mbf{G}} e^{i \mbf{r}_{\al,l} \cdot (\mbf{q} + \mbf{G} -\mbf{p}) - i \mbf{r}_{\be,l'} \cdot M^T(\mbf{q} + \mbf{G}-\mbf{p}')}  \sum_{\mbf{G}_1,\mbf{G}_2} \delta_{\mbf{q} + \mbf{G} -\mbf{p},\mbf{G}_1} \delta_{M^T(\mbf{q} + \mbf{G} -\mbf{p}'), \mbf{G}_2} \\
&=  \sum_{\mbf{G}_1,\mbf{G}_2} \frac{ t_{\mbf{p} + \mbf{G}_1} }{\Omega}e^{i \mbf{r}_{\al,l} \cdot \mbf{G}_1 - i \mbf{r}_{\be,l'} \cdot \mbf{G}_2} \delta_{\mbf{p}+\mbf{G}_1,\mbf{p}'+M^{-T} \mbf{G}_2} \ . \\
\eea
Note that $\mbf{G}_1, \mbf{G}_2$ are the graphene lattice vectors. We now use the fact that the momentum space coupling $t_{\mbf{p} + \mbf{G}}$ is rapidly decaying to cutoff the sum over lattice vectors. Let us consider what terms couple to the $\mbf{K}$ point of the top layer where $\mbf{p} = \mbf{K} + \delta \mbf{p}$. The terms that contribute are $\mbf{G} = 0, C_3 \mbf{K} - \mbf{K},C^2_3 \mbf{K} - \mbf{K}$ (all of the same magnitude due to $C_3$), because all others are outside the first BZ and are suppressed. We have
\bea
\label{eq:firsttermq}
H^{inter}_{\al l, \be l'}(\mbf{K} + \delta \mbf{p}, \mbf{p}') &= \frac{t_{\mbf{K}}}{\Omega} \sum_{\mbf{G}'} \Big( e^{- i \mbf{r}_{\be} \cdot \mbf{G}' + i \mbf{r}_{\al,l} \cdot \mbf{0}} \delta_{\mbf{K} + \delta \mbf{p}, \mbf{p}' + M^{-T} \mbf{G}'} \\ 
& \qquad \qquad \qquad + e^{- i \mbf{r}_{\be} \cdot \mbf{G}' + i \mbf{r}_{\al,l} \cdot (C_3 \mbf{K} - \mbf{K})} \delta_{C_3 \mbf{K} + \delta \mbf{p}, \mbf{p}' + M^{-T} \mbf{G}'} \\
& \qquad \qquad \qquad + e^{- i \mbf{r}_{\be}  \cdot \mbf{G}' + i \mbf{r}_{\al,l} \cdot (C^2_3 \mbf{K} - \mbf{K})} \delta_{C^2_3 \mbf{K} + \delta \mbf{p}, \mbf{p}' + M^{-T} \mbf{G}'}  \Big) \ .
\eea 
Now we consider the $\mbf{G}'$ sum. We take $\mbf{p}' =M^{-T}\mbf{K} + \delta \mbf{p}'$.
Owing to the large gap of hBN, the momentum corrections to the gap is small at the scale of moir\'e reciprocal lattice vectors.
Therefore, it is not necessarily legitimate to restrict $\delta \mbf{p}'$ to be small, and a faithful microscopic calculation must take higher moir\'e harmonics into account since they are not rapidly cut off by the hBN dispersion. 
We leave the derivation of the form of the higher harmonics to future work; in this work, we will explicitly provide the form for the first harmonic terms for concreteness. 
The delta function in the first term of \Eq{eq:firsttermq} enforces
\bea
\mbf{K} + \delta \mbf{p} &= M^{-T}\mbf{K} + \delta \mbf{p}'  +M^{-T} \mbf{G}' \\
\delta \mbf{p} - \delta \mbf{p}' &= M^{-T}\mbf{K} -\mbf{K} + M^{-T}\mbf{G}' \\
&\equiv M^{-T} \mbf{G}' - \mbf{q}_1 \ .
\eea
For the first harmonics, we have $\mbf{G}' = 0$. Repeating this for second and third terms, we eventually get
\bea
H^{inter}_{\al l, \be l'}(\mbf{K} + \delta \mbf{p},M^{-T} \mbf{K} +\delta \mbf{p}') &= \frac{t_{\mbf{K}}}{\Omega}  \Big( e^{i (\mbf{r}_{\al,l} -\mbf{r}_{\be}) \cdot \mbf{0}} \delta_{\delta \mbf{p}, \delta \mbf{p}' - \mbf{q}_1} \\ 
& \qquad \qquad  + e^{i (\mbf{r}_{\al,l} - \mbf{r}_{\be}) \cdot (C_3 \mbf{K} - \mbf{K})} \delta_{\delta \mbf{p}, \delta \mbf{p}' - \mbf{q}_2} \\
& \qquad \qquad  + e^{i (\mbf{r}_{\al,l} -\mbf{r}_{\be})  \cdot (C^2_3 \mbf{K} - \mbf{K})} \delta_{\delta \mbf{p}, \delta \mbf{p}' - \mbf{q}_3}  \Big) + ... \ ,
\eea
where ``$...$" include all the higher harmonics, and we have defined $\mbf{q}_{i+1} = C_3 \mbf{q}_1$. 

Using Eq.\,\ref{eq:ralell} for $l = 0$ in the graphene and $\mbf{r}_{\be,0} = \be \pmb{\delta}_1 - d \hat{z}, \be = 0,1$ for the hBN orbitals separated by a distance $d$ from the graphene, we find
\bea
\label{eqapp:Tequalamp}
H^{inter}_{\text{bottom}}(\mbf{K} + \delta \mbf{p}, M^{-T} \mbf{K} + \delta \mbf{p}') &= \sum_{j=1}^3 T^\dag_j \delta_{\delta \mbf{p}, \delta \mbf{p}' - \mbf{q}_{j}} + \dots ,\qquad T^\dag_j &= t_\mbf{K}(d \hat{z}) \bpm 
1 & e^{- i \frac{2\pi}{3} (j-1)}   \\
 e^{i \frac{2\pi}{3} (j-1)} & 1
\epm
\eea
for all $n$, the total number of layers, and ``$...$" contains all the higher harmonics.
We have written $t_\mbf{K}= t_\mbf{K}(d \hat{z})$ to emphasize its dependence on the inter-layer distance. We check that in the relaxed structures, the average value of $d$ over the unit cell is nearly constant (ranging from from $3.421\AA$ to $3.423\AA$ as the number of layers is increased) corresponding to $t_\mbf{K} \sim 93$meV. In doubly aligned relaxed structures, we find that the bottom layers show larger variation, with $d$ going from $3.40\AA$ to $3.43\AA$, corresponding to $98meV$ and $90$meV hoppings respectively. 

Next in hBN/R$n$G/hBN devices, we compute the coupling between the top layers. We will assume that the top and bottom hBN layers are perfectly aligned so that there is no super moir\'e pattern formed. There now appear phase factors in the $T_j$ matrices due to the shift in the position $\mbf{r}_{\al,l}$ from the rhombohedral stacking (see \Eq{eq:ralell}). We find that the top layer in an $n$ layer structure has the hopping
\bea
H^{inter}_{\text{top}}(\mbf{K} + \delta \mbf{p}, M^{-T} \mbf{K} + \delta \mbf{p}') &= \sum_{j=1}^3 e^{i \frac{2\pi (j-1)}{3} (n-1)} T^\dag_j \delta_{\delta \mbf{p}, \delta \mbf{p}' - \mbf{q}_{j}} + ... \ ,
\eea
where ``..." contains all the higher harmonics.
In position space these term can be written
\bea
\label{eq:appTb}
T_{b}(\mbf{r}) &=  \sum_{j=1}^3 e^{i \mbf{q}_j \cdot \mbf{r}} T_j+ ..., \qquad T_{t}(\mbf{r}) =  \sum_{j=1}^3 e^{- i\frac{2\pi (j-1)}{3}  (n-1)} e^{i \mbf{q}_j \cdot \mbf{r}} T_j + ... \ ,
\eea
where ``..." contains all the higher harmonics.
The two-center approximation results in all carbon-hBN hoppings have equal amplitude for the first harmonics. This is similar to how, in twisted bilayer graphene, the $AA$ and $AB$ couplings are equal in the two-center approximation, but relaxation introduces symmetry-allowed corrections beyond the two-center approximation. We now consider more general forms of the first-harmonic moir\'e coupling generalizing \Eq{eq:appTb}. Since hBN breaks all but the $C_3$ symmetry of rhombohedral graphene, the most general symmetry-allowed $\mbf{q}_j$-hopping takes the form \cite{PhysRevB.89.205414}
\begin{equation}
    T^\dag_j = \bpm t_{B,C_A} & e^{- i \frac{2\pi}{3} j} t_{N,C_A}  \\
 e^{i \frac{2\pi}{3} j}  t_{B,C_B}& t_{N,C_B} \epm \ .
\end{equation} 
We argued in the  Main Text that this hopping matrix, while containing more free parameters, does not offer a better fit to the low-energy bands than the hopping matrix in \Eq{eqapp:Tequalamp}. This is because only $t_{B,C_A}, t_{N,C_A}$ are relevant (i.e., the first row of $T_j^\dag$), as the low-energy bands are strongly sublattice polarized on the bottom layer. 

The model \Eq{eq:signelmodels} derived above can be simplified due to the separation of energy scales between the graphene (which we set to chemical potential zero) and the hBN which has eV-scale chemical potentials \cite{PhysRevX.8.031087}. Writing the Schrodinger equation (here $H_G$ is the Hamiltonian of the bottom graphene layer)
\bea
\bpm H_G & T^\dag \\ 
T & H_{BN}\epm \bpm \psi_G \\ \psi_{hBN} \epm = E \bpm \psi_G \\ \psi_{hBN} \epm
\eea
out in components yields $T^\dag \psi_{G} = (E-H_{hBN}) \psi_{hBN}$, giving
\bea
\lp H_G + T^\dag (E- H_{hBN})^{-1} T \rp \psi_{G} = E \psi_G \ .
\eea
We now assume that $E$ is an energy near the graphene Fermi energy, so we can take $(E- H_{hBN})^{-1} \sim - H_{hBN}^{-1}$ to leading order, and derive $H_G - T^\dag H_{hBN}^{-1} T \equiv H_G + V_{\text{hBN}}$ as the effective Schrodinger equation coming from integrating out the hBN.  
For the bottom layer, we have
\begin{equation}
\begin{split}
V_\text{hBN}(\mbf{r}) &=-3\bpm \frac{|t_{B,C_A}|^2}{V_B}+\frac{|t_{N,C_A}|^2}{V_N} & 0 \\ 
0 & \frac{|t_{B,C_B}|^2}{V_B}+\frac{|t_{N,C_B}|^2}{V_N} \epm \\
& -\Big\{e^{i(\mbf{q}_1-\mbf{q}_2)\cdot \mbf{r}} \bpm \frac{|t_{B,C_A}|^2}{V_B}+\omega^* \frac{|t_{N,C_A}|^2}{V_N}  & \frac{t_{B,C_A}t^*_{B,C_B}}{V_B}+\omega^*\frac{t_{N,C_A} t^*_{N,C_B}}{V_N} \\ 
\omega \frac{t_{B,C_B}t^*_{B,C_A}}{V_B}+\frac{t_{N,C_B} t^*_{N,C_A}}{V_N} & \omega\frac{|t_{B,C_B}|^2}{V_B}+\frac{|t_{N,C_B}|^2}{V_N}\epm \\
&+e^{i(\mbf{q}_2-\mbf{q}_3)\cdot \mbf{r}} \bpm \frac{|t_{B,C_A}|^2}{V_B}+\omega^* \frac{|t_{N,C_A}|^2}{V_N}  & \omega^*\frac{t_{B,C_A}t^*_{B,C_B}}{V_B}+\omega\frac{t_{N,C_A} t^*_{N,C_B}}{V_N} \\ 
\omega^*\frac{t_{B,C_B}t^*_{B,C_A}}{V_B}+\omega\frac{t_{N,C_B} t^*_{N,C_A}}{V_N} & \omega\frac{|t_{B,C_B}|^2}{V_B}+\frac{|t_{N,C_B}|^2}{V_N}\epm  \\
& +e^{i(\mbf{q}_3-\mbf{q}_1)\cdot \mbf{r}} \bpm \frac{|t_{B,C_A}|^2}{V_B}+\omega^* \frac{|t_{N,C_A}|^2}{V_N}  & \omega \frac{t_{B,C_A}t^*_{B,C_B}}{V_B}+\frac{t_{N,C_A} t^*_{N,C_B}}{V_N} \\ 
\frac{t_{B,C_B}t^*_{B,C_A}}{V_B}+\omega^*\frac{t_{N,C_B} t^*_{N,C_A}}{V_N} & \omega\frac{|t_{B,C_B}|^2}{V_B}+\frac{|t_{N,C_B}|^2}{V_N}\epm + H.c.\Big\}+ ...\ ,
\end{split}
\end{equation}
where ``..." contains terms that come from higher harmonics of $T(\bsl{r})$.
Here $\omega=e^{2i\pi/3}$. 
Since $V_\text{hBN}(\mbf{r})$ effectively couples only graphene degrees of freedom, it is reasonable for us to only keep the terms of $V_\text{hBN}(\mbf{r})$ up to the first harmonics, \ie, only keeping terms that are uniform or has spatial dependence $e^{\ii \bsl{G}_M\cdot \bsl{r}}$ with $|\bsl{G}_M|=1$.
Under this approximation, as we argue in the Main Text, only the $1,1$ entry of $V_{\text{hBN}, \text{bottom}}(\mbf{r})$ is relevant to the low energy physics.
Therefore, we can choose the effective moir\'e potential to be
\begin{equation}\label{eq:VhBNbottom}
    V_{\text{hBN}, \text{bottom}}(\mbf{r})=V_{b0} \bpm 1 & 0 \\ 0 & 1 \epm + \Big\{V_{b1} e^{i\psi_{\xi_b}}\Big[ e^{i(\mbf{q}_1-\mbf{q}_2)\cdot\mbf{r}}\bpm 1& 1 \\ \omega &\omega \epm + e^{i(\mbf{q}_2-\mbf{q}_3)\cdot\mbf{r}}\bpm 1& \omega^* \\ \omega^* &\omega \epm +e^{i(\mbf{q}_3-\mbf{q}_1)\cdot\mbf{r}}\bpm 1& \omega \\ 1 &\omega \epm \Big] + H.c. \Big\},
\end{equation} 
where $\xi_b = 0,1$ is the stacking order of the bottom layer, and 
\bea
V_{b0} = -3 \lp \frac{|t_{B,C_A}|^2}{V_B}+\frac{|t_{N,C_A}|^2}{V_N} \rp + ... , \qquad V_{1b} e^{i \psi_b}  = - \lp \frac{|t_{B,C_A}|^2}{V_B}+\omega^* \frac{|t_{N,C_A}|^2}{V_N} \rp + .. 
\eea
with ``..." containing the contribution for higher harmonics of $T(\bsl{r})$.
This simplified form of the effective moir\'e potential has been previously derived in Ref. \cite{PhysRevB.90.155406} taking $t_{B,C_A} = t_{N,C_A}$ under the first harmonic approximation of $T(\bsl{r})$. In this limit, 
\bea
t_{B,C_A} = t_{N,C_A} = t_{\mbf{q} + \mbf{G}} &= \int d^2r\, e^{- i (\mbf{q}+\mbf{G}) \cdot \mbf{r}} t_{SK}(\mbf{r}) \equiv w \ ,
\eea
the angle $\psi_b$ in \Eq{eq:psi_determination} is independent of $w$ and takes the values 
\eq{
\label{eq:psi_values}
\psi_{\xi =0} = 223.5^\circ, \quad \psi_{\xi = 1} = - \psi_{\xi = 0} - \frac{2\pi}{3} = 16.5^\circ\ .
}

For hBN/R$n$G/hBN structures, we integrate out the top hBN layer and obtain 
\begin{equation}
\label{eq:simpleform}
    V_{\text{hBN},\text{top}}(\mbf{r})=V_{t0} \bpm 1 & 0 \\ 0 & 1 \epm + \Big\{V_{t1} e^{i\psi_{\xi_t} + i \frac{2\pi}{3}(n-1)}\Big[ e^{i(\mbf{q}_1-\mbf{q}_2)\cdot\mbf{r}}\bpm 1& 1 \\ \omega &\omega \epm + e^{i(\mbf{q}_2-\mbf{q}_3)\cdot\mbf{r}}\bpm 1& \omega^* \\ \omega^* &\omega \epm +e^{i(\mbf{q}_3-\mbf{q}_1)\cdot\mbf{r}}\bpm 1& \omega \\ 1 &\omega \epm \Big] + H.c. \Big\} \  .
\end{equation}
We estimate typical values for $V_0 \sim 10$meV and $V_1 \sim 7$meV using $w = 90$meV. 

In sum, the full $2n\times 2n$ moir\'e model that we use is 
\eq{
\label{eq:2n_model}
H_{2n,\K} = H_{\K} + H_D + \bpm
V_{\text{hBN},\text{bottom}}(\mbf{r}) &   &  &   & & \\
 & 0_{2\times 2} &   &  &  \\
  &   & \ddots &  & \\
  &   &  & 0_{2\times2} & \\
&   &  &  &   V_{\text{hBN},\text{top}}(\mbf{r})
\epm \ ,
}
where $H_{\K}$ is the kinetic energy term of R$n$G in \eqnref{eq: H_K}, $H_{D}$ is the displacement field term in \eqnref{eq:H_D}.
\eqnref{eq:2n_model} works for both R$n$G/hBN ($V_{\text{hBN},\text{top}}(\mbf{r})=0$) and hBN/R$n$G/hBN structures.
Detailed discussion of the fitting is in \appref{app:fitting}.

\subsection{Effective Model}
\label{app:effmodelderivation}

Here we derive the effective model that incorporates the inter-layer AB hopping ($t_1$), the next-nearest inter-layer AB hoppings (yielding the effective velocity terms $v_3$), the inter-layer AA/BB coupling ($v_4$) as well as the next-next-layer nearest-neighbour hopping ($t_2$). The full Hamiltonian $H_n$ for R$n$G is defined in Eq. \eqref{eq:Hrhomb}, and upon projecting onto the two low-energy states (with respect to the charge neutrality) defined in Eq. \eqref{eq:lowE_basis}, we obtain (for $n \geq 3$)
\begin{equation}\label{eq:heff}
    h^\text{eff}_n (\mbf{k}) = \mathcal{U}^\dag(\mbf{k}) H_n \mathcal{U}(\mbf{k}) = \bpm H_0(\mbf{k}) + V_3(\mbf{k}) & \bar{k}^{n-3}(\beta \bar{k}^3+\gamma |k|^2 +\delta) \\ k^{n-3}(\beta k^3+\gamma |k|^2 +\delta) & H_0(\mbf{k}) -V_3(\mbf{k})\epm ,
\end{equation}
where $k=k_x+ i k_y, \bar{k} = k_x - i k_y$ is the complex coordinate for the relative momentum $\mbf{k}=(k_x,k_y)$ in the $\mbf{K}$-valley, and
\begin{equation}
    H_0(\mbf{k}) = \alpha |k|^2+ V_{ISP} \mathcal{F}_n(\frac{v_F|k|}{t_1}), \quad \text{and}\quad V_3(\mbf{k})= V \mathcal{G}_n(\frac{v_F|k|}{t_1})
\end{equation}
with the exact projection giving the following closed form results
\begin{equation}\label{eq:closedform}
    \mathcal{F}_n(x) = \begin{cases} \frac{1}{2} \left(-\frac{2 \left(x-x^n\right)^2}{\left(x^2-1\right) \left(x^{2 n}-1\right)}+n-1\right), \;\; \text{for odd $n$} \\
    \frac{1}{2} \left(n-\frac{\left(x^2+1\right) \left(x^n-1\right)}{\left(x^2-1\right) \left(x^n+1\right)}\right), \;\; \text{for even $n$}
    \end{cases} \quad\text{and}\quad \mathcal{G}_n(x) = \frac{1}{2} \left(\frac{n \left(x^{2 n}+1\right)}{x^{2 n}-1}+\frac{x^2+1}{1-x^2}\right).
\end{equation}
To be consistent in the perturbation theory, we keep terms up to $x^n$ in the expansion of $\mathcal{F}_n(x)$ and $\mathcal{G}_n(x)$, which is what we used in the main text (Eq. \eqref{eq:V3_def}). The parameter values in the projected Hamiltonian are related to the ones in the full model by
\begin{equation}\label{eq:effparameter}
    \alpha= \frac{2v_F v_4}{t_1},\quad \beta=\frac{v_F^n}{(-t_1)^{n-1}}, \quad \gamma=-\frac{(n-1)v_F^{n-2}v_3}{(-t_1)^{n-2}}-\frac{(n-2)v_F^{n-1}t_2}{(-t_1)^{n-1}}, \quad \delta=\frac{(n-2)v_F^{n-3}t_2}{(-t_1)^{n-3}}.\;\;
\end{equation}
As an example, here we explicitly work out the derivation of the $\alpha |k|^2$ term, which arises from the inter-layer hopping $v_4$ in the full Hamiltonian $H_n$ (see Eq. \eqref{eq:Hrhomb}):
\begin{equation}
\begin{split}
&\sum_{l=1}^{n-1} \sum_{a=A,B} \mathcal{U}^\dagger(\mbf{k})_{i,l,a} [H_n(\mbf{k})_{l,l+1}]_{a,a} \mathcal{U}(\mbf{k})_{j, l+1,a} + \sum_{l=2}^{n} \sum_{a=A,B} \mathcal{U}^\dagger(\mbf{k})_{i,l,a} [H_n(\mbf{k})_{l,l-1}]_{a,a} \mathcal{U}(\mbf{k})_{j, l-1,a} \\
=& \delta_{i,j} \Big[\sum_{l=0}^{n-2} (-\frac{v_F\bar{k}}{t_1})^l (-v_4 \bar{k}) (-\frac{v_F k}{t_1})^{l+1} + (-\frac{v_F \bar{k}}{t_1})^l (-v_4 k) (-\frac{v_F k}{t_1})^{l-1}\Big]/\Big[\sum_{l=0}^{n-1}(\frac{v_F|k|}{t_1})^{2l}\Big]\\
= & \delta_{i,j} 2v_F v_4 |k|^2/t_1 +\mathcal{O}(|k|^{2n}), 
\end{split}
\end{equation}
where $i,j$ label the chiral basis states, $l$ labels the graphene layer and $a$ labels the sublattice.

The above parameter values in Eq. \eqref{eq:effparameter} are further adjusted to optimize the matching between the band structure in the effective model and the DFT bands, see Table~\ref{tab:parameters_eff} in the Main Text. Note that $t_2/t_1 = 0.02$ and $v_F |\mbf{q}_1|/t_1 = 0.55$ at $\th = 0.767^\circ$ strongly suppress $\delta$ for $n > 3$. For completeness, we keep $\delta$ for all $n$, but it can be neglected with minimal changes to the band structure for $n=5$.

\begin{figure}
    \centering
    \includegraphics[width=1\columnwidth]{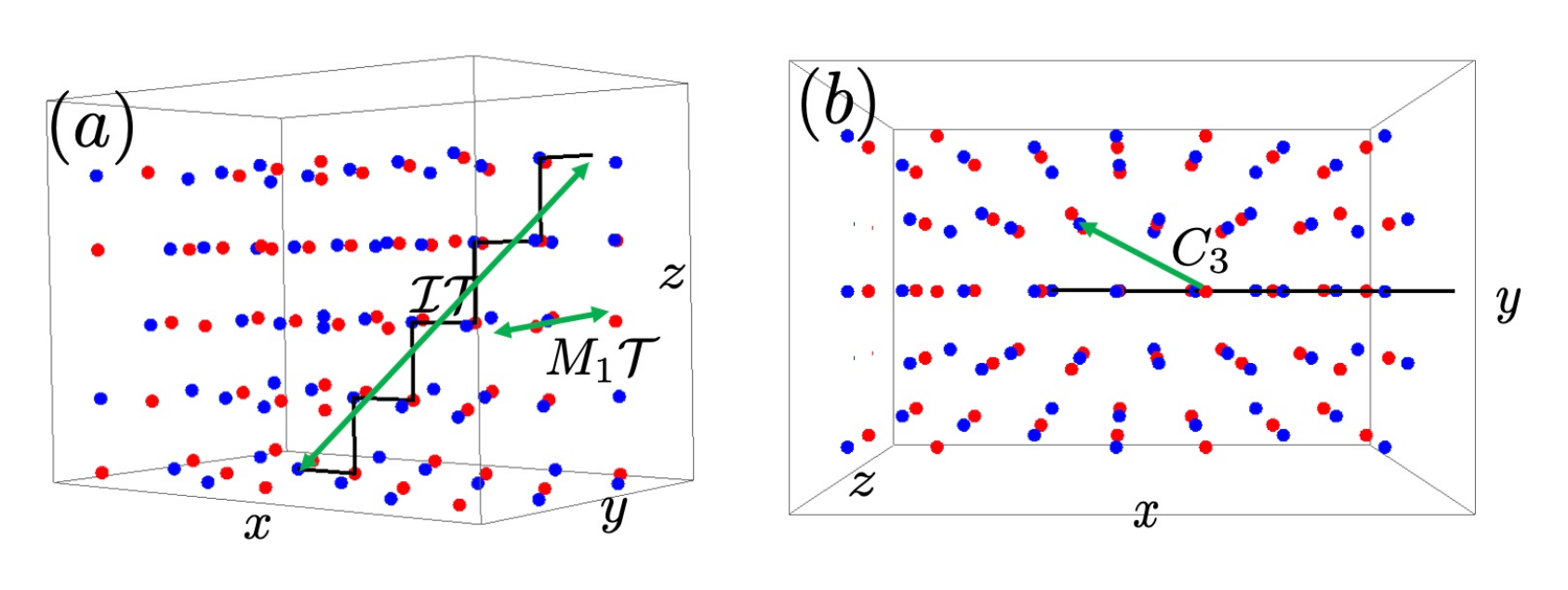}
    \caption{3D schematic of the pentalayer graphene (R5G) from the side $(a)$ and the top $(b)$. The unit cell consists of 10 carbon atoms and is connected by a line to guide the eye. Red and blue are used to color the A and B sublattices in each layer. }
    \label{fig:3dgraphic}
\end{figure}

We now discuss the symmetries of the effective model. We first consider the crystallographic symmetries $g$ obeying
\bea
D^\dag[g] h_{eff}(\mbf{r}) D[g] &= h_{eff}(g\mbf{r}) 
\eea
and $D[g]$ can be unitary or anti-unitary. The kinetic term with $V=V_0=V_1=0$ has the symmetries
\bea
D[C_3] &= e^{-\frac{2\pi}{3} i \sigma_z}, \quad  D[\mathcal{I}\mathcal{T}] = \sigma_1 \mathcal{K}, \quad  D[M_1\mathcal{T}] &= \mathcal{K} \\
\eea
where $\mathcal{I}\mathcal{T}$ is space-time inversion and $\mathcal{M}_1\mathcal{T}$ is an  anti-unitary mirror obeying $M_1 \hat{x} = - \hat{x}$, with $\mathcal{K}$ representing complex conjugation (see \Fig{fig:3dgraphic}). These symmetries preserve valley and are exact in pristine R$n$G. Note that the anti-commuting chiral symmetry $\sigma_3$ is broken by $V_{ISP}$, $D[\mathcal{I}\mathcal{T}]$ is broken by the potential difference $V$, and both $D[\mathcal{I}\mathcal{T}]$ and $M_1\mathcal{T}$ are broken by the moir\'e potential term proportional to $V_1$ in general. If $\th = 0$ so that the hBN axis is aligned with the R$n$G axis, then the $M_1\mathcal{T}$ symmetry is restored. However, we focus on $\th = 0.76715^\circ$ throughout this work. 

 Taking into account the lowest order moir\'e potential (parametrized by $V_0, V_1$ and $\psi_\xi$), as well as the effect of the out-of-plane displacement field (parametrized by $V$ the inter-layer potential energy difference), the R$n$G/hBN effective model acquires the following form:

\begin{equation}\label{eq:Heff_single}
    H^\text{eff}_{n, \xi} = h^\text{eff}_n (-i\nabla) + [V_0 + 2V_1 \sum_i \cos(\mbf{g}_j \cdot \mbf{r} + \psi_\xi)] 
\bpm 1 & 0 \\ 0 & 0 \\ \epm \ ,
\end{equation}
where the second term is obtained from $\mathcal{U}^\dagger(0) V_{\text{hBN, bottom}} (\mbf{r}) \mathcal{U}(0)$, with $\mathcal{U}(0)$ the chiral basis at $\mbf{k}=0$ (see Eq. \eqref{eq:lowE_basis}), $V_{\text{hBN, bottom}}$ defined in Eq. \eqref{eq:VhBNbottom} and $\mbf{g}_j \equiv R(\frac{2\pi}{3}(j-1)) (\mbf{q}_2 -\mbf{q}_3)$. This projection preserves the locality of the moir\'e potential. For the case of hBN/R$n$G/hBN structures, we have 
\begin{equation}\label{eq:Heff_double}
\begin{split}
    H^\text{eff}_{n,\xi_t,\xi_b} = h^\text{eff}_n (-i\nabla) +  
&\bpm V_{b0}+ 2V_{b1}\sum_i  \cos(\mbf{g}_i \cdot \mbf{r} + \psi_{\xi_b}) & 0 \\ 0 & V_{t0}+2V_{t1} \sum_i \cos(\mbf{g}_i \cdot \mbf{r} + \psi_{\xi_t}+\frac{2\pi n}{3}) \\ \epm  \ .
\end{split}
\end{equation}
For the $(\xi_t,\xi_b) = (1,0)$ configuration, we impose $\mathcal{I}\mathcal{T}$ symmetry so that $ V_{b0} =  V_{t0}$ and $-\psi_b = \psi_t+\frac{2\pi n}{3}$.

\subsection{$C_3$ eigenvalues for the continuum model R$n$G/hBN}\label{app:tripod}

\begin{figure}
    \centering
    \includegraphics[width=1\columnwidth]{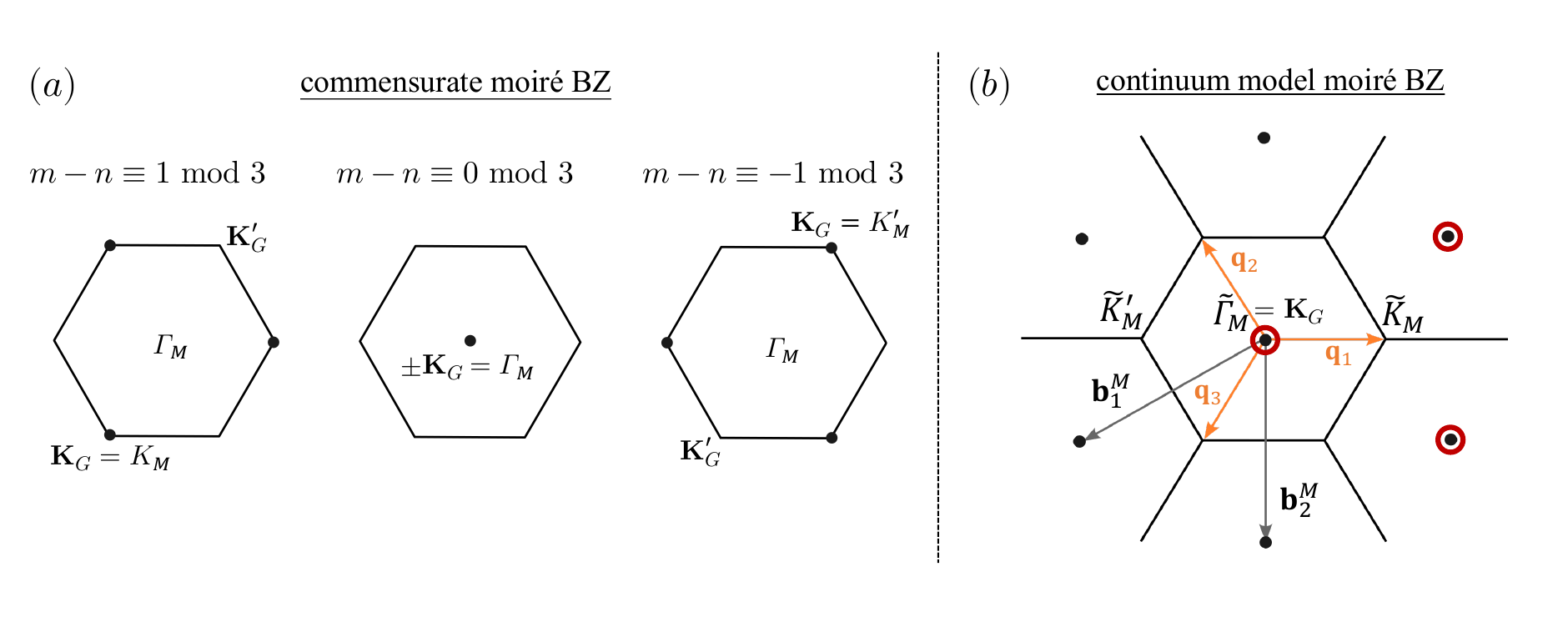}
    \caption{(a) Depiction of the moir\'e Brillouin zone (BZ) for three different commensurate moir\'e configurations (see Eq. \eqref{eq:K_G_folding}). Depending on $m-n \text{ mod }3$, with $(m ,n)\in \mathbb{Z}^2 $ labeling the commensurate configuration, the graphene $K$-point (at $\mathbf{K}_G$) is either folded onto the moir\'e $K_M$, $\Gamma_M$ or $K'_M$ point. (b) Depiction of the continuum model moir\'e BZ in which we focus on the degrees of freedom centered around the $\K$ graphene valley. The high symmetry points are labeled as $\tilde{\Gamma}_M$, $\tilde{K}_M$ and $\tilde{K}'_M$ to signify our convention of always boosting the graphene $\mathbf{K}_G$ onto $\tilde{\Gamma}_M$. The three moir\'e reciprocal points circled in red are considered in the tripod model analysis in App. \ref{app:tripod} for deducing the $C_3$ eigenvalue at $\tilde{K}_M$.}
    \label{fig:severalMBZ}
\end{figure}

Now we take the effective model in Eq. \eqref{eq:Heff_single} and analyze the $C_3$ symmetry eigenvalues at $\tilde{\Gamma}_M$, $\tilde{K}_M$ and $\tilde{K}'_M$ points in the continuum model moir\'e Brillouin zone for the graphene $K$-valley. This Brillouin zone, as well as the labeling of high symmetry points, should be distinguished from the moir\'e Brillouin zone at a commensurate twist, as depicted in Fig. \ref{fig:severalMBZ}.
At different commensurate twist angles, the graphene $\mathbf{K}_G$-point can be folded onto either $\Gamma_M$, $K_M$ or $K'_M$, as explained in Eq. \eqref{eq:K_G_folding}, while for simplicity of our discussion below, we will boost the BZ such that $\mathbf{K}_G$ is consistently situated at the BZ center, which is denote as $\tilde{\Gamma}_M$.

Let us just recall that the $D[C_3]$ operator in the $2n \times 2n$ model (c.f. Eq. \eqref{eq: SO2rotation}) takes the form
\begin{equation}
    D[C_3]_{l l'} = -\delta_{l l'} e^{i\frac{2\pi}{3}(l-1-\lfloor\frac{n}{2}\rfloor)} e^{-i\frac{\pi}{3}\sigma_3},
\end{equation}
so by projecting onto the chiral basis we obtain $D_\text{eff}[C_3]=\mathcal{U}^\dagger(0) D[C_3] \mathcal{U}(0)$ with the form
\begin{equation}\label{eq:genericC3rep}
    D_\text{eff}[C_3] = \bpm (\omega^*)^{\frac{n}{2}-1} & 0 \\ 0 & \omega^{\frac{n}{2}+1} \epm \quad \text{for even }n ,\quad D_\text{eff}[C_3]=  \bpm (\omega^*)^{\lfloor\frac{n}{2}\rfloor-1} & 0 \\ 0 & \omega^{\lfloor\frac{n}{2}\rfloor-1} \epm \quad \text{for odd }n,
\end{equation}
where $\omega=e^{2i\pi/3}$. We shall focus on the lowest conduction band, and denote the corresponding eigenvalue as $\eta_\mbf{k}$. To proceed analytically, we consider a reduced Hamiltonian that involves only the reciprocal lattice points closest to the high symmetry point of interest. Recall that the effective Hamiltonian in the full reciprocal space takes the following form:
\begin{equation}
    H^\text{eff}_n(\mbf{k})_{\mbf{G},\mbf{G}'} = \delta_{\mbf{G}, \mbf{G}'} h^\text{eff}_n(\mbf{k}-\mbf{G}) + \sigma_{11} (\delta_{\mbf{G},\mbf{G}'} V_0 
 +\delta_{\mbf{G}-\mbf{G}',-\mbf{g}_j}V_1 e^{i\psi_\xi}+\delta_{\mbf{G}-\mbf{G}',\mbf{g}_j}V_1 e^{-i\psi_\xi})
\end{equation}
where $\sigma_{11} \equiv (\sigma_0 +\sigma_3)/2$.

At $\tilde{\Gamma}_M$ (i.e. $\mbf{k}=0$), there is one closest reciprocal lattice point (i.e., $\mbf{G}=0$), from which we obtain a 2 $\times$ 2 model describing the perturbation of the zero modes
\begin{equation}
    H^\text{eff}_n(\tilde{\Gamma}_M) = (2V_{ISP}+\frac{V_0}{2}) \mathds{1} + (\frac{V_0}{2}-2V)\sigma_3
\end{equation} 
and we can neglect higher shells due to their large kinetic energy. Below we consider $n=5$ for concreteness, though it is straightforward to generalize our results to arbitrary $n$ by implementing the appropriate representation of $C_3$. 
Following Eq. \eqref{eq:genericC3rep}, the projected $C_3$ symmetry operator is $D_\text{eff}[C_3]=\text{diag}(\omega^*,\omega)$ for $n=5$. Thus,
\begin{equation}
    \eta_{\tilde{\Gamma}_M} = \begin{cases} \omega^*, \quad \text{for} \quad \frac{V_0}{2}-2V>0, \\ \omega, \quad \text{for}\quad \frac{V_0}{2}-2V<0.\end{cases}
\end{equation}

At $\tilde{K}_M$ (i.e., $\mbf{k}=\mbf{q}_1$), there are three closest reciprocal lattice vectors, namely $\mbf{G}=0,\mbf{q}_1-\mbf{q}_2, \mbf{q}_1-\mbf{q}_3$, see Fig. \ref{fig:severalMBZ}(b). Other reciprocal points like $\mbf{q}_2-\mbf{q}_1$ and $\mbf{q}_3-\mbf{q}_1$ are at larger distances and will be neglected in our first-order degenerate perturbation analysis.  Notice that $\tilde{\mbf{K}}_M-(\mbf{q}_1-\mbf{q}_2)=C_3 \tilde{\mbf{K}}_M$ and $\tilde{\mbf{K}}_M-(\mbf{q}_1-\mbf{q}_3)=C^2_3 \tilde{\mbf{K}}_M$, we obtain the following 6 $\times$ 6 Hamiltonian
\begin{equation}
    H^\text{eff}_n(\tilde{\mbf{K}}_M) = \bpm h^\text{eff}_n(\tilde{\mbf{K}}_M)+ V_0\sigma_{11}  & V_1 e^{i\psi_\xi} \sigma_{11} & V_1 e^{-i\psi_\xi} \sigma_{11} \\ V_1 e^{-i\psi_\xi} \sigma_{11} & h^\text{eff}_n(C_3 \tilde{\mbf{K}}_M) + V_0\sigma_{11} & V_1 e^{i\psi_\xi} \sigma_{11} \\ V_1 e^{i\psi_\xi} \sigma_{11} & V_1 e^{-i\psi_\xi} \sigma_{11} &h^\text{eff}_n(C^2_3 \tilde{\mbf{K}}_M) + V_0\sigma_{11} \epm , \qquad \sigma_{11} \equiv (\sigma_0 +\sigma_3)/2 \ .
\end{equation}\
The $C_3$ operator that commutes with $H^\text{eff}_n(\tilde{\mbf{K}}_M)$ takes the form 
\begin{equation}
    \bpm 0 & 0 & 1\\ 1 & 0 & 0 \\ 0 & 1 & 0 \epm \otimes D_\text{eff}[C_3], 
\end{equation}
which can be used to diagonalize  $H^\text{eff}_n$ into three 2 $\times$ 2 blocks that correspond to symmetry eigenvalues $\omega^j$ ($j=0,1,2$). The highest energy in each of these symmetry sectors are found to be 
\begin{subequations}\label{eq:Kenergy}
    \begin{align}
        \epsilon^{(c)}_{j}(\tilde{K}_M) &= f_0 + V_1 \cos(\psi_\xi-(j+1)\frac{2\pi}{3})+\sqrt{f^2_+ +\Big(\frac{V_0}{2}+V_3 (|\tilde{K}_M|)+V_1 \cos(\psi_\xi-(j+1)\frac{2\pi}{3})\Big)^2}
    \end{align}
\end{subequations}
where $f_0 = H_0(|\tilde{K}_M|)+V_0/2$, see Eq. \eqref{eq:closedform}, and we define
\begin{equation}
    f_\pm^2 = \abs{\tilde{K}_M}^4 \big[\abs{\tilde{K}_M}^6 \beta^2+(|\tilde{K}_M|^2\gamma+\delta)^2\pm 2\abs{\tilde{K}_M}^3 \beta (|\tilde{K}_M|^2\gamma+\delta) \cos 3\theta_K \big], \quad \text{with}\quad \theta_K = \tan^{-1} \frac{\tilde{K}_{M,y}}{\tilde{K}_{M,x}}.
\end{equation}
Notice that $f_\pm$ is always non-negative. A similar analysis for the $\tilde{K}_M'$ point leads to
\begin{subequations}\label{eq:Kpenergy}
    \begin{align}
        \eps^{(c)}_j(\tilde{K}_M') &= f_0 + V_1 \cos(\psi_\xi+(j+1)\frac{2\pi}{3})+\sqrt{f^2_-+\Big(\frac{V_0}{2}+V_3(|\tilde{K}_M'|) +V_1 \cos(\psi_\xi+(j+1)\frac{2\pi}{3})\Big)^2}
    \end{align}
\end{subequations}
We consider certain parameter regime where it is easy to identify the energy of the lowest conduction band at $\tilde{K}_M$ and $\tilde{K}_M'$: (I) with negative (or small positive) displacement field such that $\frac{V_0}{2}+V_3(|\tilde{K}_M|) >0$, it is clear that the lowest conduction state is identified by $\min\{\cos(\psi_\xi-2\pi/3), \cos(\psi_\xi+2\pi/3), \cos\psi_\xi\}$; (II) with very large and positive displacement field where $|\frac{V_0}{2}+V_3(|\tilde{K}_M|)| \gg f_{\pm}, V_1$, one can Taylor expand the square root and realize that the lowest conduction state corresponds to maximizing $|\frac{V_0}{2}+V_3+V_1\cos(\psi_\xi+\frac{2\pi j}{3})|$. Since $\frac{V_0}{2}+V_3$ acquires a large negative value in this case, the lowest conduction state again corresponds to choosing $\min\{\cos(\psi_\xi-2\pi/3), \cos(\psi_\xi+2\pi/3), \cos\psi_\xi\}$. For reference, using SK values of pristine graphene parameters, we have $f_{\pm} \sim 20$ meV for experimental twist angle $\theta=0.77^\circ$, while $V_0, V_1 \sim 5$ meV as determined by our band-structure fitting analysis.

\eqnref{eq:simpleform} implies $\psi_{\xi=0} = 223.5^\circ$, while $\psi_{\xi=1} = 16.5^\circ$, which suggest
\begin{equation}
    \eta_{\K} = \begin{cases}
        \omega^*, \quad\text{for} \quad \xi=0 \\
        \omega, \quad\text{for} \quad \xi=1
    \end{cases}\quad\text{and}\quad\quad      \eta_{\K'} = \begin{cases}
        \omega^*, \quad\text{for} \quad \xi=0 \\
        1, \quad\text{for} \quad \xi=1
    \end{cases}.
\end{equation}
Irrespective of the value of $\psi_\xi$ (and hence the stacking configuration), we consistently have $\eta_K \eta_{\K'} = \omega$. 
Altogether, using the relationship between point-group symmetry eigenvalues and Chern number \cite{PhysRevB.86.115112}, we find
\begin{equation}
    \exp(\frac{2i\pi C_c}{3}) = \begin{cases} 1, \quad \text{for} \quad \frac{V_0}{2}-2V>0, \\ \omega^*, \quad \text{for}\quad \frac{V_0}{2}-2V<0.\end{cases}
\end{equation}
Furthermore, for very large \textit{positive} displacement field where $-(\frac{V_0}{2}+V_3(|\tilde{K}_M|)) \gg f_{\pm}$ and $V_1$, the three highest energies in the conduction band $(\epsilon^{(c)}_0, \epsilon^{(c)}_1, \epsilon^{(c)}_2)$ would stick together, which corresponds to band-folding with an ``empty" moir\'e superlattice (empty lattice approximation). This can be understood from the fact that the moir\'e potential provided by the bottom hBN has negligible effects on the conduction band electrons localized on the top layer (see \Fig{fig:main_2n}). 

\section{Lattice structure and Structural relaxation}
\label{app:relaxation}

\begin{figure}
    \centering
\includegraphics[width=1.0\linewidth]{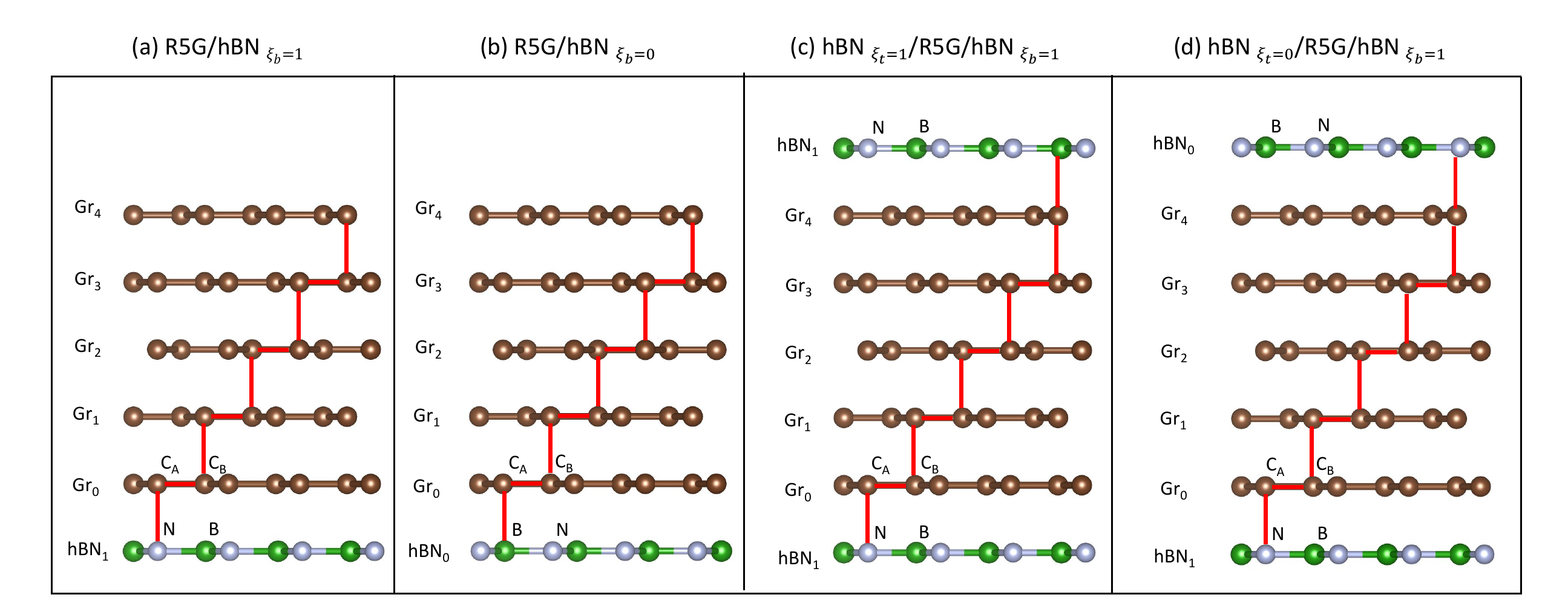}
    \caption{ Rigid R$5$G/hBN and hBN/R$5$G/hBN lattice structures with zero twist. (a),(b) are pentalayer graphene stack on hBN, (a) with a carbon atom strictly colinear with a boron atom in bottom layer along z-axis, (b) with a carbon atom strictly colinear with a nitrogen atom in bottom layer along z-axis. (c),(d) are pentalayer graphene between two hBN layers. (a) and (b) are connected by applying $C_{2}$ rotation (with axis along z-axis) on hBN only, (d) has an approximate inversion symmetry while (c) breaks the inversion symmetry stronly. 
    } 
    \label{lattice_define}
\end{figure}

There are two types of structures R$n$G/hBN. 
The two structures are distinguished by their stackings in the AA regions of the moir\'e structure. (See the example of zero twist in \figref{lattice_define}(a),(b).)
In one structure, the carbon A/B in the bottom layer of graphene is aligned with nitrogen/boron in the AA region, which we call $\xi=1$ configuration; the other 
 $\xi=0$ configuration corresponds to that the carbon A/B in the bottom layer of graphene align with boron/nitrogen.
The symmetries of the $\xi=0, 1$ structures are reduced compared to the pristine R$n$G as hBN breaks inversion and nonzero small twist angles breaks mirrors.
Only the three-fold rotation $C_{3}$ (with axis perpendicular to the sample) can be preserved for all twist angles. 
The $\xi=0, 1$ structures are related by a $C_{2}$ rotation (about the axis perpendicular to the sample) on the hBN only. 

We also consider hBN/R$n$G/hBN structures with hBN nearly aligned on top and on bottom. 
We choose the alignments of two hBNs such that there is only one moir\'e pattern and $C_3$ symmetry (with axis perpendicular to the sample) is preserved.
The stackings of the top and bottom hBNs (relative to graphene) are labeled by $\xi_t, \xi_b$, respectively.
For simplicity, we in this work only study $(\xi_b, \xi_t) = (1,1)$ (\figref{lattice_define}(c))  and $(\xi_b, \xi_t) = (1,0)$ (\figref{lattice_define}(d)).
We do not need to study $(\xi_b, \xi_t) = (0,0)$, since it is related to $(\xi_b, \xi_t) = (1,1)$ by a $C_2$ rotation.
We note that $(\xi_b, \xi_t) = (1,0), (0,1)$ can restore the inversion symmetry if the two hBNs are related by inversion and their inversion center is aligned with that of the R$n$G; we choose $(\xi_b, \xi_t) = (1,0)$ as a representative as this type of structure.
In our DFT calculation for $(\xi_b, \xi_t) = (1,0)$ structure, the two inversion centers are nearly aligned with inversion symmetry preserved up to energy error about 0.1meV, which will be discussed in \appref{app:band_structure}.
Certainly, inversion symmetry can be broken by the displacement field. 

The rigid lattice constant for Graphene is 0.246nm and for hBN is 0.25nm, rigid layer distance for both Graphene-Graphene and Graphene-hBN  are 0.336nm. Our calculations mainly focus on the $0.77^\circ$ twisted angle, which the moir\'e lattice constant is approximately 11.6nm, and the number of carbon atom in the moir\'e superlattice is $4472$ per layer, the number of boron and nitrogen atom in the moir\'e superlattice are both $2163$ per layer. 

We perform classical structure relaxation implemented in LAMMPS\cite{thompson_lammps_2022}. During the relaxation, we fix each hBN layer to simulate a thick substrate and  keep the moir\'e unit cell unchanged. We used empirical inter-atomic potentials in LAMMPS to perform relaxation.  In graphene and hBN systems, the inter-layer interaction acts differently than intra-layer one due to the van der Waals interaction. Therefore, they are usually treated with different empirical potentials in classical molecular dynamics simulations. For intra-layer interactions within graphene layers, we used the reactive empirical bond order potential\cite{brenner_second-generation_2002}. For inter-layer interactions, we used an inter-layer potential developed for graphene and hBN systems\cite{ouyang_nanoserpents_2018}. This combination of empirical potentials has been well tested in \refcite{ouyang_nanoserpents_2018}, and have good agreement with DFT results.

 \begin{figure}
     \centering
     \includegraphics[width=1.0\linewidth]{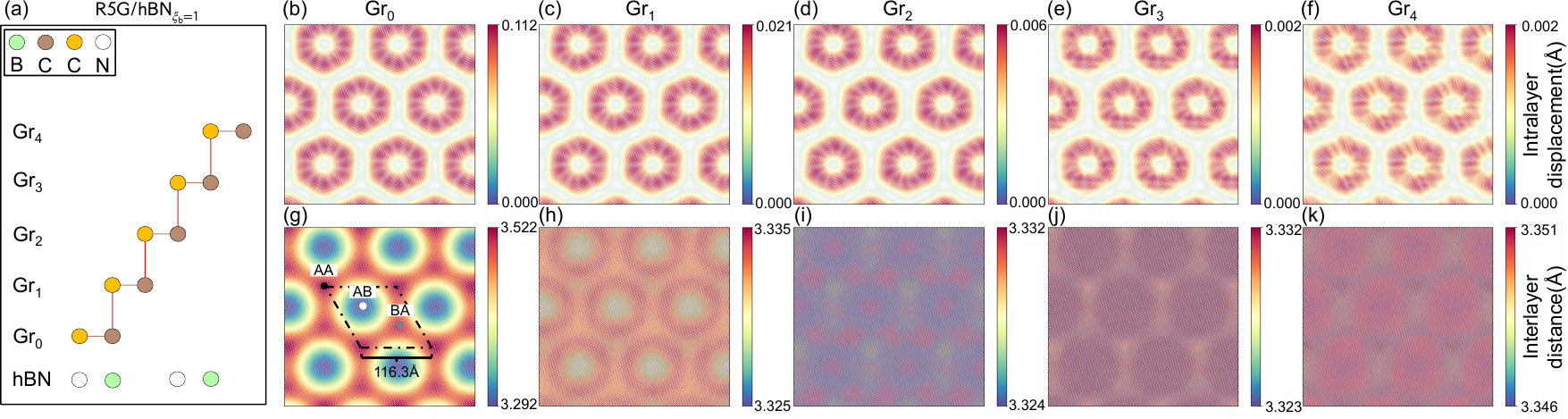}
     \caption{Relaxation results of 0.767 $^\circ$ R5G/hBN$_{\xi_b=1}$. $(a)$ Rigid structure of R5G/hBN$_{\xi_b=1}$ for $\xi=1$. $(b)$-$(f)$ show the intra-layer displacement of each layer. $(g)$-$(k)$ show the inter-layer distance of each layer. The Gr$_0$ layer is the lowest layer, which lies on the hBN substrate. The intra-layer displacement is the in-plane displacement from rigid position to relaxed position of an atom in the corresponding layer. The inter-layer distance indicates the distance between the corresponding layer and the layer under it.}
     \label{rlx_hBN0_C5}
 \end{figure}

 \begin{figure}
     \centering
     \includegraphics[width=1.0\linewidth]{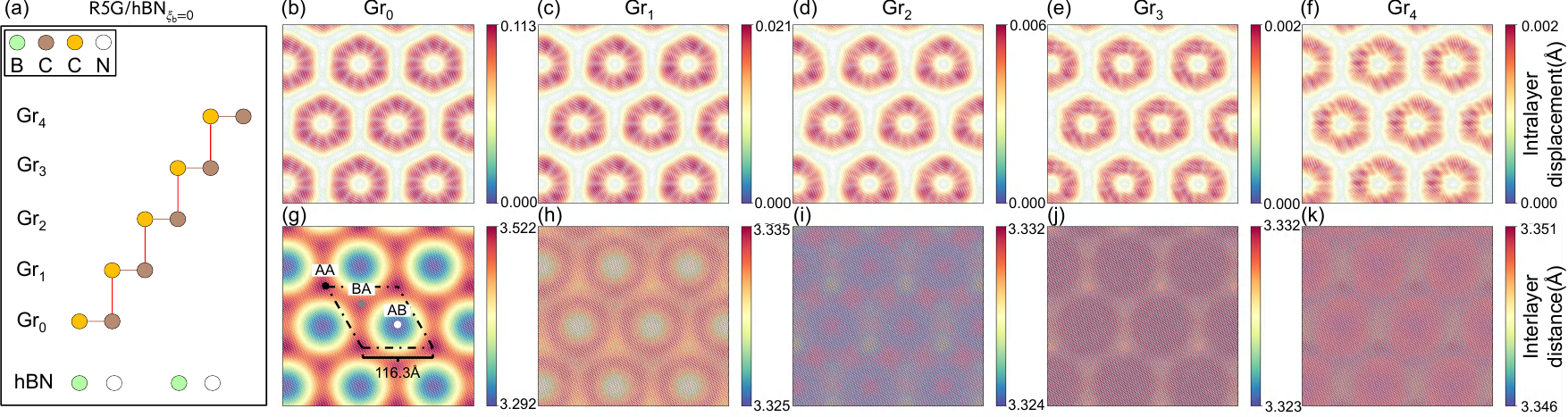}
     \caption{Relaxation results of 0.767 $^\circ$ R5G/hBN$_{\xi_b=0}$. (a), structure  of R5G/hBN$_{\xi_b=0}$. (b)-(f), intra-layer displacement of each layer. (g)-(k), inter-layer distance of each layer. The Gr$_0$ layer is the lowest layer, which lies on the hBN substrate. }
     \label{rlx_hBN180_C5}
 \end{figure}

The relaxation results for R$n$G/hBN configurations are listed in \figref{rlx_hBN0_C5} and \figref{rlx_hBN180_C5}. The interlayer distance between the Gr$_0$ layer and hBN becomes the largest(3.52$\AA$) in the AA region and the smallest(3.29$\AA$) in the AB region. In upper graphene layers, the inter-layer distance between graphene is around 3.33$\AA$ in all regions. On the other hand, the magnitude of intra-layer displacement also decreases with the layer number, which is reasonable since the hBN substrate has less effect on the upper layers. The atoms near the AB region tend to rotate in a clockwise direction, against the global twist (counterclockwise). This will enlarge the AB region, where the local stacking energy reaches the minimum. 

\begin{figure}[H]
     \centering
     \includegraphics[width=1.0\linewidth]{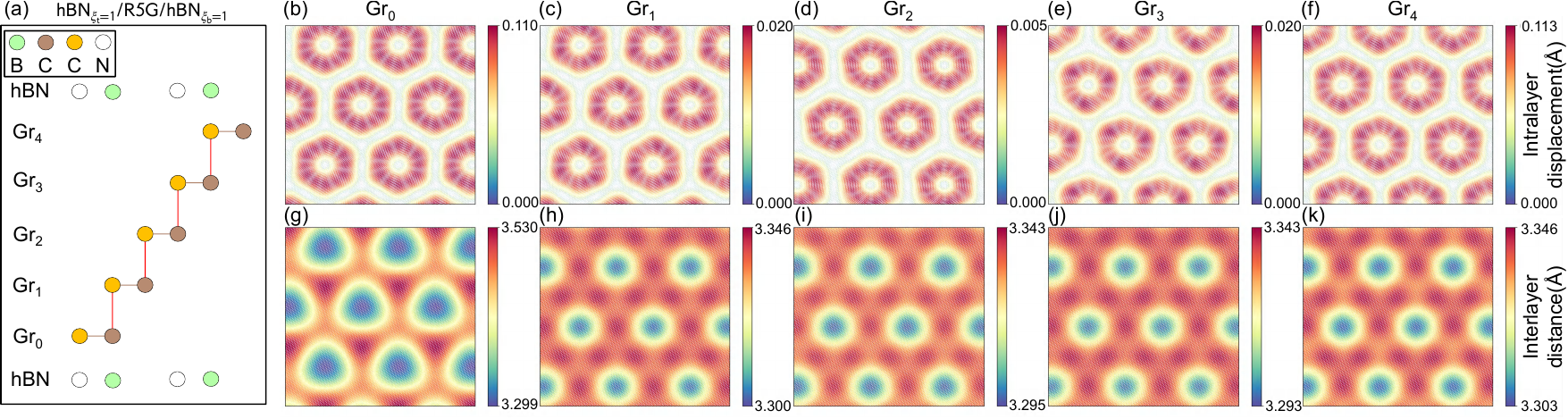}
     \caption{Relaxation results of 0.767 $^\circ$ hBN$_{\xi_t=1}$/R5G/hBN$_{\xi_b=1}$. (a), structure of hBN$_{\xi_t=1}$/R5G/hBN$_{\xi_b=1}$. (b)-(f), intra-layer displacement of each layer. (g)-(k), inter-layer distance of each layer. The Gr$_0$ layer is the lowest, and Gr$_4$ is the highest layer. The Gr$_0$ and Gr$_4$ layers are near hBN substrates.}
     \label{rlx_hBN0_C5_hBN0}
 \end{figure}

\begin{figure}[H]
     \centering
     \includegraphics[width=1.0\linewidth]{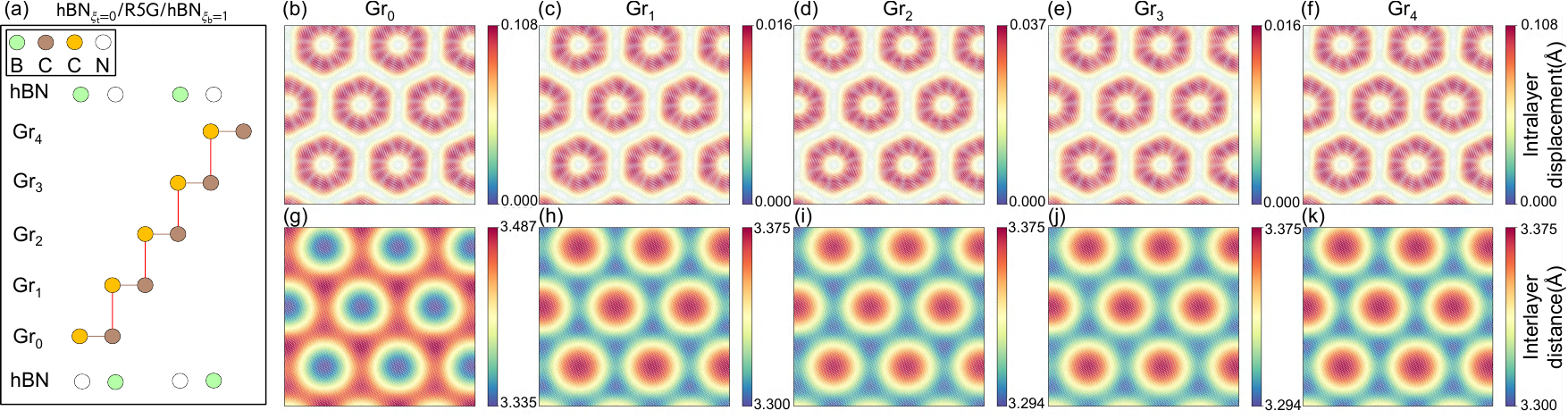}
     \caption{Relaxation results of 0.767 $^\circ$ hBN$_{\xi_t=0}$/R5G/hBN$_{\xi_b=1}$. (a), structure of hBN$_{\xi_t=0}$/R5G/hBN$_{\xi_b=1}$. (b)-(f), intra-layer displacement of each layer. (g)-(h), inter-layer distance of each layer. The Gr$_0$ layer is the lowest layer, and Gr$_4$ is the highest layer. The Gr$_0$ and Gr$_4$ layers are near hBN substrates.}
     \label{rlx_hBN0_C5_hBN180}
\end{figure}

The relaxation results for hBN/R$5$G/hBN are also listed in \figref{rlx_hBN0_C5_hBN0} and \figref{rlx_hBN0_C5_hBN180}. In these configurations, the distance between upper and lower hBN is fixed at 22.66$\AA$.

\section{Slater-Koster method}
\label{app:band_structure}

\begin{figure}
    \centering
    \includegraphics[width=1.0\linewidth]{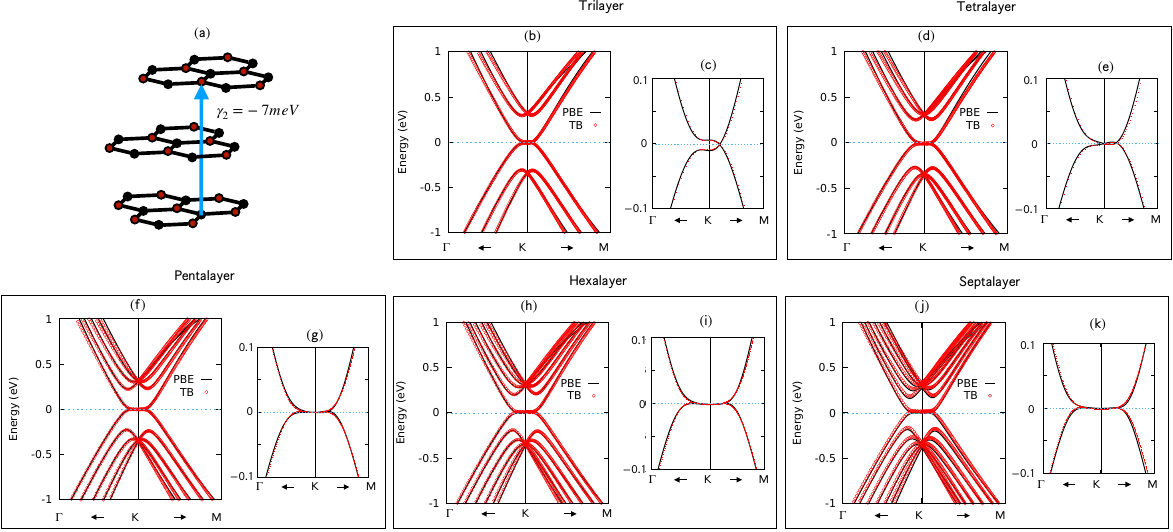}
    \caption{  Comparison of the tight-binding Hamiltonian (red circles) and ab initio DFT (black lines) band structure calculations along the $\Gamma-\text{K}-\text{M}$ path for ABC-stacked graphene: (a) Illustration of the $\gamma_2$  bond correction in the SK tight-binding model. Panels (b) and (c) for trilayer, (d) and (e) for tetralayer, (f) and (g) for pentalayer, (h) and (i) for hexalayer, and (j) and (k) for septalayer graphene depict band structures near the $\K$ point within two energy windows, from -0.1 eV to 0.1 eV and from -1 eV to 1 eV, using a set value of $V_{\mathrm{pp}\pi}^{0}=-2.81$ eV and an internal symmetrical polarization (ISP) of 5mV/\AA. }
    \label{layer-graphene-vasp}
\end{figure}

The band structure are obtained within the SK tight-binding model using the parameters listed in \eqnref{eq:SK}. The diagonalization is performed with an open-source software WannierTools\cite{WU2018405}. In our study, we consider an internal symmetrical polarization (ISP) as 5mV/\AA \ and $V_{\mathrm{pp}\pi}^{0}$ as -2.81 eV~\cite{SKTB-Wu2020}, a value exceeding the conventional -2.7 eV. This adjustment is made to enhance the compatibility with our DFT calculations of ABC-stacked graphene with different number layers in \figref{layer-graphene-vasp}.

\begin{figure}
    \centering
    \includegraphics[width=1.0\linewidth]{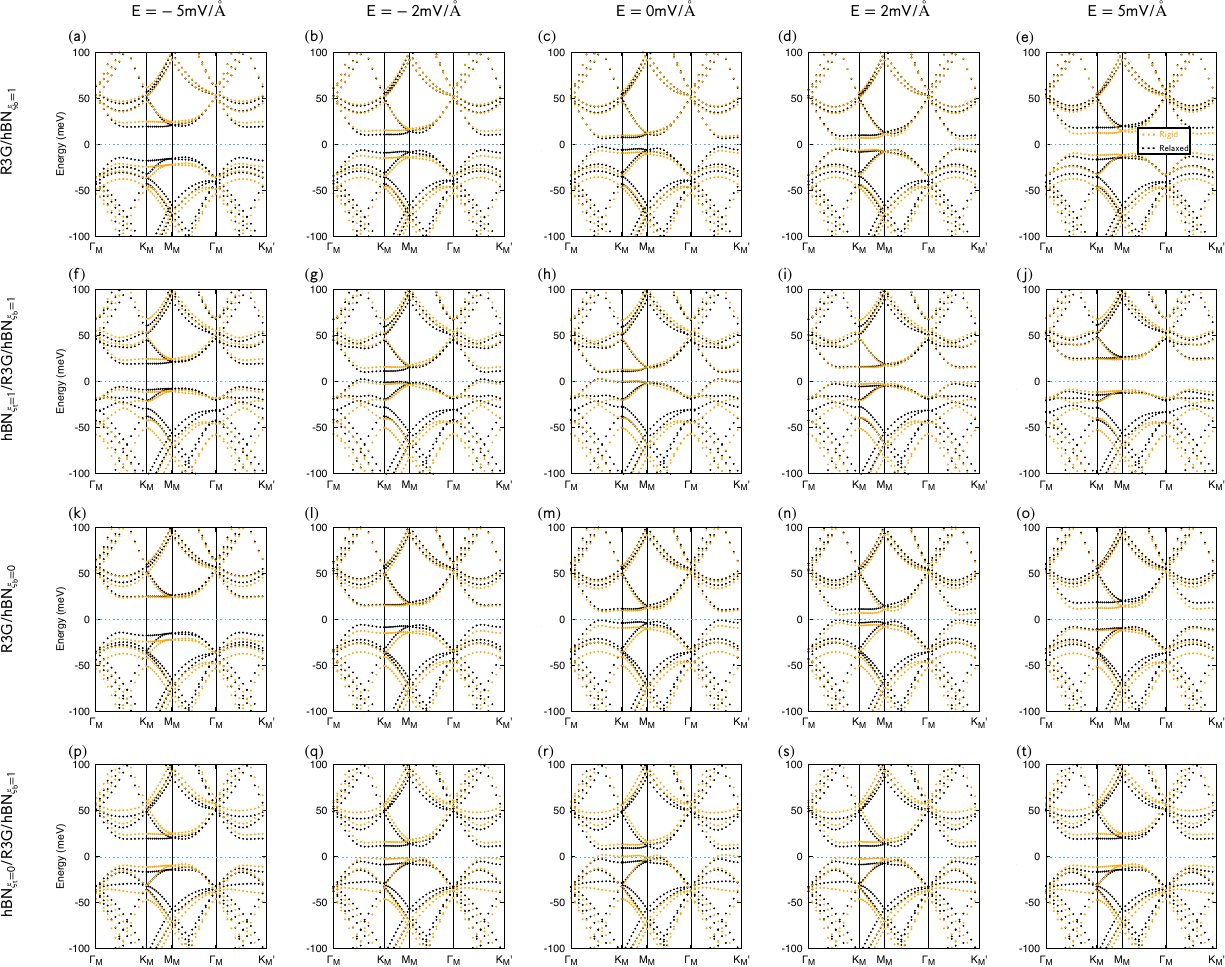}
    \caption{Comparison of band structures between rigid and relaxed structures of R3G and 0.76715$^\circ$ twisted angle setting ISP=5mV/\AA. E is an applied electrical ﬁeld and ISP is an internal symmetrical polarization due to the diﬀerent chemical environment of outer and inner atoms in trilayer graphene. The positive direction of E and ISP are shown in \figref{relaxation}. The structures corresponding to the first row to the fourth row are R3G/$\mathrm{hBN}_{\xi_b=1}$, $\mathrm{hBN}_{\xi_t=1}$/R3G/$\mathrm{hBN}_{\xi_b=1}$, R3G/$\mathrm{hBN}_{\xi_b=0}$ and $\mathrm{hBN}_{\xi_t=0}$/R3G/$\mathrm{hBN}_{\xi_b=1}$ respectively, and the applied electrical fields corresponding to the first column to the fifth column are -5, -2 0, 2 and 5mV/\AA \ respectively.  The band structures are  depicted with orange dotted lines for rigid structures and black dotted lines for relaxed structures. Here, both $K$ valley and $K'$ valley bands are included.}
    \label{compare-rigid-relaxed-3Gr-with-t2.pdf}
\end{figure}

\begin{figure}
    \centering
    \includegraphics[width=1.0\linewidth]{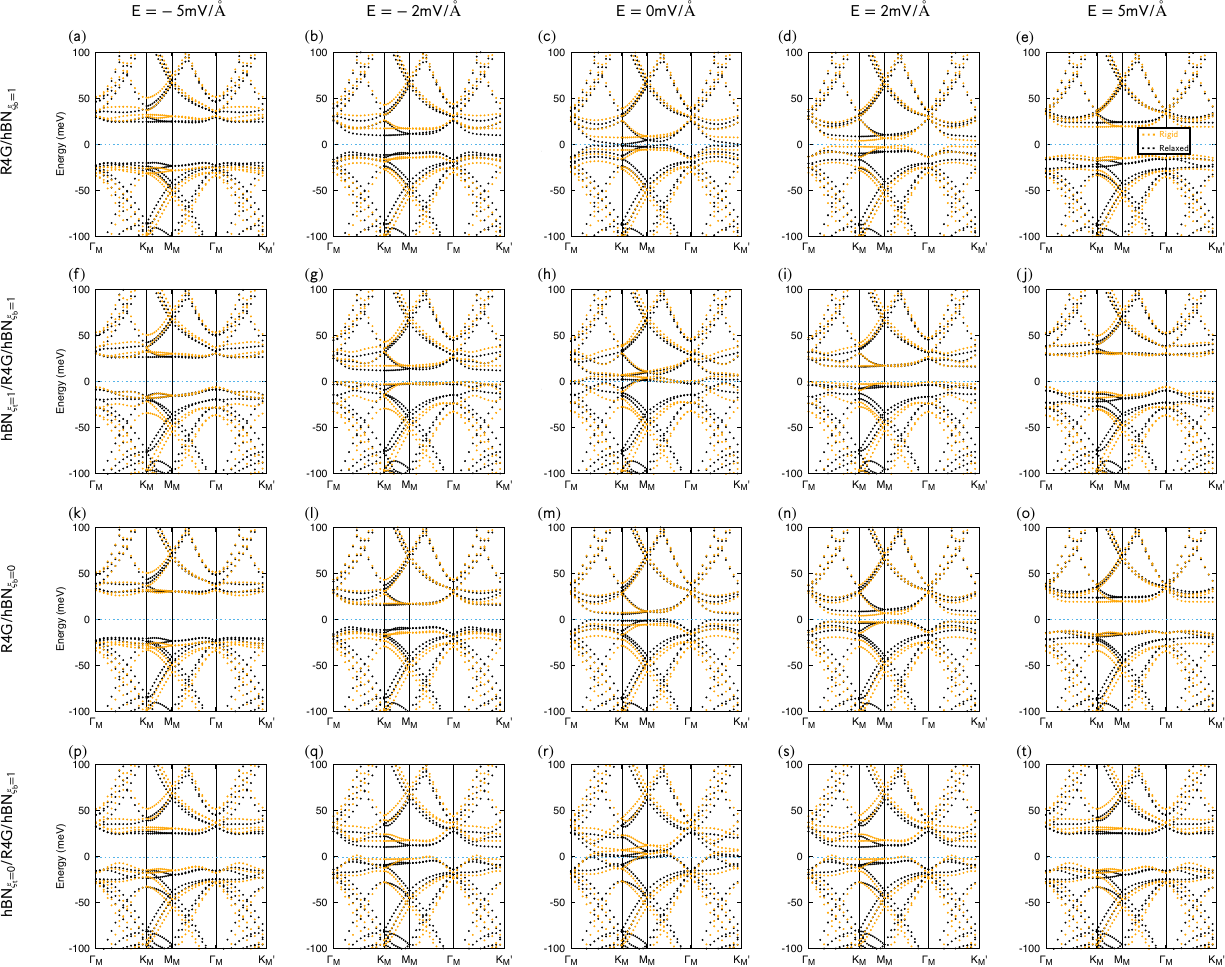}
    \caption{Comparison of band structures between rigid and relaxed structures of R4G and 0.76715$^\circ$ twisted angle setting ISP=5mV/\AA. E is an applied electrical ﬁeld and ISP is an internal symmetrical polarization due to the diﬀerent chemical environment of outer and inner atoms in tetralayer graphene. The positive direction of E and ISP are shown in \figref{relaxation}.  The structures corresponding to the first row to the fourth row are R4G/$\mathrm{hBN}_{\xi_b=1}$, $\mathrm{hBN}_{\xi_t=1}$/R4G/$\mathrm{hBN}_{\xi_b=1}$, R4G/$\mathrm{hBN}_{\xi_b=0}$ and $\mathrm{hBN}_{\xi_t=0}$/R4G/$\mathrm{hBN}_{\xi_b=1}$ respectively, and the applied electrical fields corresponding to the first column to the fifth column are -5, -2 0, 2 and 5mV/\AA \ respectively. The band structures are  depicted with orange dotted lines for rigid structures and black dotted lines for relaxed structures. Here, both $K$ valley and $K'$ valley bands are included.}
    \label{compare-rigid-relaxed-4Gr-with-t2.pdf}
\end{figure}

\begin{figure}
    \centering
    \includegraphics[width=1.0\linewidth]{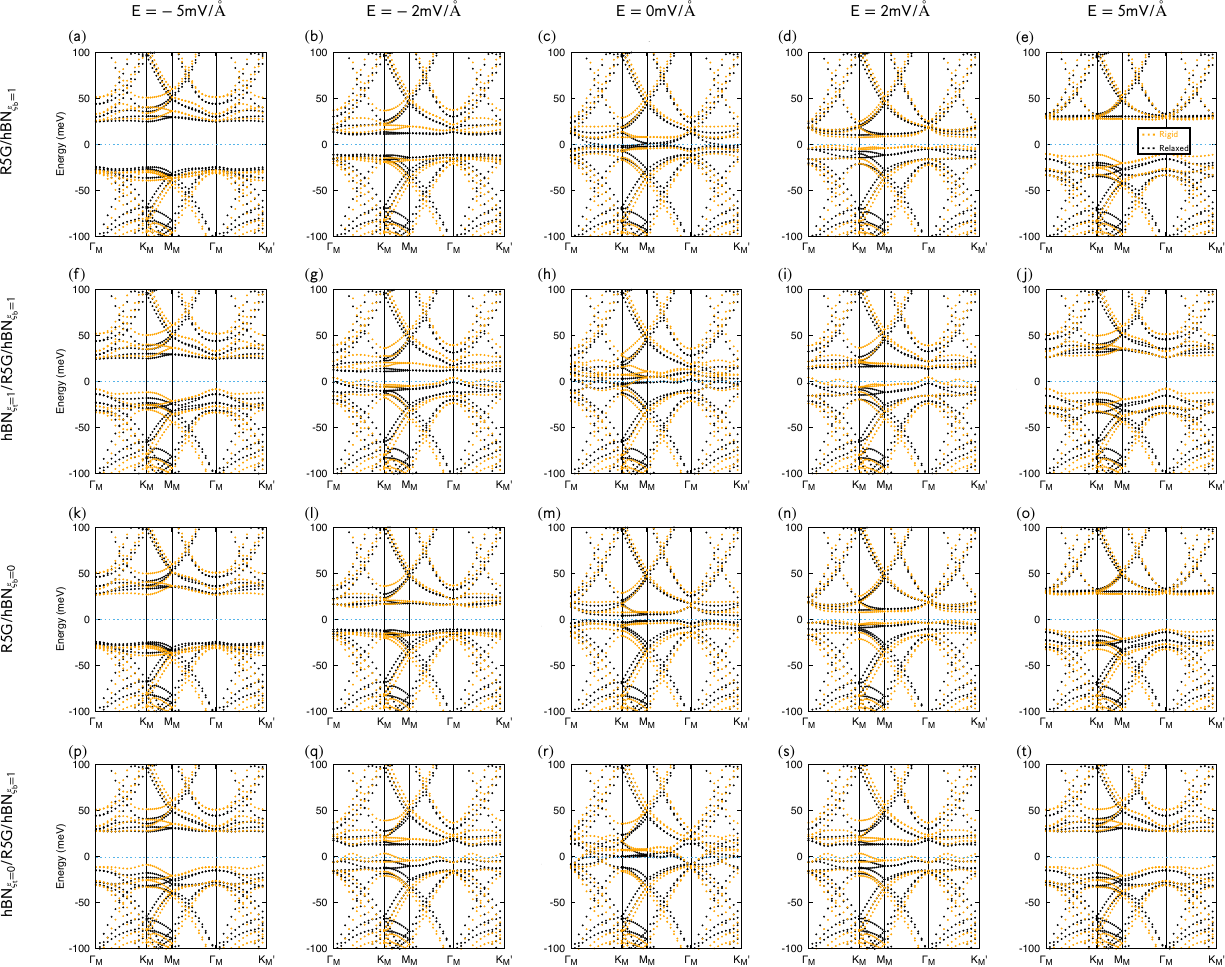}
    \caption{Comparison of band structures between rigid and relaxed structures of R5G and 0.76715$^\circ$ twisted angle setting ISP=5mV/\AA. E is an applied electrical ﬁeld and ISP is an internal symmetrical polarization due to the diﬀerent chemical environment of outer and inner atoms in pentalayer graphene. The positive direction of E and ISP are shown in \figref{relaxation}.   The structures corresponding to the first row to the fourth row are R5G/$\mathrm{hBN}_{\xi_b=1}$, $\mathrm{hBN}_{\xi_t=1}$/R5G/$\mathrm{hBN}_{\xi_b=1}$, R5G/$\mathrm{hBN}_{\xi_b=0}$ and $\mathrm{hBN}_{\xi_t=0}$/R5G/$\mathrm{hBN}_{\xi_b=1}$ respectively, and the applied electrical fields corresponding to the first column to the fifth column are -5, -2 0, 2 and 5mV/\AA \ respectively. The band structures are depicted with orange dotted lines for rigid structures and black dotted lines for relaxed structures. Here, both $K$ valley and $K'$ valley bands are included.}
    \label{compare-rigid-relaxed-5Gr-with-t2.pdf}
\end{figure}

\begin{figure}
    \centering
    \includegraphics[width=1.0\linewidth]{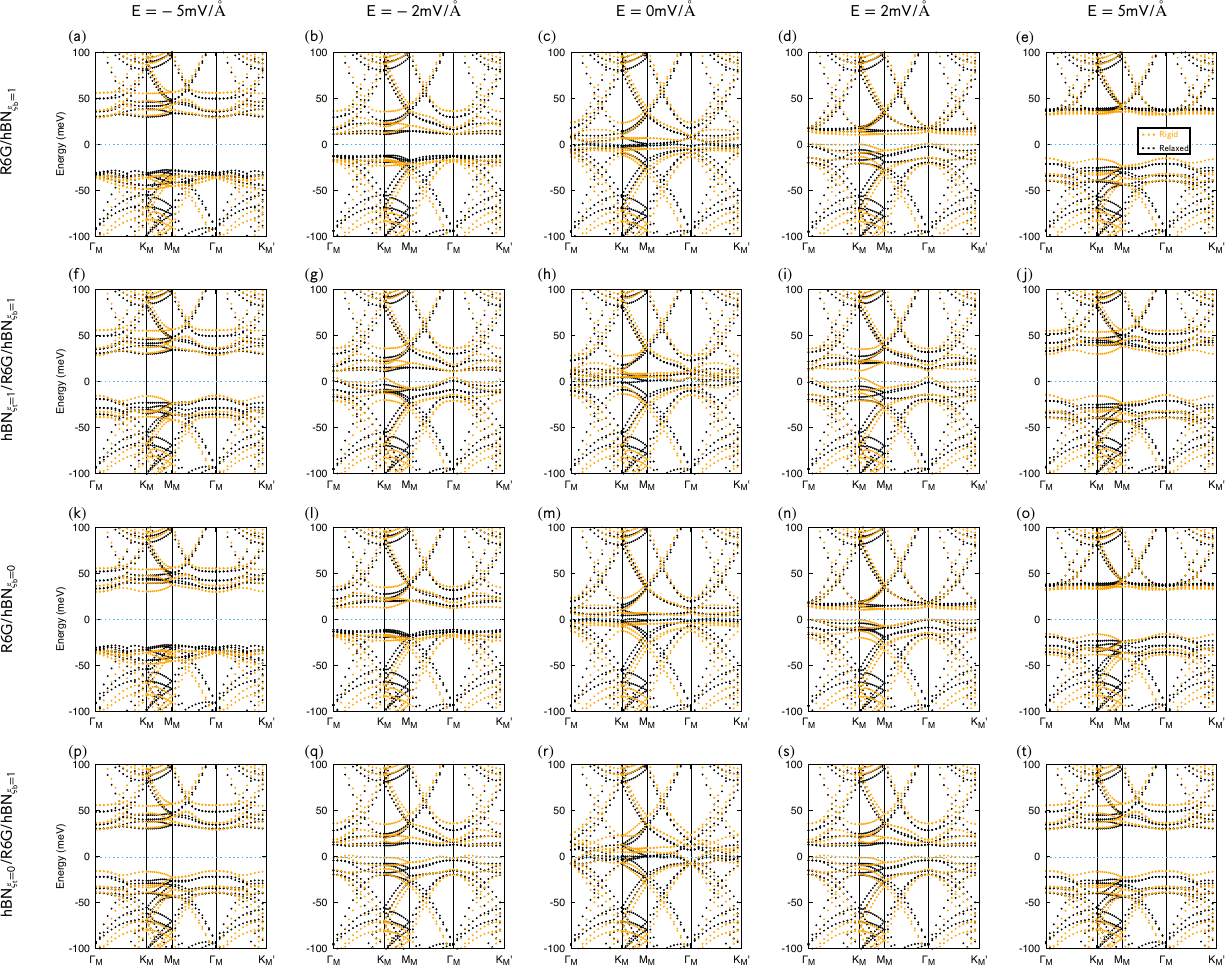}
    \caption{Comparison of band structures between rigid and relaxed structures of R6G and 0.76715$^\circ$ twisted angle setting ISP=5mV/\AA. E is an applied electrical ﬁeld and ISP is an internal symmetrical polarization due to the diﬀerent chemical environment of outer and inner atoms in hexalayer graphene. The positive direction of E and ISP are shown in \figref{relaxation}.  The structures corresponding to the first row to the fourth row are R6G/$\mathrm{hBN}_{\xi_b=1}$, $\mathrm{hBN}_{\xi_t=1}$/R6G/$\mathrm{hBN}_{\xi_b=1}$, R6G/$\mathrm{hBN}_{\xi_b=0}$ and $\mathrm{hBN}_{\xi_t=0}$/R6G/$\mathrm{hBN}_{\xi_b=1}$ respectively, and the applied electrical fields corresponding to the first column to the fifth column are -5, -2 0, 2 and 5mV/\AA \ respectively. The band structures are  depicted with orange dotted lines for rigid structures and black dotted lines for relaxed structures. Here, both $K$ valley and $K'$ valley bands are included.}
    \label{compare-rigid-relaxed-6Gr-with-t2.pdf}
\end{figure}

\begin{figure}
    \centering
    \includegraphics[width=1.0\linewidth]{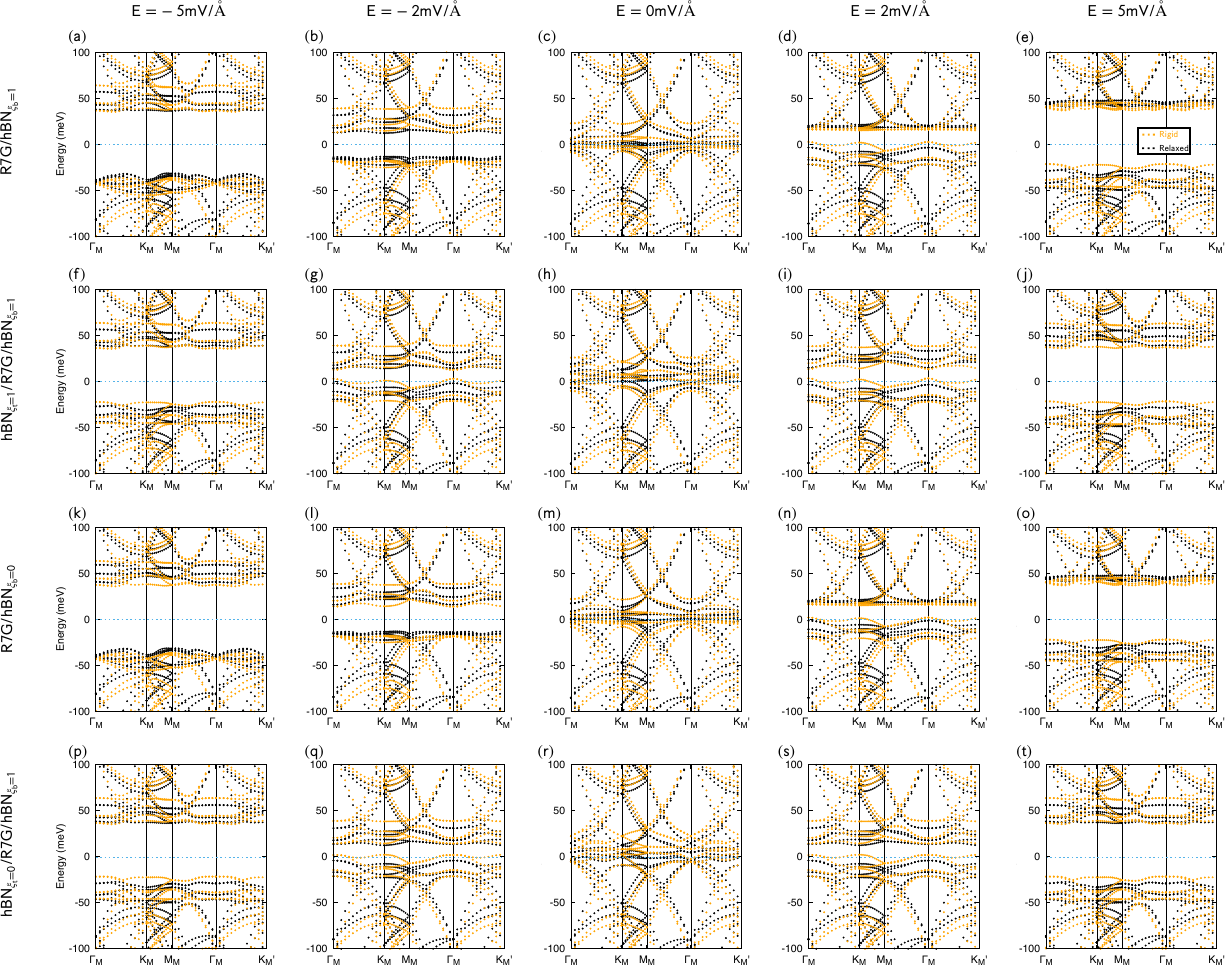}
    \caption{Comparison of band structures between rigid and relaxed structures of R7G and 0.76715$^\circ$ twisted angle setting ISP=5mV/\AA. E is an applied electrical ﬁeld and ISP is an internal symmetrical polarization due to the diﬀerent chemical environment of outer and inner atoms in septalayer graphene. The positive direction of E and ISP are shown in \figref{relaxation}.   The structures corresponding to the first row to the fourth row are R7G/$\mathrm{hBN}_{\xi_b=1}$, $\mathrm{hBN}_{\xi_t=1}$/R7G/$\mathrm{hBN}_{\xi_b=1}$, R7G/$\mathrm{hBN}_{\xi_b=0}$ and $\mathrm{hBN}_{\xi_t=0}$/R7G/$\mathrm{hBN}_{\xi_b=1}$ respectively, and the applied electrical fields corresponding to the first column to the fifth column are -5, -2 0, 2 and 5mV/\AA \ respectively. The band structures are  depicted with orange dotted lines for rigid structures and black dotted lines for relaxed structures. Here, both $K$ valley and $K'$ valley bands are included.}
    \label{compare-rigid-relaxed-7Gr-with-t2.pdf}
\end{figure}

Our study delves into the distinctive characteristics of the low-energy valence and conduction bands. By comparing the relaxed band structure and the rigid band structure in \figsref{compare-rigid-relaxed-3Gr-with-t2.pdf}, \ref{compare-rigid-relaxed-4Gr-with-t2.pdf}, \ref{compare-rigid-relaxed-5Gr-with-t2.pdf}, \ref{compare-rigid-relaxed-6Gr-with-t2.pdf} and \ref{compare-rigid-relaxed-7Gr-with-t2.pdf}, we can find significant changes in the band structure in some specific cases whose gap decrease at charge neutrality. In order to delve deeper into this pattern, we systematically investigated the flatter bandwidths across different numbers of graphene layers, ranging from 3 to 7. The outcomes of our analysis are succinctly presented in \figsref{hBN-3Gr-isp5-0.77}, \ref{hBN-4Gr-isp5-0.77}, \ref{hBN-5Gr-isp5-0.77}, \ref{hBN-6Gr-isp5-0.77} and \ref{hBN-7Gr-isp5-0.77} whose panels (a)-(t) showcase the band structures of R$n$G/$\mathrm{hBN}_{\xi_b=1}$, $\mathrm{hBN}_{\xi_t=1}$/R$n$G/$\mathrm{hBN}_{\xi_b=1}$, R$n$G/$\mathrm{hBN}_{\xi_b=0}$ and $\mathrm{hBN}_{\xi_t=0}$/R$n$G/$\mathrm{hBN}_{\xi_b=1}$  respectively. Applying displacement field would help generate flat bands.

\clearpage
\section{Valley-Resolved Band Structure}
\label{app:valley_resolved}
In superlattices form by R$n$G and hBN, the valleys of R$n$G emerge as a good quantum number in the low energy part of bands. Inherited from the gapless Dirac dispersion band structure at $\K,\K'$ valleys in untwisted case, the low energy bands of R$n$G-hBN superlattice either belongs to $K$ valley or $K'$ valley. Thus, We can construct an operator to distinguish the valley degree of freedom with eigenvalue $+1$ for the states at one valley and eigenvalue $-1$ for the states at the other valley, which is in the same spirit as the spin-z operator in basis $\{\ket{\uparrow},\ket{\downarrow}\}$. 

\begin{figure}[H]
    \centering
    \includegraphics[width=1.0\linewidth]{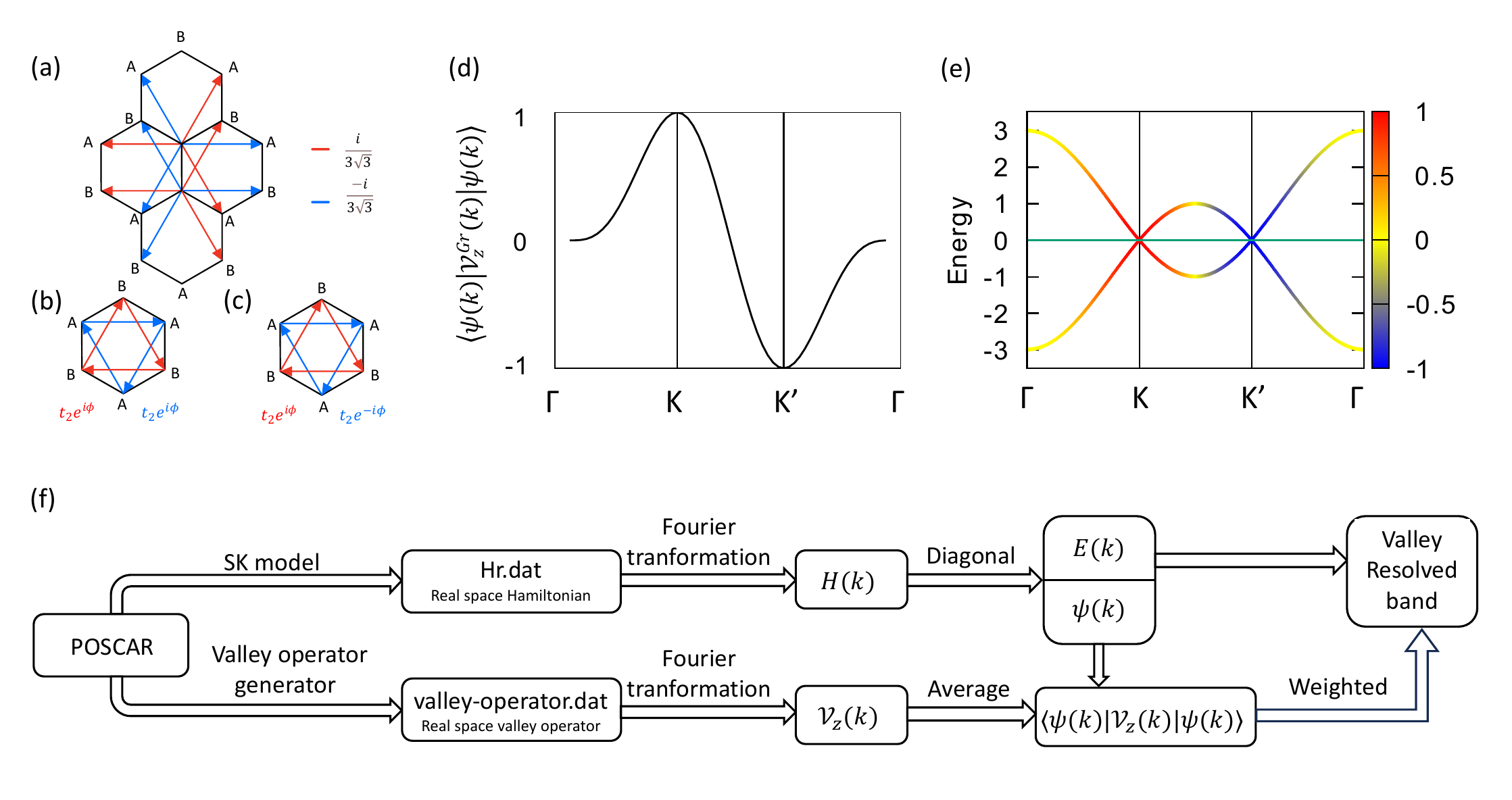}
    \caption{ Valley operator for single layer graphene. (a) shows the valley operator in real space for one unit cell; (b) shows the Haldane's NNN hopping term with local magnetic flux and (c) shows the corresponding NNN hopping term with valley flux; (d) shows the average value of valley operator $\bra{\psi(k)} \mathcal{V}_z^{Gr}(k)  \ket{\psi(k)}$ in high symmetry point path of reciprocal space. (e) is the energy band of SLG follow the same path as (d), weighted by $\bra{\psi(k)} \mathcal{V}_z^{Gr}(k)  \ket{\psi(k)}$.
    (f) is the flow diagram of valley resolved bands calculation, this flow is applicable for both SLG and hBN-R$n$G.}
    \label{fig:single_layer_valley}
\end{figure}

The valley operator for Dirac fermion model --- single layer graphene(SLG) in real space reads \cite{PhysRevResearch.2.033357}
\begin{equation}    
    \mathcal{ V }_z^{Gr} = \frac{i}{3\sqrt{3}}\sum_{\ll i,j\gg,s} \eta_{ij}\sigma_z^{i\,j}c^{\dagger}_{i,s}c_{j,s}
\end{equation}
where $\ll i,j\gg$ denotes the next-nearest-neighbor(NNN) hopping term, $s$ denote spin, $\eta_{ij}=\pm 1$ for clockwise or counterclockwise hopping and $\sigma_z$ is in the sublattice degree of freedom, meaning that $\sigma_z^{i\,j}= 1$ if both $i,j$ are on the $A$ sublattice, $\sigma_z^{i\,j}= -1$ if both $i,j$ are on the $B$ sublattice and $\sigma_z^{i\,j}= 0$ otherwise. The schematic diagram of valley operator in real space for an unit cell is shown in \figref{fig:single_layer_valley}(a). The factor $\eta_{ij} \sigma_z^{ij}$ forms the valley flux, which distinguishes the NNN hopping in $K$ direction and $K'$ direction. To specify the valley flux, we compare it with Haldane's local magnetic flux(\figref{fig:single_layer_valley}(b),(c)). While the latter holds the same flux with both sublattice A and B, valley flux means sublattice A/B holds opposite local magnetic flux. After performing Fourier transformation, the valley operator of single layer graphene in sublattice is proportional to the identity matrix in reciprocal space, $\mathcal{ V }_z^{Gr}(k)=\left ( \begin{matrix}
 f(k) & \\
  & f(k)
\end{matrix} \right ) $. $f(k)$ is an odd function in Brillouin Zone $f(k)=-f(-k)$, which satisfies $\bra{\psi(K)} \mathcal{V}_z(K)  \ket{\psi(K)} = 1$ and $\bra{\psi(K')} \mathcal{V}_z(K')  \ket{\psi(K')} = -1$(\figref{fig:single_layer_valley}(d),(e)).

Since valley operator in real space only contains intra-layer hopping term, it can be easily generalised to R$n$G by sum up valley operator of each layer together $\mathcal{V}_z=\sum_{l}\mathcal{V}_z^{l}$, where $l$ denotes the layer index. And for twisted hBN-R$n$G system, 
valley operator of each layer can be obtained numerically by considering all the NNN hopping term multiplied by valley flux factor $\eta_{ij}\sigma_z^{ij}$ for the atoms in superlattice and this layer. The process to obtain real space valley operator for moir\'e system corresponds to valley band folding in momentum space.

\begin{figure}[H]
    \centering
    \includegraphics[width=1.0\linewidth]{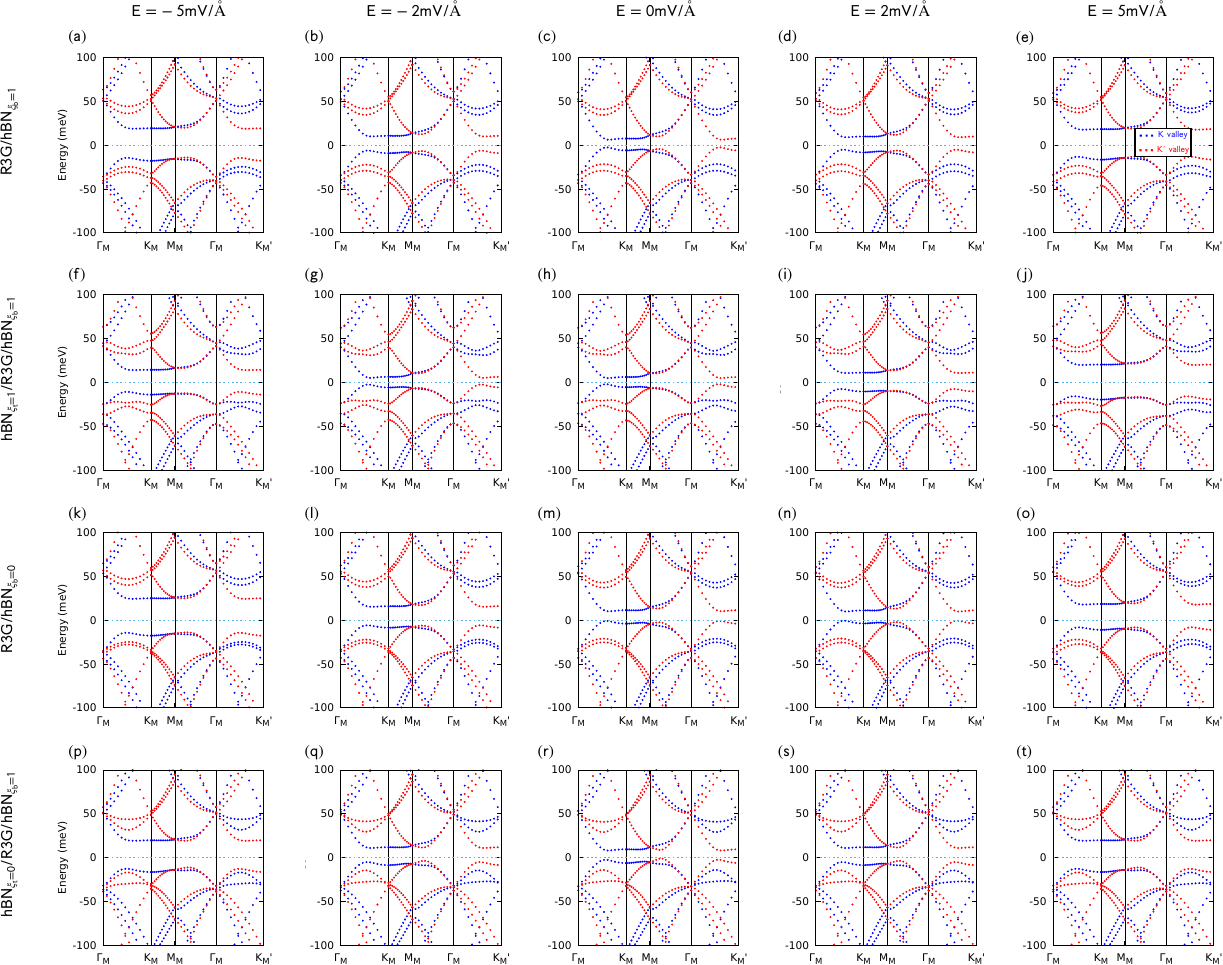}
    \caption{Tight-binding band structures of R3G and 0.76715$^\circ$ twist angle setting ISP=$5$meV/\AA. E is an applied electrical ﬁeld, and ISP is an internal symmetrical polarization due to the diﬀerent chemical environment of outer and inner atoms in trilayer graphene. The positive direction of E and ISP are shown in \figref{relaxation}. The structures corresponding to the first row to the fourth row are R3G/$\mathrm{hBN}_{\xi_b=1}$, $\mathrm{hBN}_{\xi_t=1}$/R3G/$\mathrm{hBN}_{\xi_b=1}$, R3G/$\mathrm{hBN}_{\xi_b=0}$ and $\mathrm{hBN}_{\xi_t=0}$/R3G/$\mathrm{hBN}_{\xi_b=1}$ respectively, and the applied electrical fields corresponding to the first column to the fifth column are -5, -2 0, 2 and 5mV/\AA \ respectively. The band structures are  depicted with blue dotted lines for $K$ valley and red dotted lines for $K'$ valley. }
    \label{hBN-3Gr-isp5-0.77}
\end{figure}

\begin{figure}[H]
    \centering
    \includegraphics[width=1.0\linewidth]{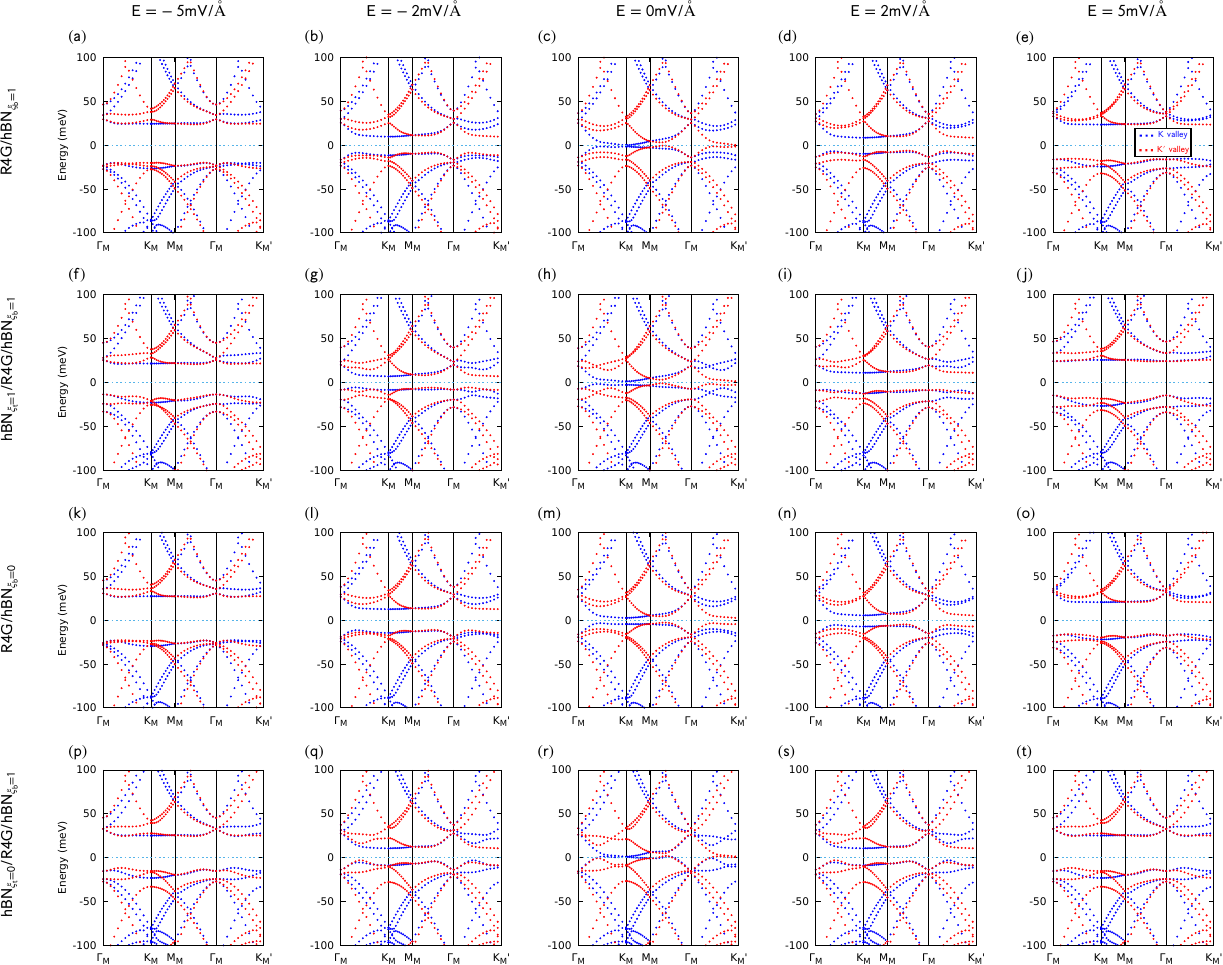}
    \caption{Tight-binding band structures of R4G and 0.76715$^\circ$ twisted angle setting ISP=5mV/\AA. E is an applied electrical ﬁeld and ISP is an internal symmetrical polarization due to the diﬀerent chemical environment of outer and inner atoms in tetralayer graphene. The positive direction of E and ISP are shown in \figref{relaxation}.
     The structures corresponding to the first row to the fourth row are R4G/$\mathrm{hBN}_{\xi_b=1}$, $\mathrm{hBN}_{\xi_t=1}$/R4G/$\mathrm{hBN}_{\xi_b=1}$, R4G/$\mathrm{hBN}_{\xi_b=0}$ and $\mathrm{hBN}_{\xi_t=0}$/R4G/$\mathrm{hBN}_{\xi_b=1}$ respectively, and the applied electrical fields corresponding to the first column to the fifth column are -5, -2, 0, 2 and 5mV/\AA \ respectively. The band structures are  depicted with blue dotted lines for $K$ valley and red dotted lines for $K'$ valley. }
    \label{hBN-4Gr-isp5-0.77}
\end{figure}

\begin{figure}[H]
    \centering
    \includegraphics[width=1.0\linewidth]{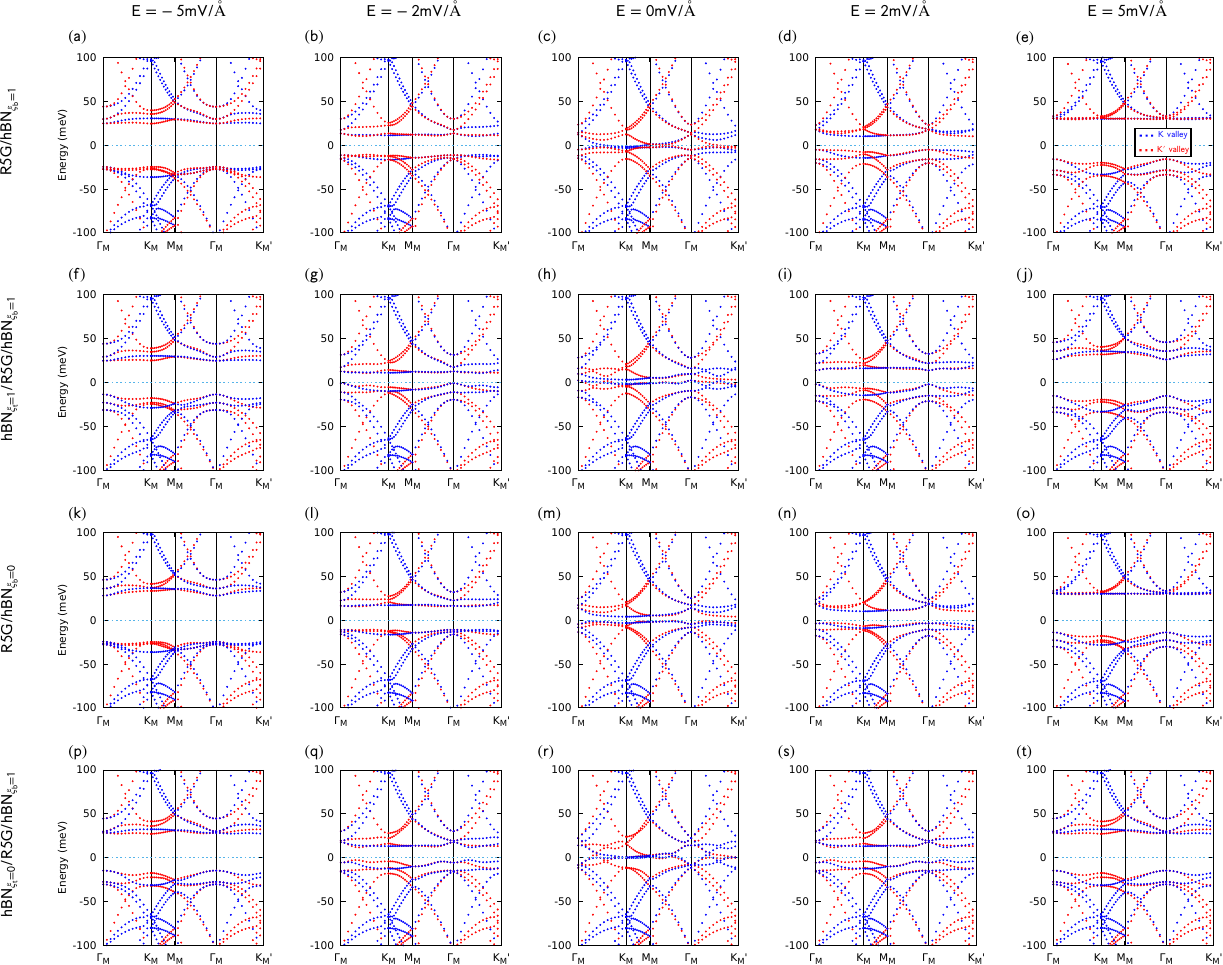}
    \caption{Tight-binding band structures of R5G and 0.76715$^\circ$ twisted angle setting ISP=5mV/\AA. E is an applied electrical ﬁeld and ISP is an internal symmetrical polarization due to the diﬀerent chemical environment of outer and inner atoms in pentalayer graphene. The positive direction of E and ISP are shown in \figref{relaxation}. 
     The structures corresponding to the first row to the fourth row are R5G/$\mathrm{hBN}_{\xi_b=1}$, $\mathrm{hBN}_{\xi_t=1}$/R5G/$\mathrm{hBN}_{\xi_b=1}$, R5G/$\mathrm{hBN}_{\xi_b=0}$ and $\mathrm{hBN}_{\xi_t=0}$/R5G/$\mathrm{hBN}_{\xi_b=1}$ respectively and the applied electrical fields corresponding to the first column to the fifth column are -5, -2, 0, 2 and 5mV/\AA \ respectively. The band structures are  depicted with blue dotted lines for $K$ valley and red dotted lines for $K'$ valley. }
    \label{hBN-5Gr-isp5-0.77}
\end{figure}

\begin{figure}[H]
    \centering
    \includegraphics[width=1.0\linewidth]{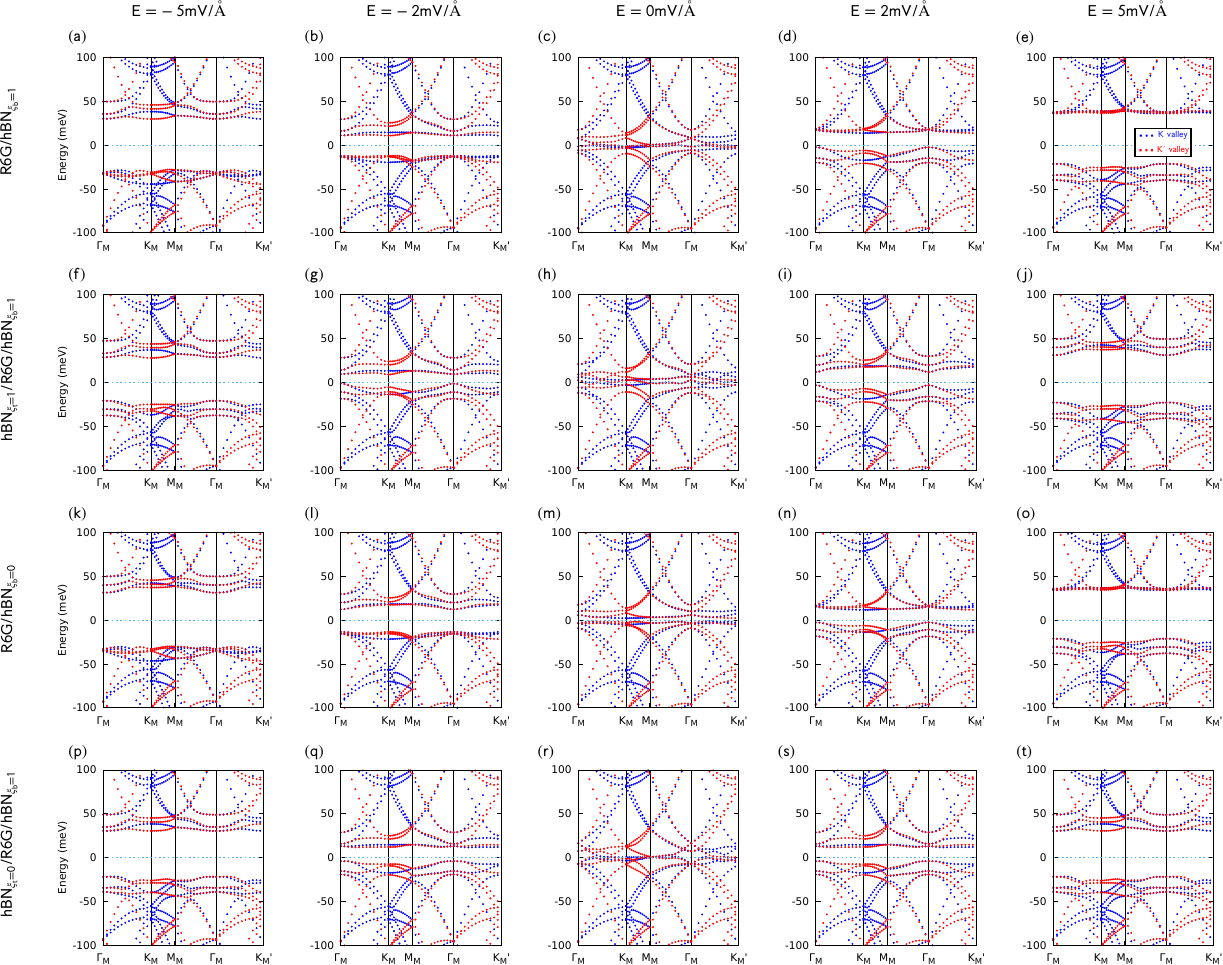}
    \caption{Tight-binding band structures of R6G and 0.76715$^\circ$ twisted angle setting ISP=5mV/\AA. E is an applied electrical ﬁeld and ISP is an internal symmetrical polarization due to the diﬀerent chemical environment of outer and inner atoms in hexalayer graphene. The positive direction of E and ISP are shown in \figref{relaxation}. 
     The structures corresponding to the first row to the fourth row are R6G/$\mathrm{hBN}_{\xi_b=1}$, $\mathrm{hBN}_{\xi_t=1}$/R6G/$\mathrm{hBN}_{\xi_b=1}$, R6G/$\mathrm{hBN}_{\xi_b=0}$ and $\mathrm{hBN}_{\xi_t=0}$/R6G/$\mathrm{hBN}_{\xi_b=1}$ respectively, and the applied electrical fields corresponding to the first column to the fifth column are -5, -2, 0, 2 and 5mV/\AA \ respectively. The band structures are  depicted with blue dotted lines for $K$ valley and red dotted lines for $K'$ valley. }
    \label{hBN-6Gr-isp5-0.77}
\end{figure}

\begin{figure}[H]
    \centering
    \includegraphics[width=1.0\linewidth]{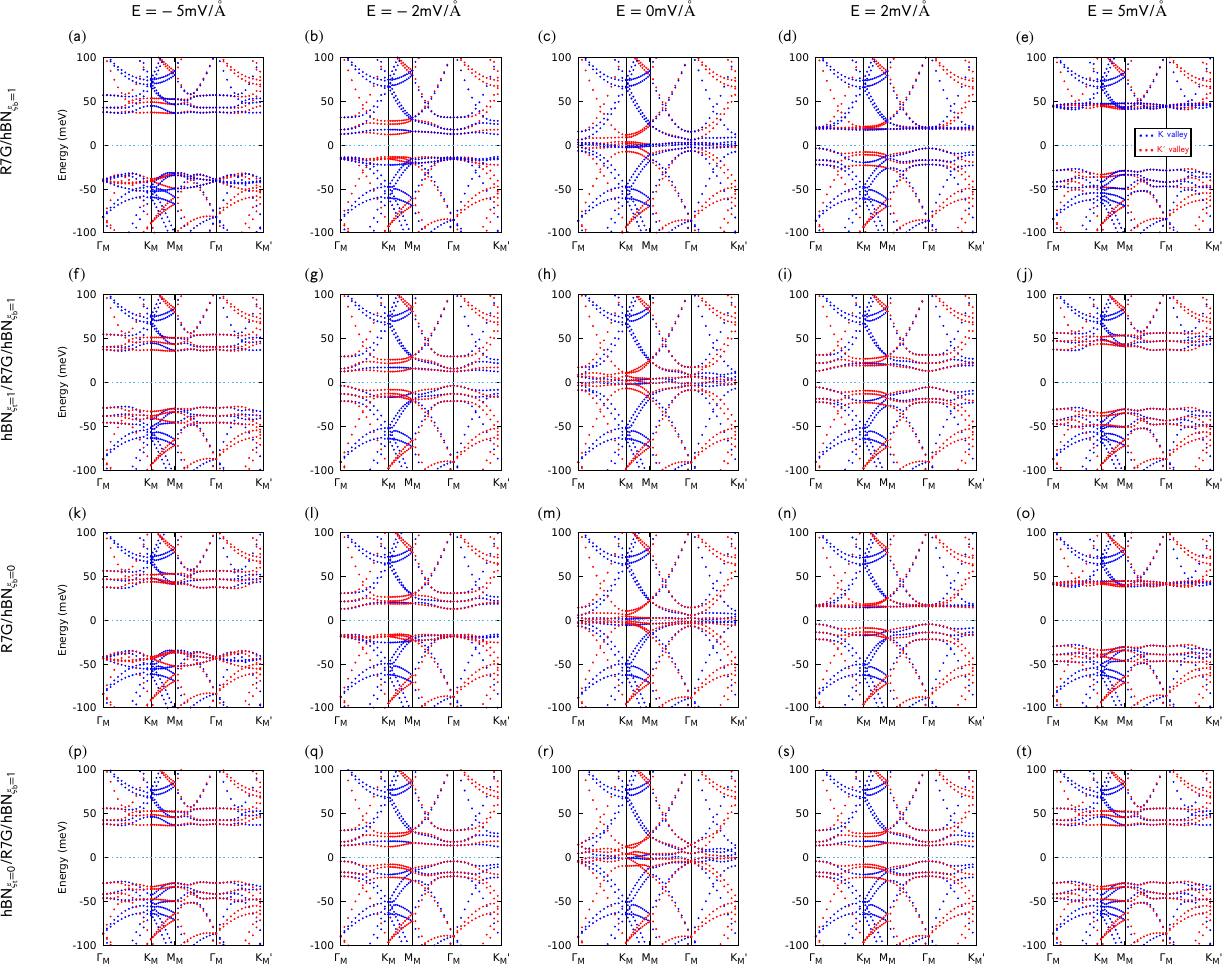}
    \caption{Tight-binding band structures of R7G and 0.76715$^\circ$ twisted angle setting ISP=5mV/\AA. E is an applied electrical ﬁeld and ISP is an internal symmetrical polarization due to the diﬀerent chemical environment of outer and inner atoms in septalayer graphene. The positive direction of E and ISP are shown in \figref{relaxation}. The structures corresponding to the first row to the fourth row are R7G/$\mathrm{hBN}_{\xi_b=1}$, $\mathrm{hBN}_{\xi_t=1}$/R7G/$\mathrm{hBN}_{\xi_b=1}$, R7G/$\mathrm{hBN}_{\xi_b=0}$ and $\mathrm{hBN}_{\xi_t=0}$/R7G/$\mathrm{hBN}_{\xi_b=1}$ respectively, and the applied electrical fields corresponding to the first column to the fifth column are -5, -2, 0, 2 and 5mV/\AA \ respectively. The band structures are  depicted with blue dotted lines for $K$ valley and red dotted lines for $K'$ valley. }
    \label{hBN-7Gr-isp5-0.77}
\end{figure}

To identify the valley quantum number of the energy bands in the momentum space, we solve the valley operator of hBN-R$n$G $\mathcal{V}_z$ in reciprocal space. Using the average value of $\mathcal{V}_z(k)$ under the eigenvectors of $H(k)$, we can distinguish the valley components of each state.
And in the low energy region, it satisfies $\bra{\psi(k)} \mathcal{V}_z(k)  \ket{\psi(k)} \simeq 1$ for $K$ valley and $\bra{\psi(k)} \mathcal{V}_z(k)  \ket{\psi(k)} \simeq -1$ for $K'$ valley. Based on the eigenvalue of each state, we can make a bipartition for bands with different valley components and obtain the valley-resolved bands. The calculation steps for valley-resolved bands are shown in \figref{fig:single_layer_valley}(f). The calculations are implemented in WannierTools package\cite{WU2018405}. The valley-resolved bands of hBN-R$n$G(n=3,4,5,6,7) are shown in \figsref{hBN-3Gr-isp5-0.77}, \ref{hBN-4Gr-isp5-0.77}, \ref{hBN-5Gr-isp5-0.77}, \ref{hBN-6Gr-isp5-0.77} and \ref{hBN-7Gr-isp5-0.77}.

\section{Parameter Fitting and Phase diagrams}
\label{app:phase_diagrams}

\label{app:fitting}

In this part, we discuss the fitting of the $2n\times 2n$ continuum model (\eqnref{eq:2n_model}) and the $2\times 2$ effective continuum model (\eqnref{eq:Heff_double}).
We will also discuss the single-particle phase diagram given by the $2n\times 2n$ continuum model.
Without loss of generality, we focus on the $\K$ valley.

We discuss the $2n\times 2n$ continuum model first.
From SK method, we can directly derive the parameter values for R$n$G ($n\geq 3$), which read
\eq{
\label{eq:RnG_para}
v_F= 542.1 \meV\cdot \nm\ ,\ t_1 = 355.16 \meV\ ,\ v_3 = 34.5 \meV\cdot \nm \ ,\ t_2 = -7 \meV \ .
}
Regarding to the moir\'e potential parameters in \eqnref{eq:VhBNbottom} and \Eq{eq:simpleform}, we adopt the values of $\psi_{\xi_b}$ and $\psi_{\xi_t}$ based on the perturbative expression in \eqnref{eq:psi_determination}, while treating $V_{b0}$, $V_{b1}$, $V_{t0}$ and $V_{t1}$ as independent parameters.

\begin{figure}
    \centering
    \includegraphics[width=1.0\columnwidth]{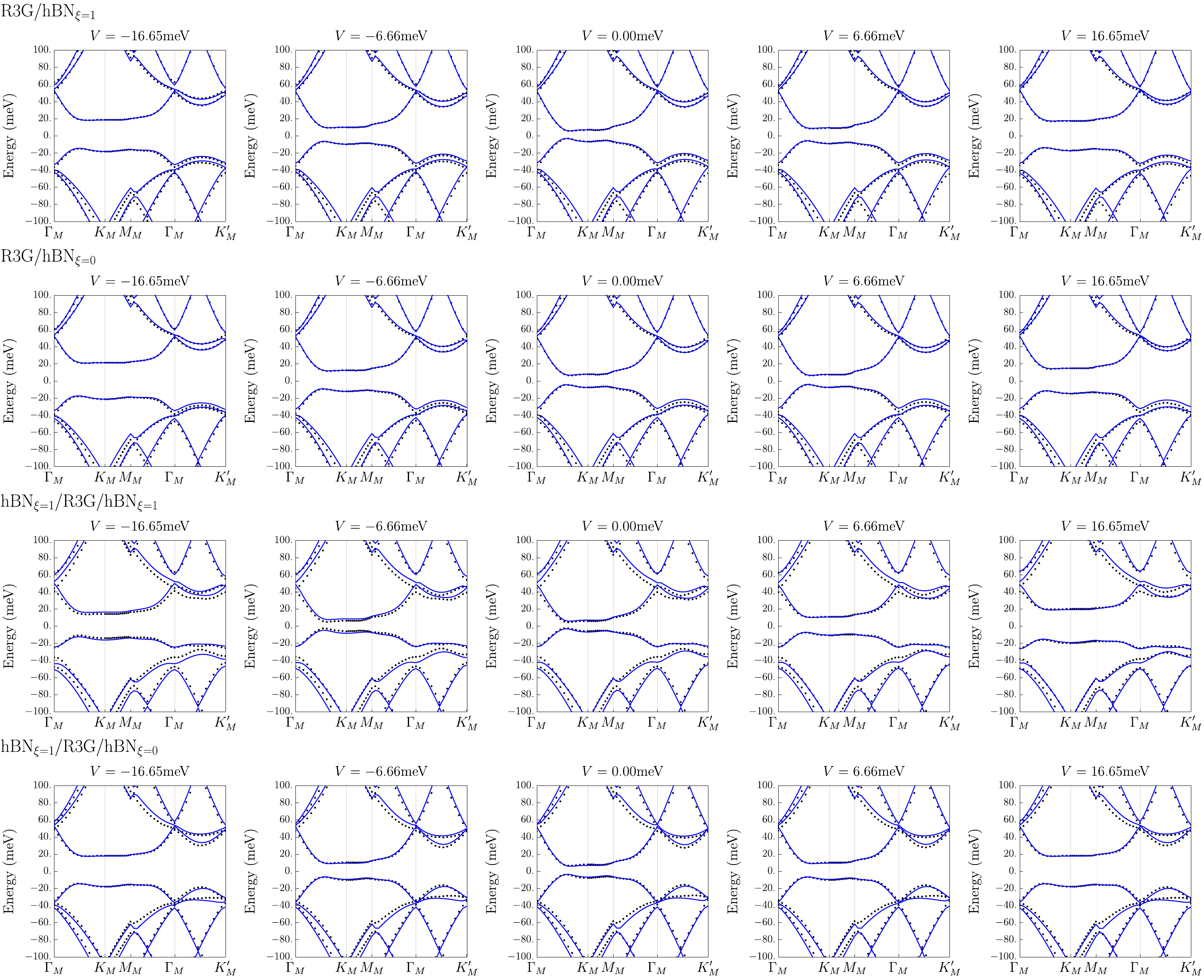}
    \caption{
    The comparison between the DFT+SK bands (black) and the bands from the $2n\times 2n$ continuum model (blue lines) in \eqnref{eq:2n_model} for $n=3$.
    The parameter values are listed in \tabref{tab:parameters_full} and \tabref{table:double_parameters} of Main Text.
    }
    \label{fig:fitting_3}
\end{figure}

\begin{figure}
    \centering
    \includegraphics[width=1.0\columnwidth]{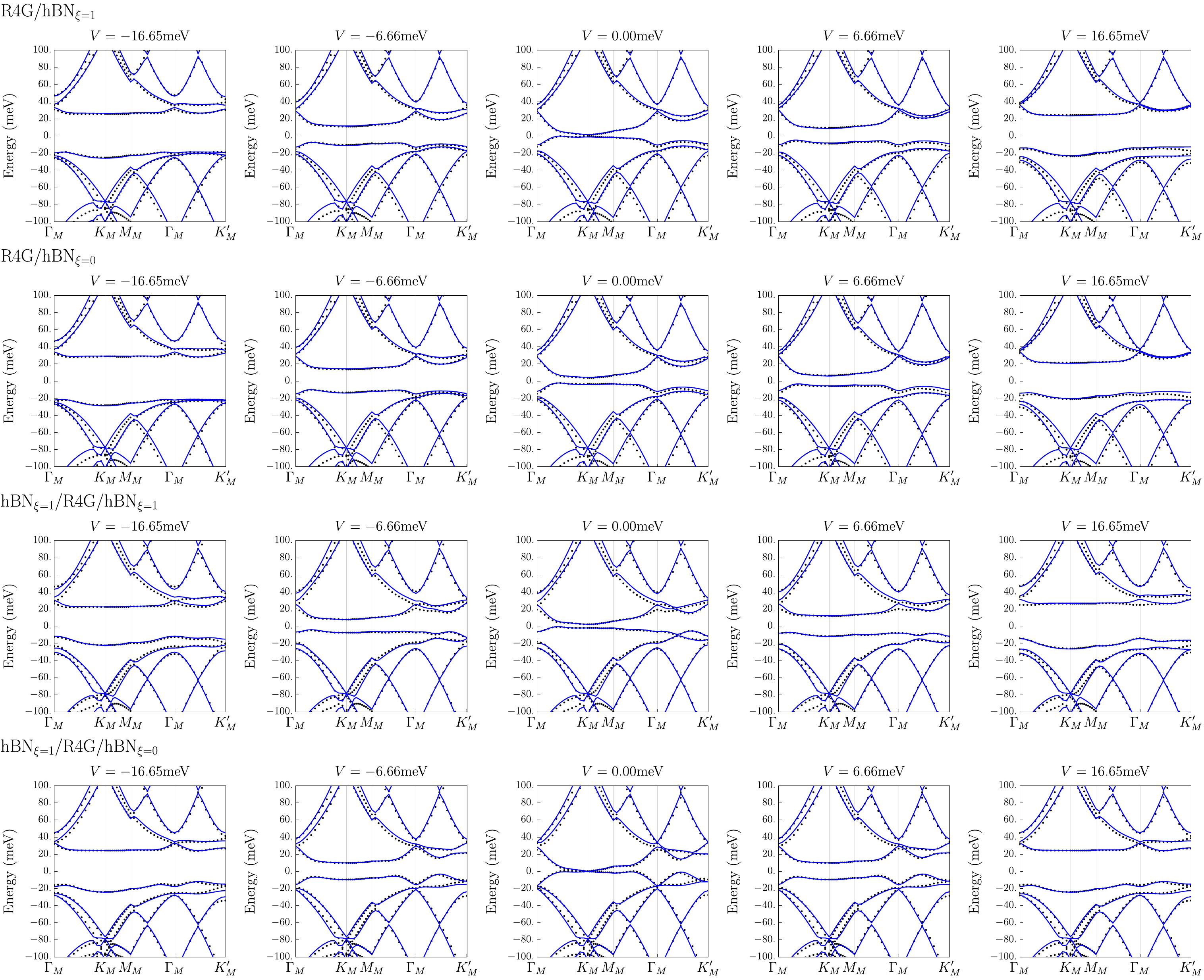}
    \caption{
    The comparison between the DFT+SK bands (black) and the bands from the $2n\times 2n$ continuum model (blue lines) in \eqnref{eq:2n_model} for $n=4$.
    The parameter values are listed in \tabref{tab:parameters_full} and \tabref{table:double_parameters} of Main Text.
    }
    \label{fig:fitting_4}
\end{figure}

\begin{figure}
    \centering
    \includegraphics[width=1.0\columnwidth]{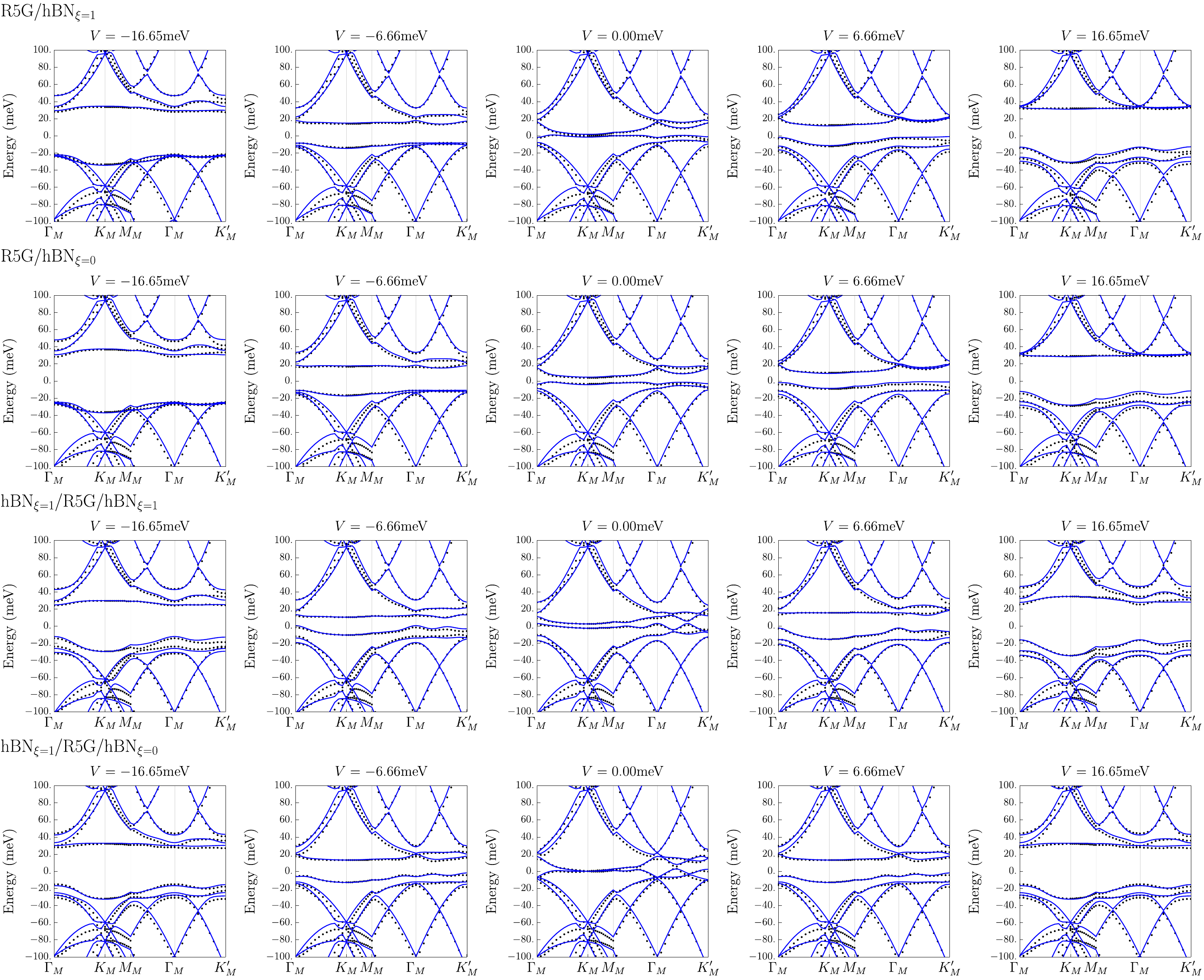}
   \caption{
    The comparison between the DFT+SK bands (black) and the bands from the $2n\times 2n$ continuum model (blue lines) in \eqnref{eq:2n_model} for $n=5$.
    The parameter values are listed in \tabref{tab:parameters_full} and \tabref{table:double_parameters} of Main Text.
    }
    \label{fig:fitting_5}
\end{figure}

\begin{figure}
    \centering
    \includegraphics[width=1.0\columnwidth]{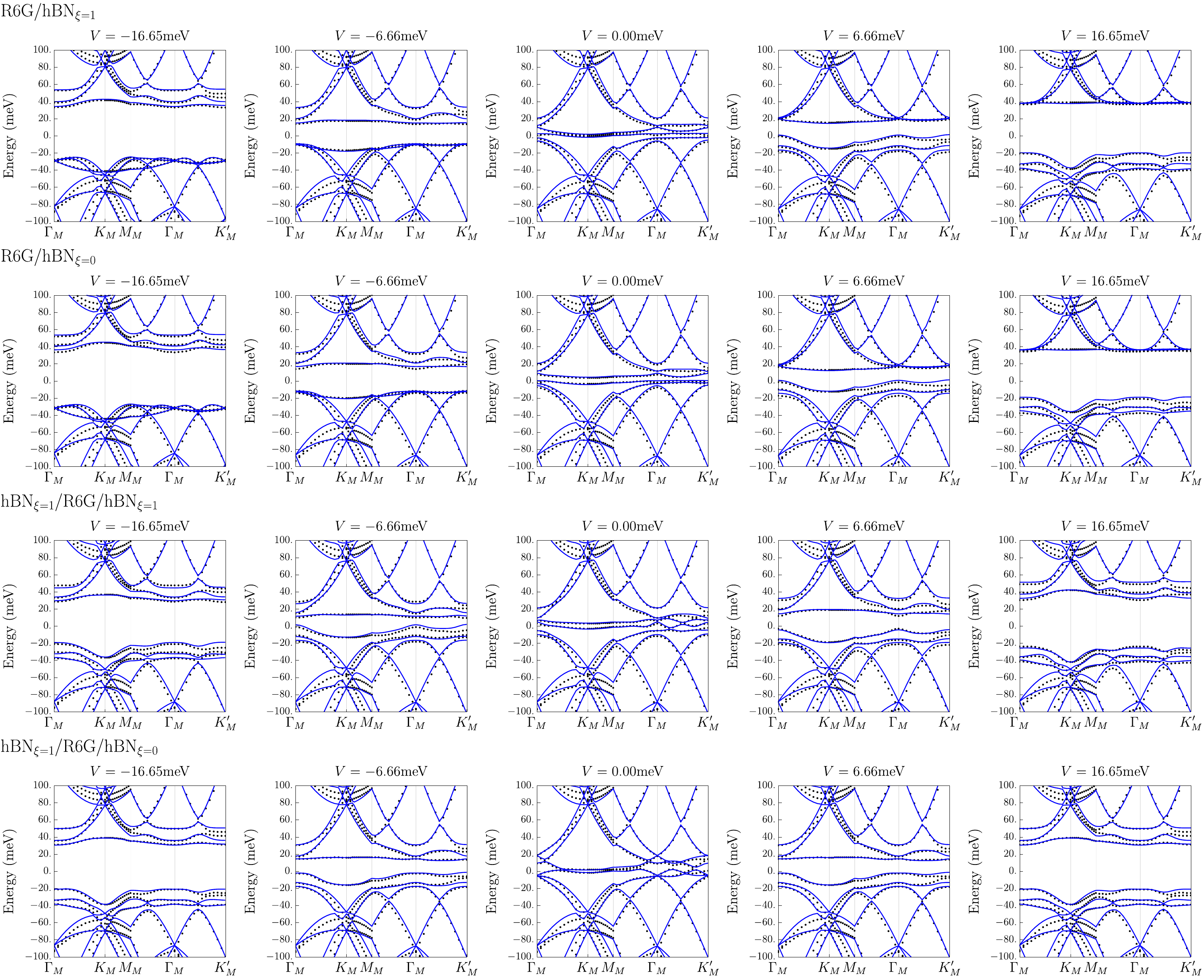}
    \caption{
    The comparison between the DFT+SK bands (black) and the bands from the $2n\times 2n$ continuum model (blue lines) in \eqnref{eq:2n_model} for $n=6$.
    The parameter values are listed in \tabref{tab:parameters_full} and \tabref{table:double_parameters} of Main Text.
    }
    \label{fig:fitting_6}
\end{figure}

\begin{figure}
    \centering
    \includegraphics[width=1.0\columnwidth]{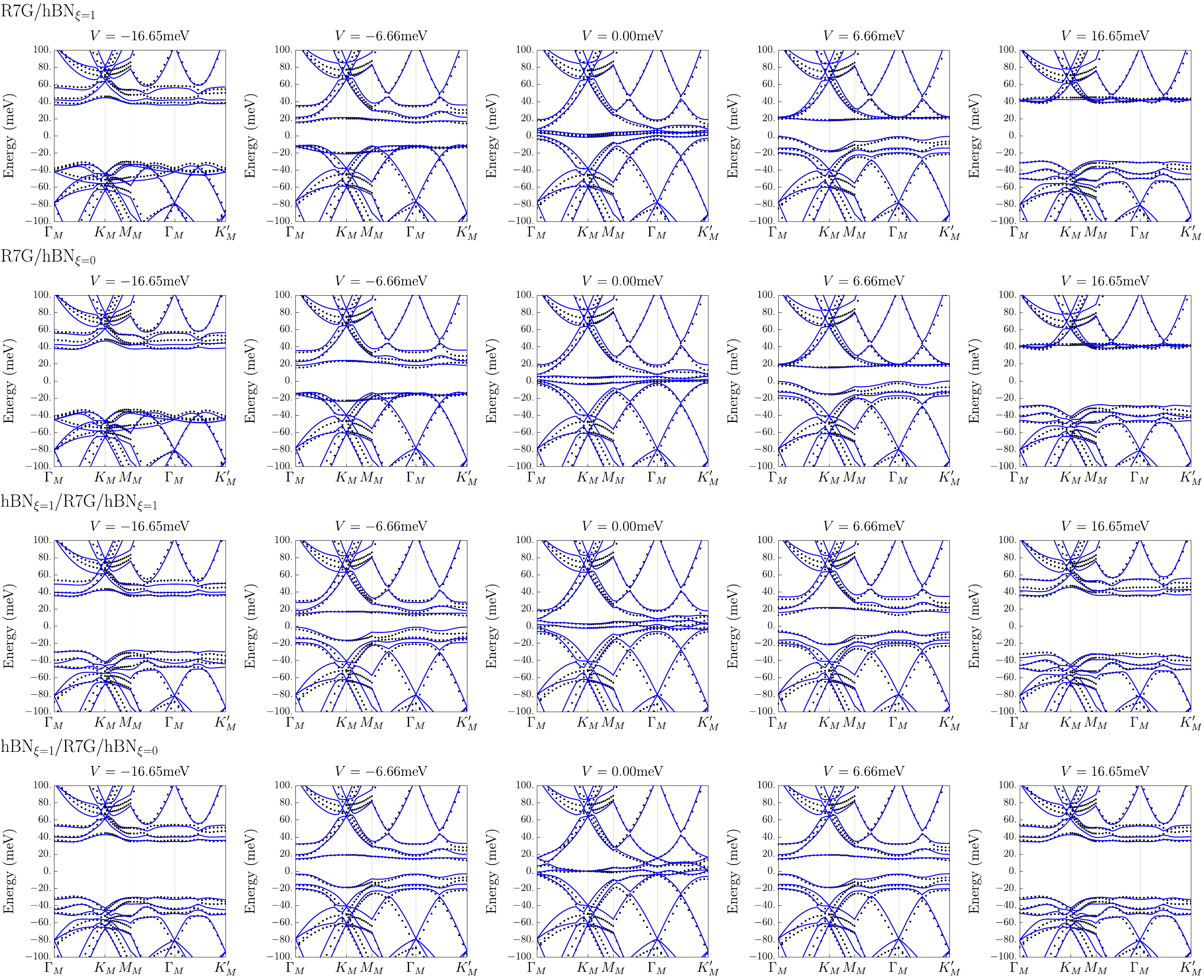}
    \caption{
    The comparison between the DFT+SK bands (black) and the bands from the $2n\times 2n$ continuum model (blue lines) in \eqnref{eq:2n_model} for $n=7$.
    The parameter values are listed in \tabref{tab:parameters_full} and \tabref{table:double_parameters} of Main Text.
    }
    \label{fig:fitting_7}
\end{figure}

For R$n$G/hBN structures, we fix $V_{t0}=0$ and $V_{t1}=0$, and fit  $V_{b0}$ and $V_{b1}$ to the DFT+SK band structure in the $\K$ valley; according to the effective model in \eqnref{eq:onlyA}, $|V_{b0}|$ (or $\sqrt{V_{b0}^2 + 4 t_2^2}$ for $n=3$) is just the gap of the bands at $\K_M$, which can be directly read out form the DFT+SK bands, and we adopt $V_{b0}>0$ owing to  \eqnref{eq:psi_determination}. (If the gap at $\K_M$ is smaller than $|2t_2|=14$meV, we just choose $V_{b0}=0$.)
Therefore, for single hBN, we effectively only need to optimize one parameter which is $V_{b1}$.
The parameter values after the optimization are listed in \tabref{tab:parameters_full}, where  $V_{b0}$,  $V_{b1}$ and  $\psi_{\xi_b}$ are labeled as $V_0$, $V_1$ and $\psi$, respectively, in short.
The fitting is remarkably good as shown in \figsref{fig:fitting_3}-\ref{fig:fitting_7}.

For hBN/R$n$G/hBN structures with configuration $(\xi_b,\xi_t)=(1,1)$, we have four tuning parameters $V_{b0}$, $V_{b1}$, $V_{t0}$ and $V_{t1}$.
According to the effective model in \eqnref{eq:onlyA}, $(V_{b0}+V_{t0})/2$ is just a total energy shift for low-energy bands, while $|V_{b0}-V_{t0}|$ (or $\sqrt{|V_{b0}-V_{t0}|^2 + 4 t_2^2}$ for $n=3$) corresponds to the gap at $\K_M$ in the $\K$ valley; thus, we just choose $V_{b0}$ to be the corresponding R$n$G/hBN value, and determine $V_{t0}$ from the gap.
(Here we choose $V_{t0}>V_{b0}$ since the opposite gives considerably worse fitting, and we choose $V_{t0}=V_{b0}$ for $n=3$ if the gap at $\K_M$ in the $\K$ valley is larger than $|2t_2|=14$meV.
Therefore, we need to optimize two parameters $V_{b1}$ and $V_{t1}$ in this case.
For double hBN with configuration $(\xi_b,\xi_t)=(1,0)$, the extra inversion symmetry requires  $V_{b0}=V_{t0}$ and $V_{b1}=V_{t1}$.
According to the effective model \eqnref{eq:onlyA},   $V_{b0}=V_{t0}$ is just effectively a total energy shift for low-energy bands, which we can just choose $V_{b0}$ to be the corresponding R$n$G/hBN value.
As a result, we only have one parameter to optimize which is $V_{b1}$.
The parameter values for double hBN after the optimization are listed in \tabref{table:double_parameters}, and the fitting is good as shown in \figsref{fig:fitting_3}-\ref{fig:fitting_7}.

\begin{figure}
    \centering
    \includegraphics[width=1.0\linewidth]{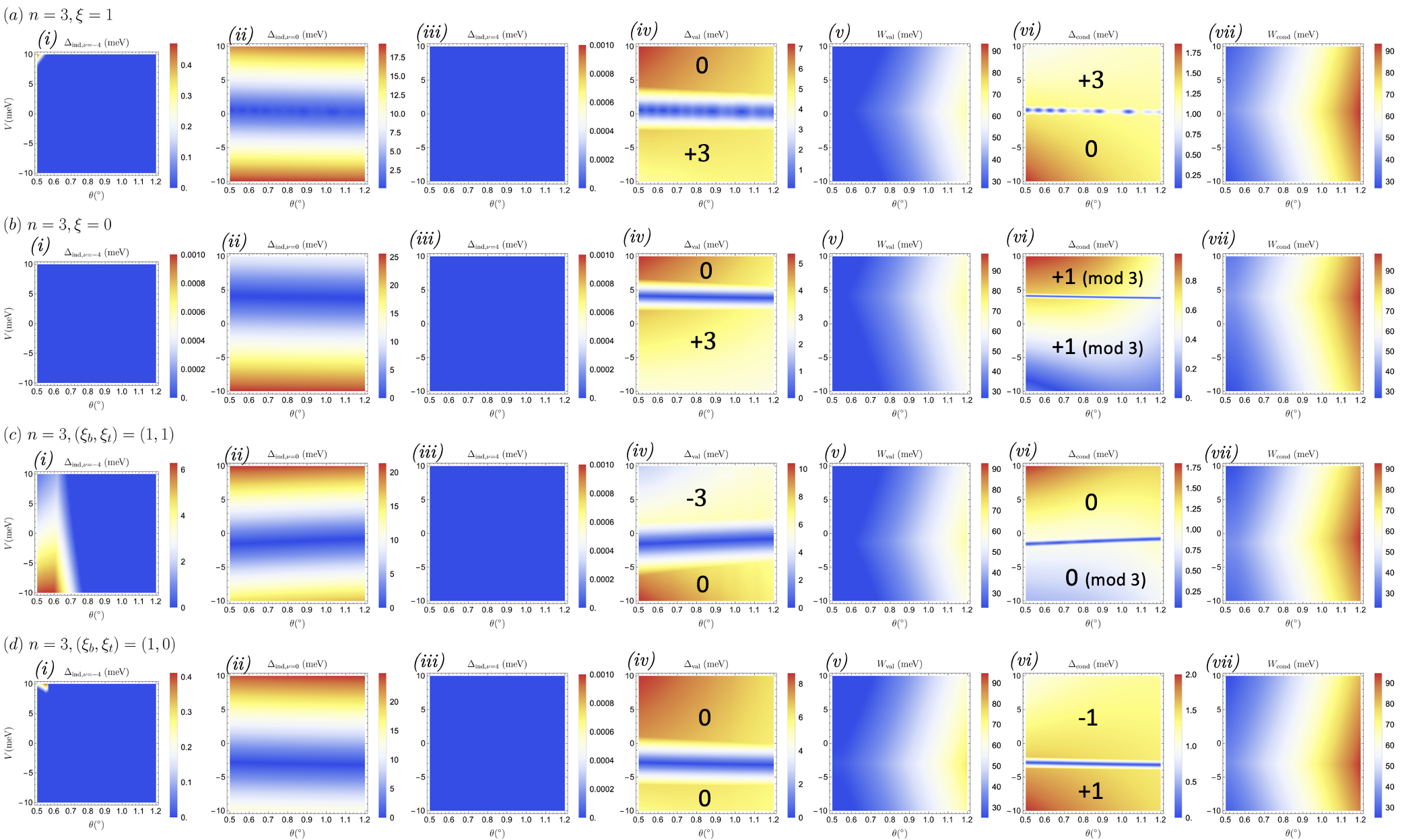}
    \caption{Phase diagrams of R3G-hBN superlattices of various stacking configurations: (a) $\xi=1$ , (b) $\xi=0$, (c) $(\xi_b,\xi_t)=(1,1)$, and (d) $(\xi_b,\xi_t)=(1,0)$. Panel (i) shows the single-particle indirect gap $\Delta_{\text{ind}, \nu=-4}$ at filling $\nu=-4$; (ii) shows the indirect gap $\Delta_{\text{ind}, \nu=0}$ at filling $\nu=0$; (iii) shows the indirect gap $\Delta_{\text{ind},\nu=4}$ at filling $\nu=4$; (iv) shows the minimal direct gap $\Delta_{\text{val}}$ around the highest valence band in one valley; (v) shows the bandwidth $W_{\text{val}}$ of the highest valence band; (vi) shows the minimal direct gap $\Delta_{\text{cond}}$ around the lowest conduction band in one valley; (vii) shows the bandwidth $W_{\text{cond}}$ of the lowest conduction band. Chern numbers of the highest valence band and the lowest conduction band in $\K$ valley are indicated on panel (iv) and panel (vi), respectively, where boundaries of topologically distinct phases can be seen as the direct gap closes. }
    \label{3LG_phasediag}
\end{figure}

\begin{figure}
    \centering
    \includegraphics[width=1.0\linewidth]{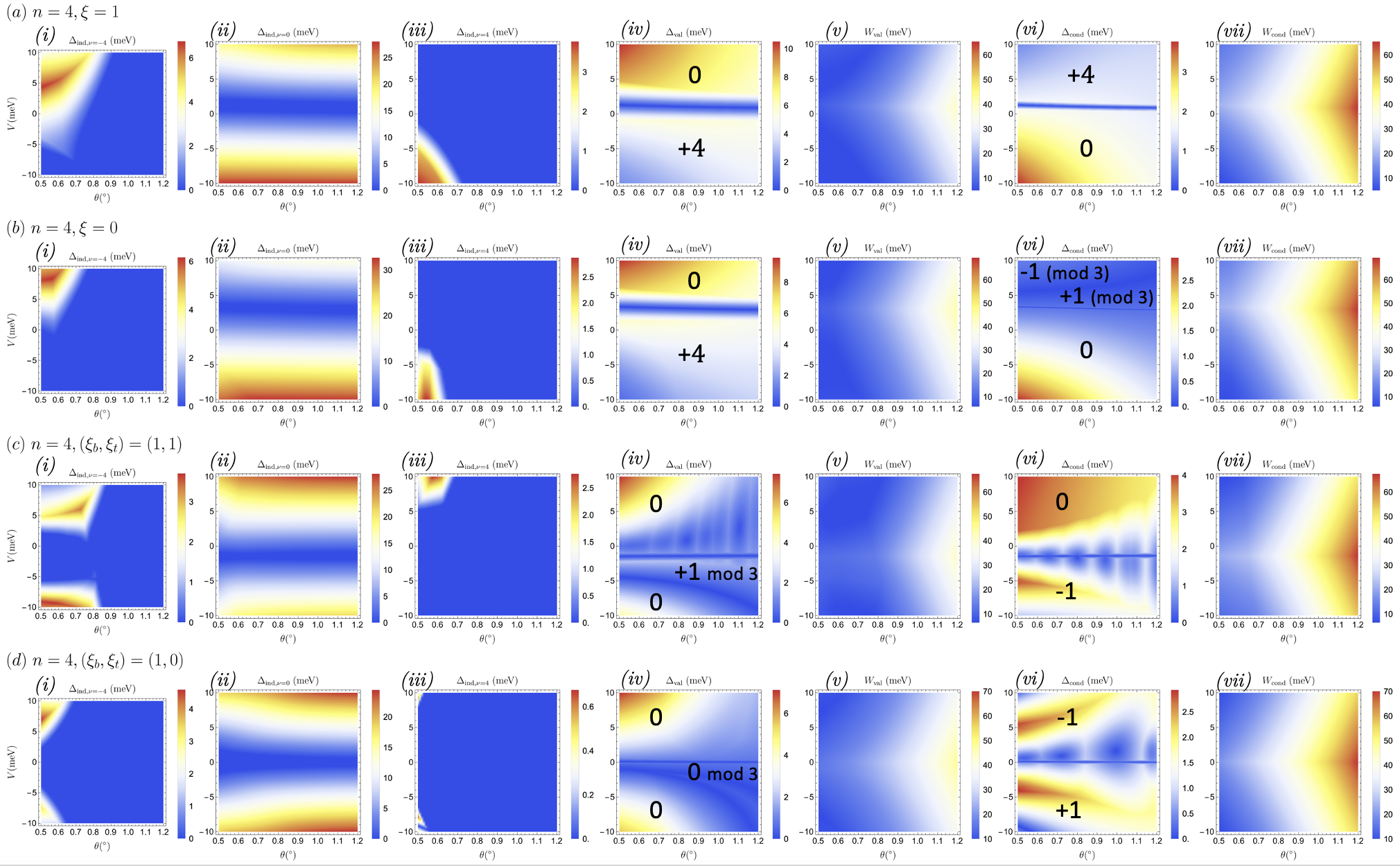}
     \caption{Phase diagrams of R4G-hBN superlattices of various stacking configurations: (a) $\xi=1$ , (b) $\xi=0$, (c) $(\xi_b,\xi_t)=(1,1)$, and (d) $(\xi_b,\xi_t)=(1,0)$. Panel (i) shows the single-particle indirect gap $\Delta_{\text{ind}, \nu=-4}$ at filling $\nu=-4$; (ii) shows the indirect gap $\Delta_{\text{ind}, \nu=0}$ at filling $\nu=0$; (iii) shows the indirect gap $\Delta_{\text{ind},\nu=4}$ at filling $\nu=4$; (iv) shows the minimal direct gap $\Delta_{\text{val}}$ around the highest valence band in one valley; (v) shows the bandwidth $W_{\text{val}}$ of the highest valence band; (vi) shows the minimal direct gap $\Delta_{\text{cond}}$ around the lowest conduction band in one valley; (vii) shows the bandwidth $W_{\text{cond}}$ of the lowest conduction band. Chern numbers of the highest valence band and the lowest conduction band in $\K$ valley are indicated on panel (iv) and panel (vi), respectively, where boundaries of topologically distinct phases can be seen as the direct gap closes. }
    \label{4LG_phasediag}
\end{figure}

\begin{figure}
    \centering
    \includegraphics[width=1.0\linewidth]{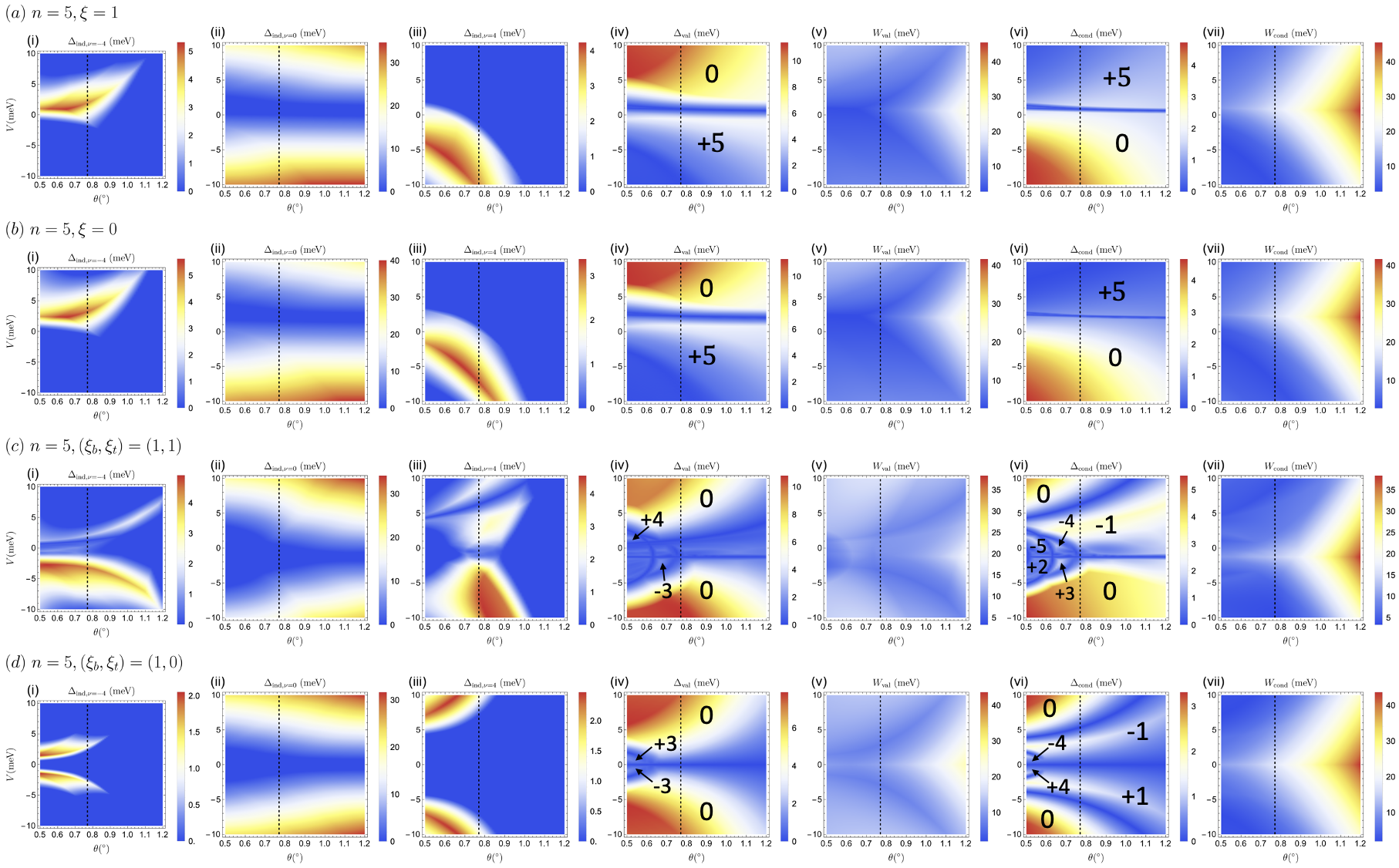}
     \caption{Phase diagrams of R5G-hBN superlattices of various stacking configurations: (a) $\xi=1$ , (b) $\xi=0$, (c) $(\xi_b,\xi_t)=(1,1)$, and (d) $(\xi_b,\xi_t)=(1,0)$. Panel (i) shows the single-particle indirect gap $\Delta_{\text{ind}, \nu=-4}$ at filling $\nu=-4$; (ii) shows the indirect gap $\Delta_{\text{ind}, \nu=0}$ at filling $\nu=0$; (iii) shows the indirect gap $\Delta_{\text{ind},\nu=4}$ at filling $\nu=4$; (iv) shows the minimal direct gap $\Delta_{\text{val}}$ around the highest valence band in one valley; (v) shows the bandwidth $W_{\text{val}}$ of the highest valence band; (vi) shows the minimal direct gap $\Delta_{\text{cond}}$ around the lowest conduction band in one valley; (vii) shows the bandwidth $W_{\text{cond}}$ of the lowest conduction band. Chern numbers of the highest valence band and the lowest conduction band in $\K$ valley are indicated on panel (iv) and panel (vi), respectively, where boundaries of topologically distinct phases can be seen as the direct gap closes.  }
    \label{5LG_phasediag}
\end{figure}

\begin{figure}
    \centering
    \includegraphics[width=1.0\linewidth]{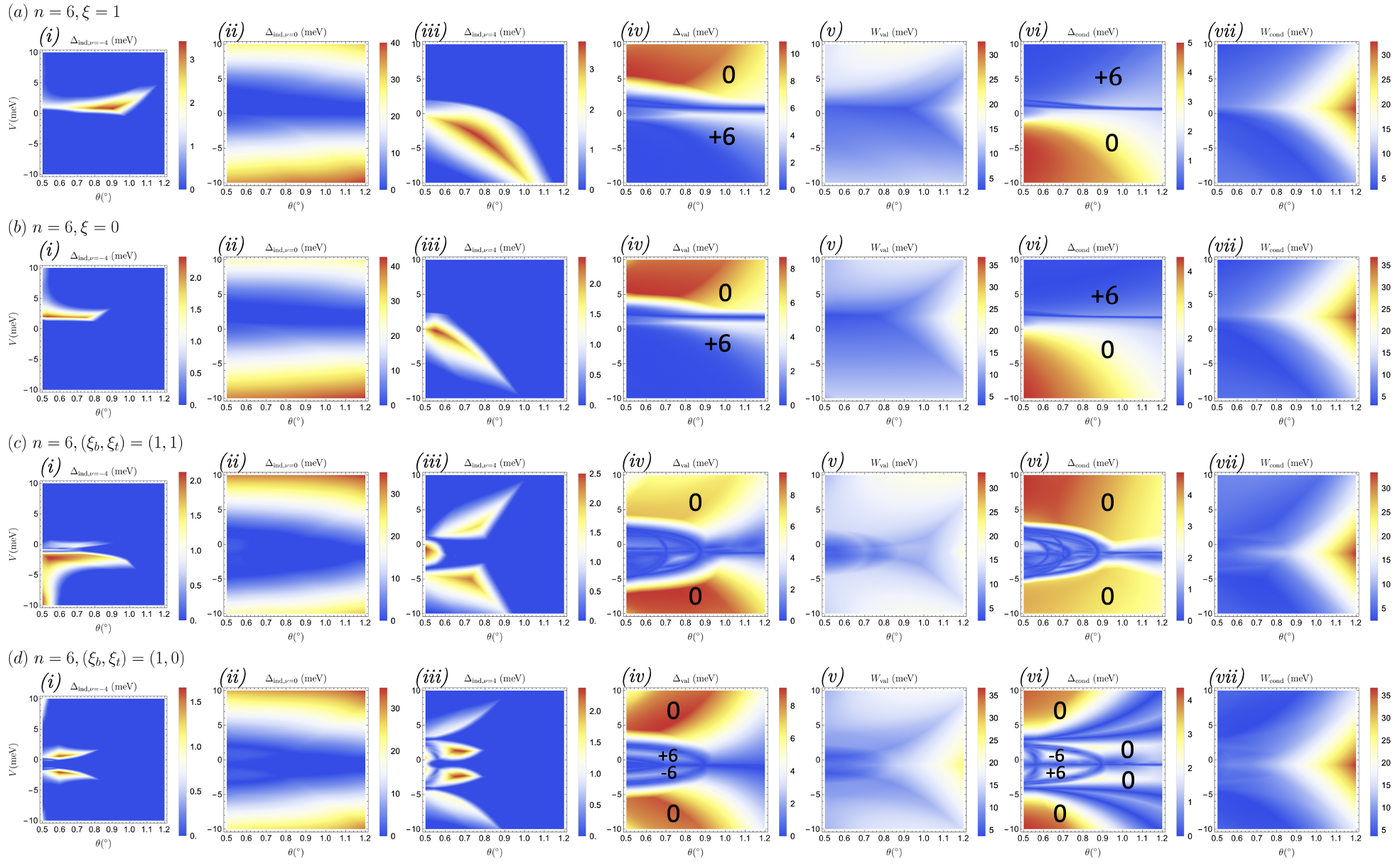}
     \caption{Phase diagrams of R6G-hBN superlattices of various stacking configurations: (a) $\xi=1$ , (b) $\xi=0$, (c) $(\xi_b,\xi_t)=(1,1)$, and (d) $(\xi_b,\xi_t)=(1,0)$. Panel (i) shows the single-particle indirect gap $\Delta_{\text{ind}, \nu=-4}$ at filling $\nu=-4$; (ii) shows the indirect gap $\Delta_{\text{ind}, \nu=0}$ at filling $\nu=0$; (iii) shows the indirect gap $\Delta_{\text{ind},\nu=4}$ at filling $\nu=4$; (iv) shows the minimal direct gap $\Delta_{\text{val}}$ around the highest valence band in one valley; (v) shows the bandwidth $W_{\text{val}}$ of the highest valence band; (vi) shows the minimal direct gap $\Delta_{\text{cond}}$ around the lowest conduction band in one valley; (vii) shows the bandwidth $W_{\text{cond}}$ of the lowest conduction band. Chern numbers of the highest valence band and the lowest conduction band in $\K$ valley are indicated on panel (iv) and panel (vi), respectively, where boundaries of topologically distinct phases can be seen as the direct gap closes.  }
    \label{6LG_phasediag}
\end{figure}

\begin{figure}
    \centering
    \includegraphics[width=1.0\linewidth]{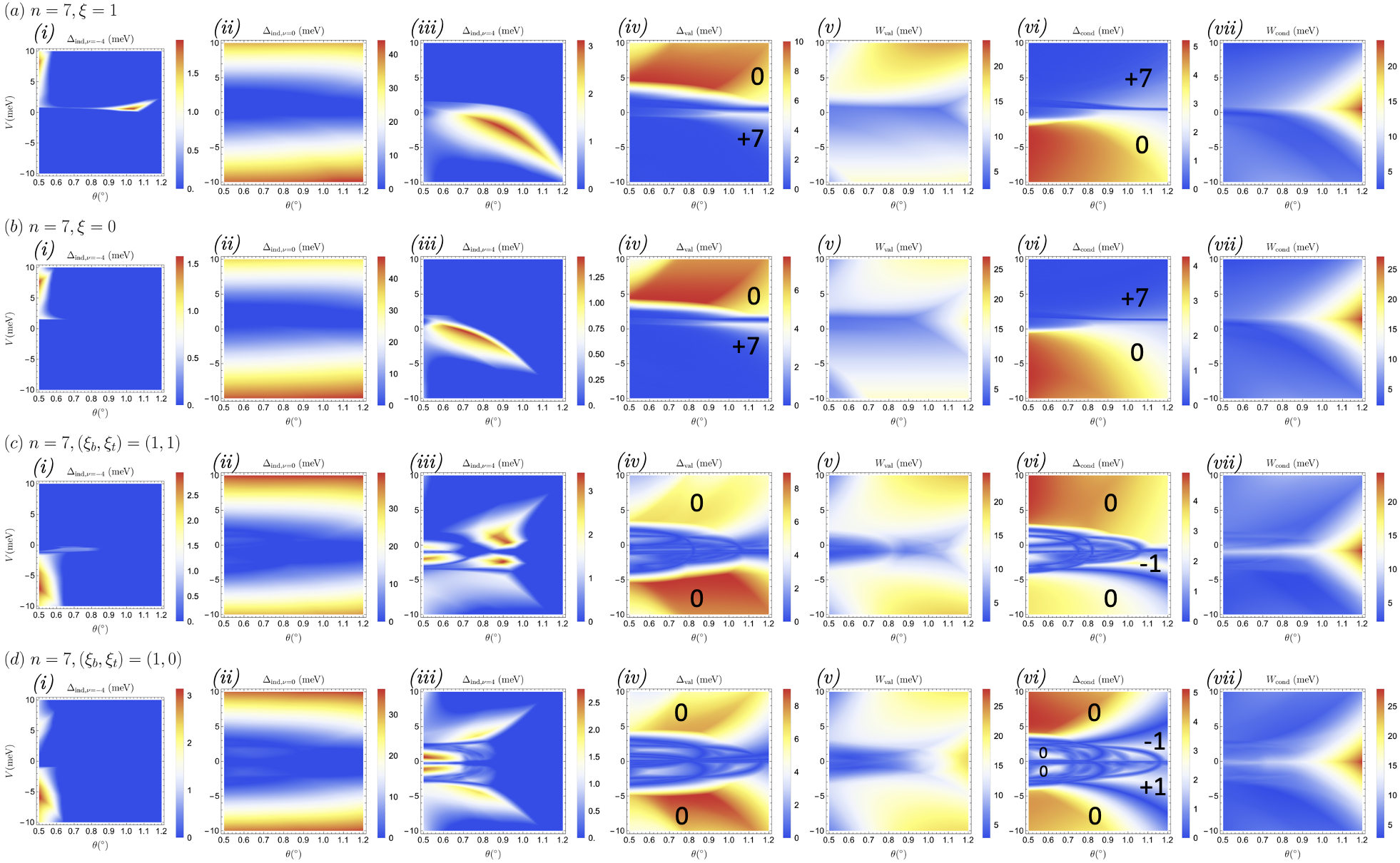}
     \caption{Phase diagrams of R7G-hBN superlattices of various stacking configurations: (a) $\xi=1$ , (b) $\xi=0$, (c) $(\xi_b,\xi_t)=(1,1)$, and (d) $(\xi_b,\xi_t)=(1,0)$. Panel (i) shows the single-particle indirect gap $\Delta_{\text{ind}, \nu=-4}$ at filling $\nu=-4$; (ii) shows the indirect gap $\Delta_{\text{ind}, \nu=0}$ at filling $\nu=0$; (iii) shows the indirect gap $\Delta_{\text{ind},\nu=4}$ at filling $\nu=4$; (iv) shows the minimal direct gap $\Delta_{\text{val}}$ around the highest valence band in one valley; (v) shows the bandwidth $W_{\text{val}}$ of the highest valence band; (vi) shows the minimal direct gap $\Delta_{\text{cond}}$ around the lowest conduction band in one valley; (vii) shows the bandwidth $W_{\text{cond}}$ of the lowest conduction band. Chern numbers of the highest valence band and the lowest conduction band in $\K$ valley are indicated on panel (iv) and panel (vi), respectively, where boundaries of topologically distinct phases can be seen as the direct gap closes.  }
    \label{7LG_phasediag}
\end{figure}

Based on the $2n\times 2n$ continuum model, we can plot the single-particle phase diagrams as a function of twisted angle $\theta$ and the inter-layer potential energy difference $V$ for all the configurations, as shown in \figsref{3LG_phasediag}-\ref{7LG_phasediag}.
In the phase diagrams, we only consider the Chern numbers of lowest conduction band and the highest valence band.
In sum, we find that the R$n$G/hBN configuration always achieve Chern numbers $0$ or $n$ when the direct gap of those bands is larger than 1meV, while $\pm1$ states can happen in the hBN/R$n$G/hBN configuration with considerable direct gap ($>$2meV).

\begin{figure}
    \centering
    \includegraphics[width=1.0\columnwidth]{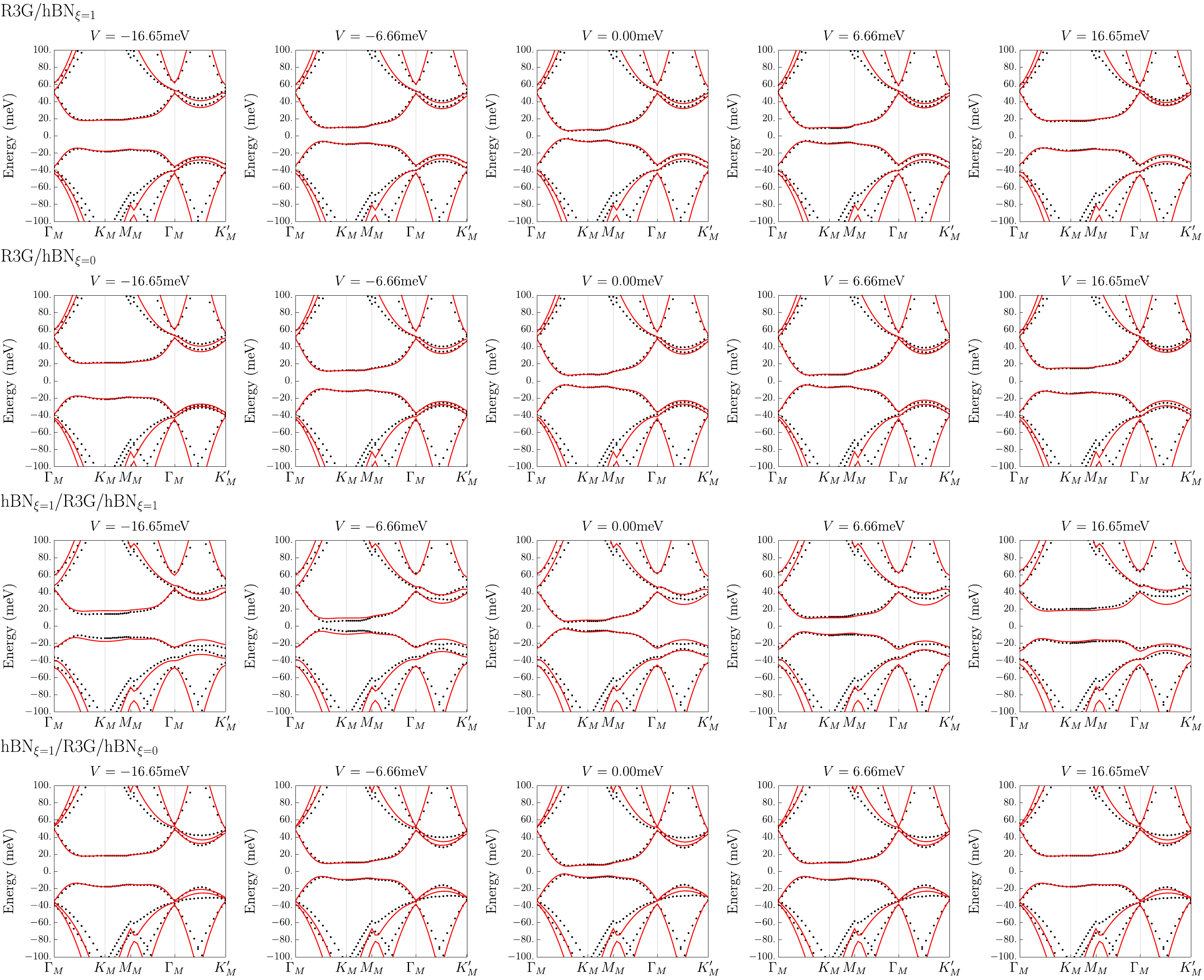}
    \caption{
    The comparison between the DFT+SK bands (black) and the bands from the $2\times 2$ effective continuum model (red lines) in \eqnref{eq:Heff_double} for $n=3$.
    The parameter values are listed in \tabref{tab:parameters_eff} and \tabref{tab:parameters_eff_double} of Main Text.
    }
    \label{fig:fitting_eff_3}
\end{figure}

We now turn to the $2\times 2$ effective continuum model in \eqnref{eq:Heff_double}.
In principle, all parameters in the $2\times 2$ effective model can be derived from the $2n\times 2n$ model in Eq.\,\ref{eq:2n_model}, as discussed in \appref{app:effmodelderivation}.
However, as shown in \figref{fig:main_eff} of the Main Text,  directly using the values derived from the perturbation theory can only partly well match the low-energy features, mainly around $\K_M$ point. (Recall that we focus on the $\K$ valley.)
To achieve a better match to the band structure, we treat $\{\alpha,\beta,\gamma, V_{1b}\}$  as tuning parameters for R$n$G/hBN structures ($\{\alpha,\beta,\gamma, V_{1b}, V_{1t}\}$ for hBN/R$n$G/hBN structures), and optimize them around their values derived from the perturbation theory, while preserving the form of the model.
Other parameters take the values derived from the $2n\times 2n$ continuum model.
With those fitting parameter values listed in \tabref{tab:parameters_full} and \tabref{tab:parameters_eff_double}, we can improve the matching away from $\K_M$ point, especially at $\Gamma_M$, and capture well the low-energy features of the DFT+SK bands, as shown in \figsref{fig:fitting_eff_3}-\ref{fig:fitting_7}.

\begin{table*}[t]
\centering
\begin{tabular}{c | cccc| cc| cc}
 & $\alpha$ & $\beta$ & $\gamma$ & $\delta$  & $V_{b0}$ & $V_{b1}$ & $V_{t0}$ & $V_{t1}$  \\ \hline
 \text{n=3, $(\xi _b,\xi _t)$=(1,1)} & \text{59.61(103.79)} & \text{864.58(1263.00)} & \text{81.70(120.10)} & -7.00 & 0 & \text{0.09(3.20)} & 0 & \text{12.03(11.09)} \\
 \text{n=4, $(\xi _b,\xi _t)$=(1,1)} & \text{55.21(103.79)} & \text{-1208.20(-1927.70)} & \text{-197.86(-287.42)} & 21.37 & 1.44 & \text{0.92(5.76)} & 5.40 & \text{8.87(7.08)} \\
 \text{n=5, $(\xi _b,\xi _t)$=(1,1)} & \text{49.37(103.79)} & \text{1879.20(2942.40)} & \text{405.76(597.60)} & -48.92 & 1.50 & \text{6.27(7.29)} & 6.48 & \text{6.46(7.91)} \\
 \text{n=6, $(\xi _b,\xi _t)$=(1,1)} & \text{34.99(103.79)} & \text{-3000.00(-4491.10)} & \text{-650.25(-1154.70)} & 99.57 & 1.56 & \text{5.70(6.02)} & 7.52 & \text{5.53(7.78)} \\
 \text{n=7, $(\xi _b,\xi _t)$=(1,1)} & \text{36.69(103.79)} & \text{4113.40(6855.10)} & \text{15.67(2132.70)} & -189.97 & 1.47 & \text{4.90(5.45)} & 5.79 & \text{5.49(7.79)} \\
 \hline
 \text{n=3, $(\xi _b,\xi _t)$=(1,0)} & \text{69.25(103.79)} & \text{870.13(1263.00)} & \text{92.72(120.10)} & -7.00 & 0 & \text{1.76(6.72)} & 0 & \text{1.76(6.72)} \\
 \text{n=4, $(\xi _b,\xi _t)$=(1,0)} & \text{58.73(103.79)} & \text{-1324.00(-1927.70)} & \text{-137.21(-287.42)} & 21.37 & 1.44 & \text{3.78(7.65)} & 1.44 & \text{3.78(7.65)} \\
 \text{n=5, $(\xi _b,\xi _t)$=(1,0)} & \text{52.95(103.79)} & \text{1914.20(2942.40)} & \text{414.48(597.60)} & -48.92 & 1.50 & \text{4.14(5.43)} & 1.50 & \text{4.14(5.43)} \\
 \text{n=6, $(\xi _b,\xi _t)$=(1,0)} & \text{53.64(103.79)} & \text{-3357.80(-4491.10)} & \text{-89.39(-1154.70)} & 99.57 & 1.56 & \text{4.78(7.80)} & 1.56 & \text{4.78(7.80)} \\
 \text{n=7, $(\xi _b,\xi _t)$=(1,0)} & \text{48.46(103.79)} & \text{4619.20(6855.10)} & \text{1876.40(2132.70)} & -189.97 & 1.47 & \text{5.11(7.22)} & 1.47 & \text{5.11(7.22)} \\
\end{tabular}
 \caption{Parameter values of the $2\times 2$ effective moir\'e model (\eqnref{eq:Heff_double}) for the hBN/R$n$G/hBN structure of $n=3,4,5,6,7$ layers. Here $\alpha,\beta,\gamma,\delta$ are reported in meV$\cdot$nm$^2$,meV$\cdot$nm$^n$, meV$\cdot$nm$^{n-1}$ and meV, respectively, while $V_{b0}$, $V_{b1}$, $V_{t0}$ and $V_{t1}$ are in meV. If the parameter values are difference from those directly derived from the perturbation theory of $2n$-band model, the latter will be provided in the brackets.
 The values of $\psi_t$ and $\psi_b$ are the same as those in Tab.\,\ref{table:double_parameters}; we omit those values here due to the length limit.
 }
    \label{tab:parameters_eff_double}
 \end{table*}


\begin{figure}
    \centering
    \includegraphics[width=1.0\columnwidth]{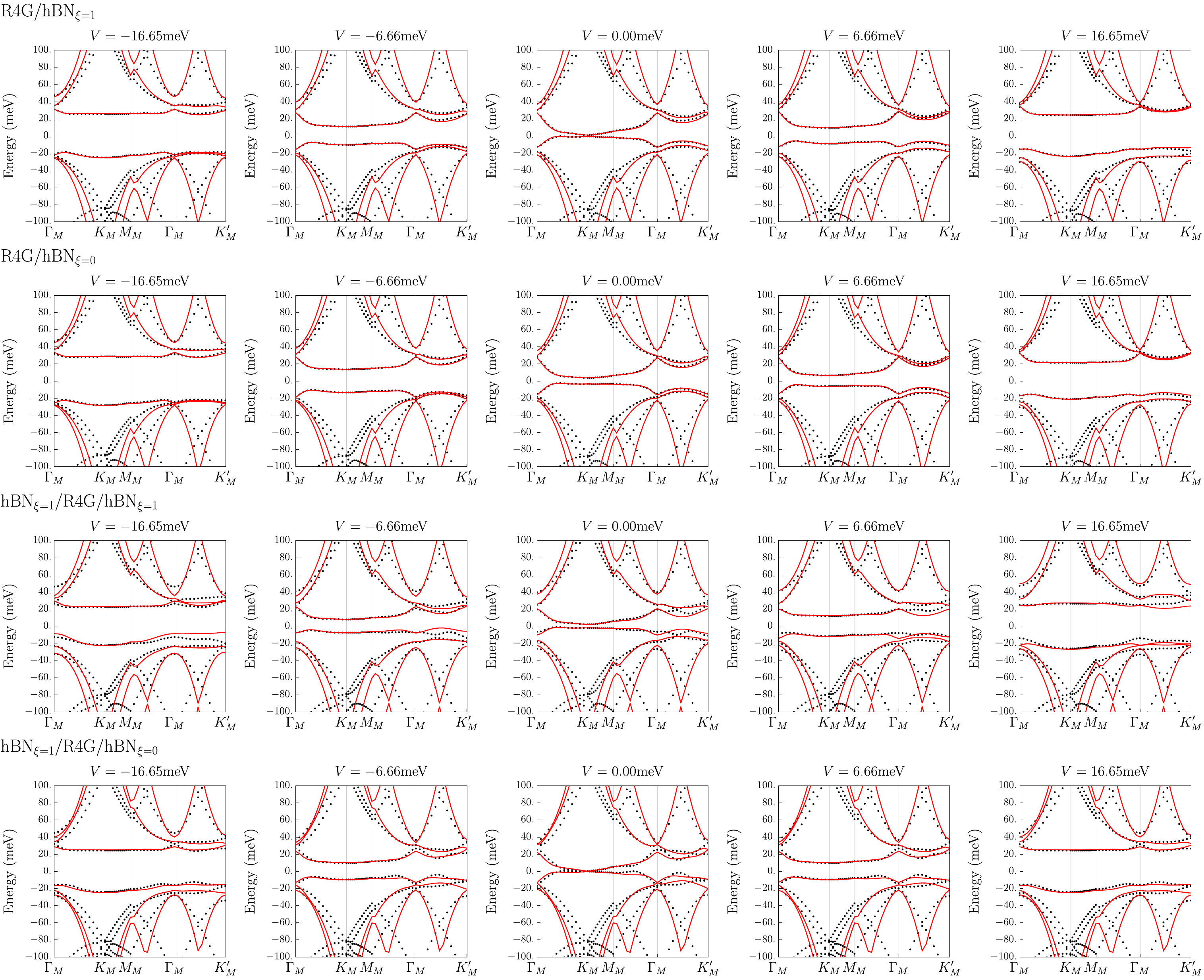}
    \caption{
    The comparison between the DFT+SK bands (black) and the bands from the $2\times 2$ effective continuum model (red lines) in \eqnref{eq:Heff_double} for $n=4$.
    The parameter values are listed in \tabref{tab:parameters_eff} and \tabref{tab:parameters_eff_double} of Main Text.
    }
    \label{fig:fitting_eff_4}
\end{figure}

\begin{figure}
    \centering
    \includegraphics[width=1.0\columnwidth]{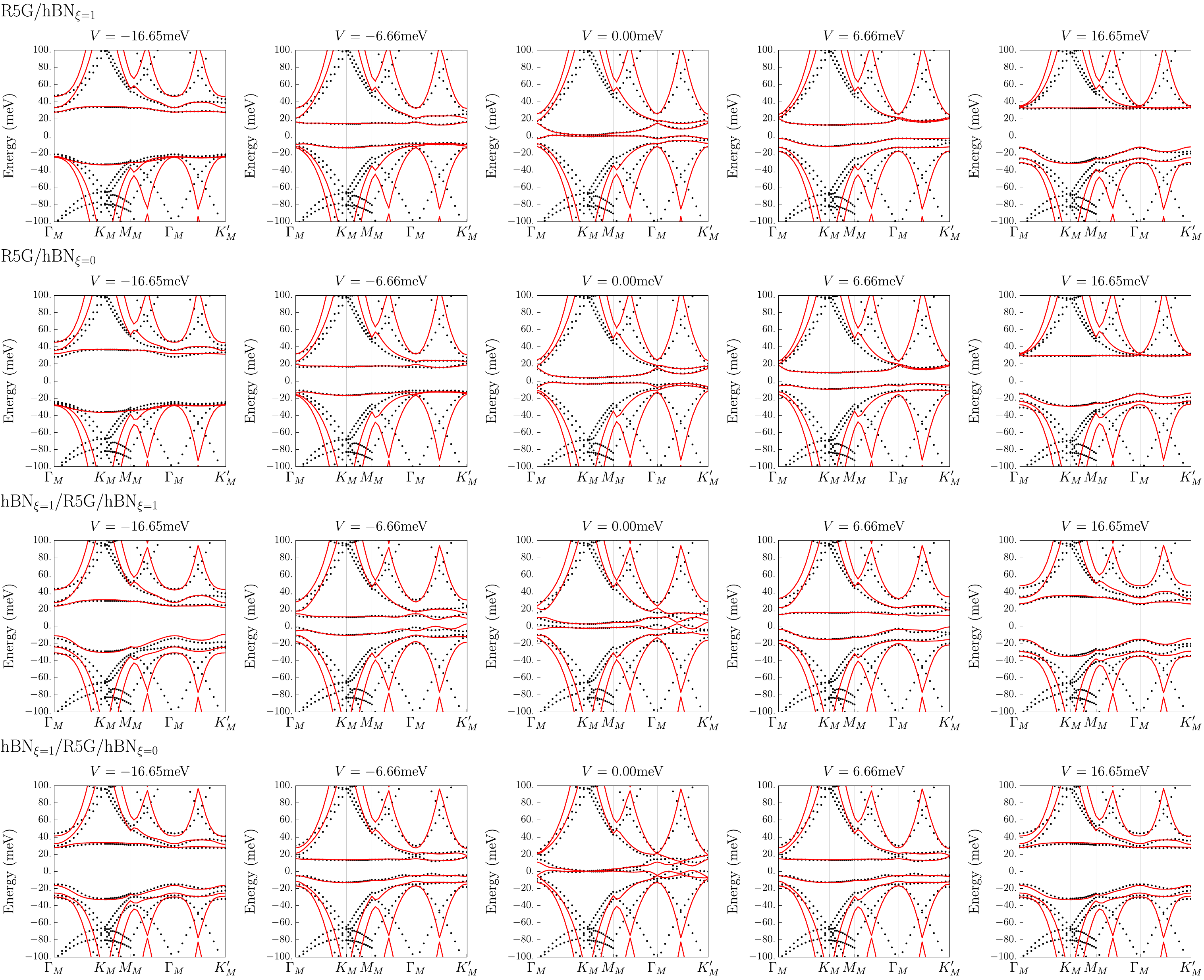}
    \caption{
    The comparison between the DFT+SK bands (black) and the bands from the $2\times 2$ effective continuum model (red lines) in \eqnref{eq:Heff_double} for $n=5$.
    The parameter values are listed in \tabref{tab:parameters_eff} and \tabref{tab:parameters_eff_double} of Main Text.
    }
    \label{fig:fitting_eff_5}
\end{figure}

\begin{figure}
    \centering
    \includegraphics[width=1.0\columnwidth]{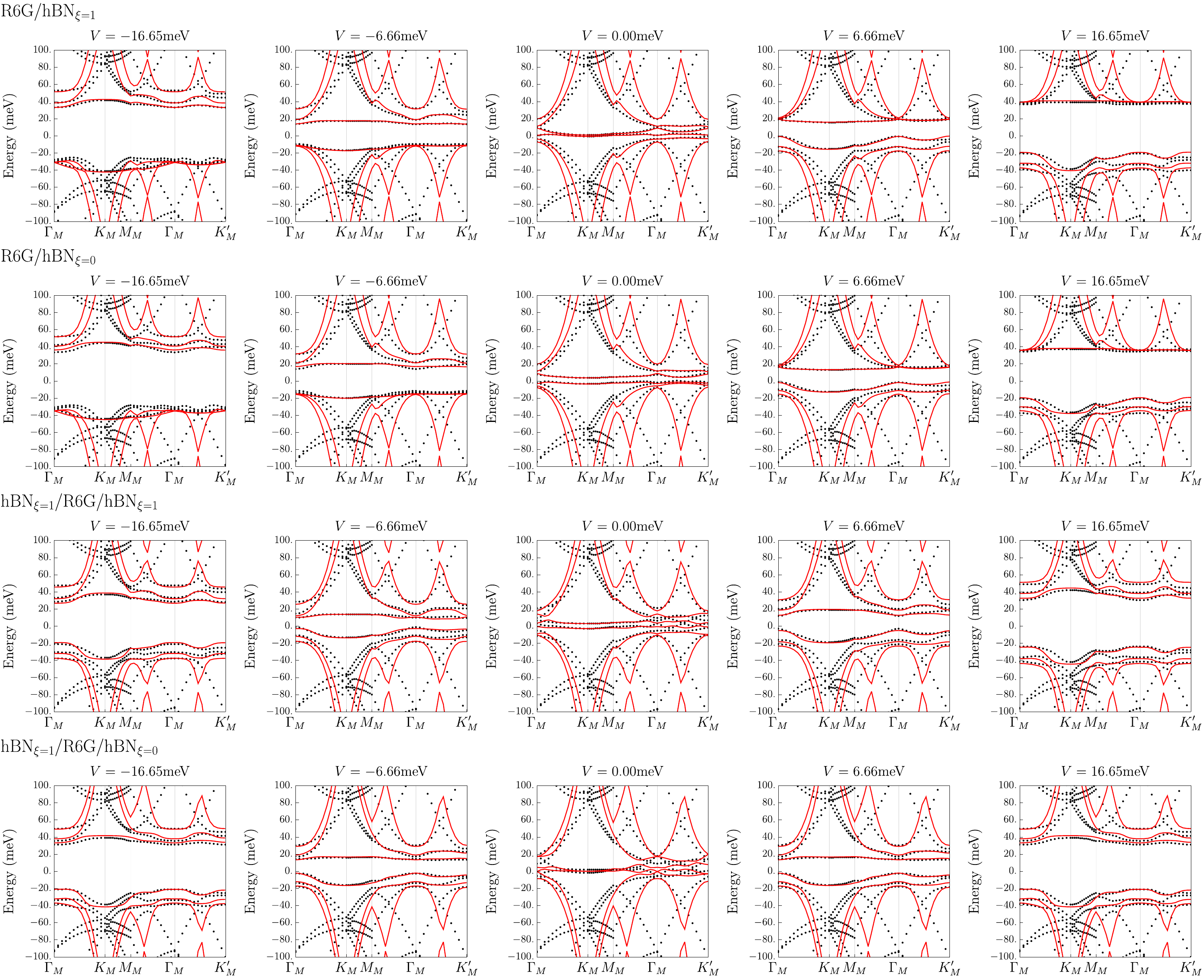}
    \caption{
    The comparison between the DFT+SK bands (black) and the bands from the $2\times 2$ effective continuum model (red lines) in \eqnref{eq:Heff_double} for $n=6$.
    The parameter values are listed in \tabref{tab:parameters_eff} and \tabref{tab:parameters_eff_double} of Main Text.
    }
    \label{fig:fitting_eff_6}
\end{figure}

\begin{figure}
    \centering
    \includegraphics[width=1.0\columnwidth]{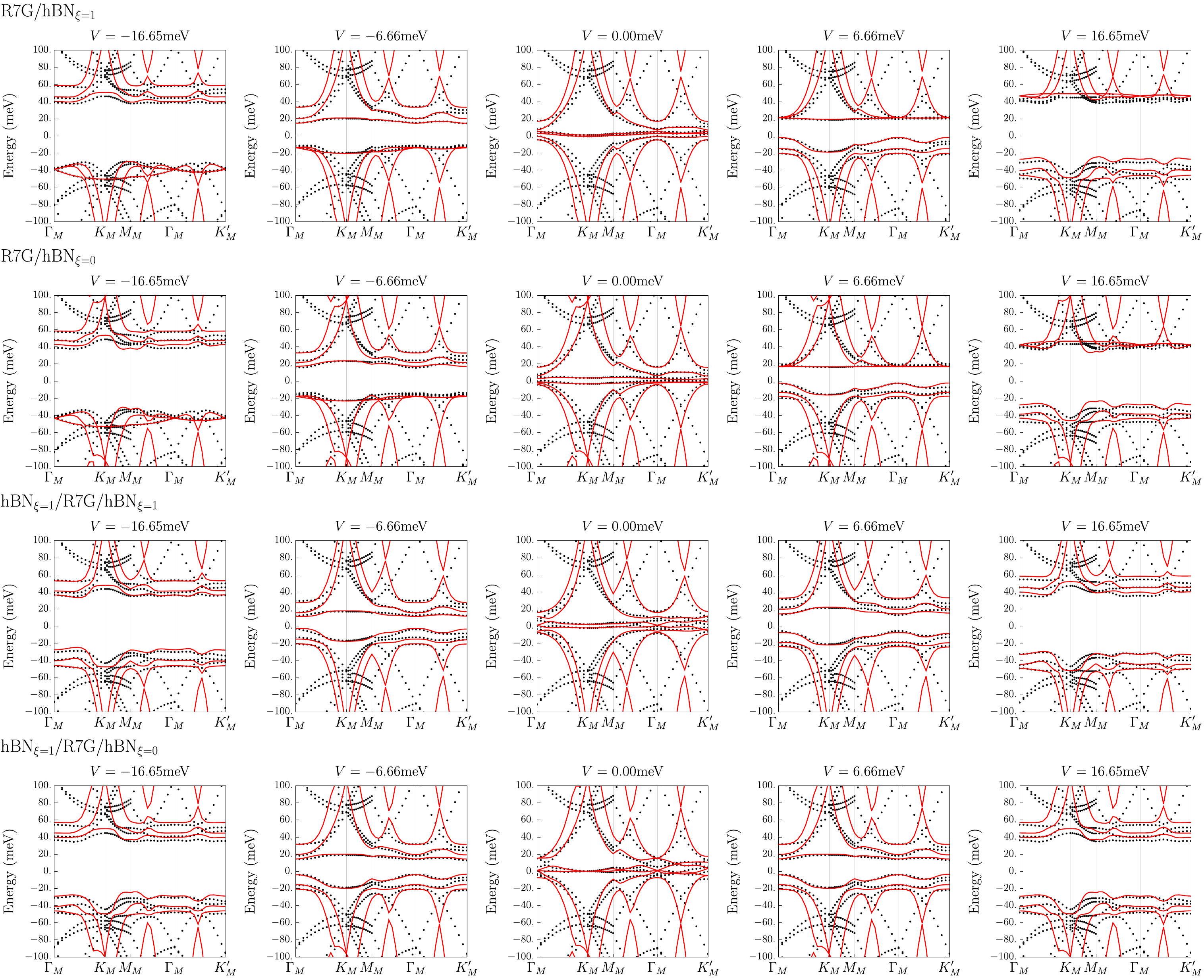}
    \caption{
    The comparison between the DFT+SK bands (black) and the bands from the $2\times 2$ effective continuum model (red lines) in \eqnref{eq:Heff_double} for $n=7$.
    The parameter values are listed in \tabref{tab:parameters_eff} and \tabref{tab:parameters_eff_double} of Main Text.
    }
    \label{fig:fitting_eff_7}
\end{figure}

\end{document}